\documentclass[a4paper,11pt]{article}
\pdfoutput=1 

\usepackage{jinstpub} 
\usepackage{xurl}
\usepackage{placeins}

\title{A Microwave Blackbody Target for Cosmic Microwave Background Spectral Measurements in the 10--20\,GHz range}

\author[a,b,f,1]{P. Alonso-Arias,\note{Corresponding author.}}
\author[c]{F. Cuttaia,}
\author[c]{L. Terenzi,}
\author[d]{A. Simonetto,}
\author[a,e]{P.~A. Fuerte-Rodríguez,}
\author[a]{R. Hoyland, }
\author[a,b]{and J.~A. Rubiño-Martín}

\affiliation[a]{Instituto de Astrofísica de Canarias,\\Calle Vía Láctea sn, ES38205 Santa Cruz de Tenerife, Spain}
\affiliation[b]{Universidad de La Laguna,\\Calle Padre Herrera sn, ES38205 Santa Cruz de Tenerife, Spain}
\affiliation[c]{Istituto Nazionale di Astrofisica,\\Via Piero Gobetti 93, 40129 Bologna , Italia}
\affiliation[d]{Consiglio Nazionale delle Ricerche - ISTP,\\Via R. Cozzi 53, 20125 Milano,Italia}
\affiliation[e]{European Southern Observatory, \\Karl-Schwarzschild Strasse 2, D-85748 Garching, Germany}
\affiliation[f]{European Organization for Nuclear Research (CERN), \\Espl. des Particules 1, 1211 Meyrin, Switzerland}

\emailAdd{paz.alonso.arias@cern.ch}

\abstract{The Tenerife Microwave Spectrometer (TMS) is a ground-based radio-spectrometer that will take absolute measurements of the sky between 10--20\,GHz. To ensure the sensitivity and immunity to systematic errors of these measurements, TMS includes an internal calibration system  optimised for the TMS band, and cooled down to 4\,K. It consists of an Aluminium core, composed of a baseplate and a bed of pyramidal elements 
coated with an absorber material and a metallic shield. The absorber coating is made of a commercial resin ECCOSORB CR/MF\,117. To achieve the high stability ($\mathrm{\pm 1\,mK/h}$), temperature homogeneity (thermal gradients $\mathrm{\Delta T\leq 25\,mK}$), and emissivity ($\mathrm{e\geq 0.999}$)  requirements of the reference unit, careful consideration has been given to the RF and thermal properties of the materials, as well as their geometry. In summary, this paper presents a comprehensive account of the design, characterisation, and test results of the TMS reference system.}


\begin{document}
\maketitle
\flushbottom

\section{Introduction}\label{sec:intro}

The spectral density of the Cosmic Microwave  Background (CMB) is a rich source of information on the thermal history of the early Universe. The CMB absolute spectrum was first measured by the COBE FIRAS experiment, \cite{fixen1994}, in the wavelength range from 1 to 95\,$\mathrm{cm^{-1}}$. These initial results showed the blackbody form of the CMB spectrum, but lacked the sensitivity needed to detect the spectral distortions  expected according to the $\mathrm{\Lambda}$CDM model, \cite{sunyaev2013,chluba2014}. The technological difficulty of obtaining measurements with sufficient accuracy ($\mathrm{\mu}$K variations) to detect these distortions led to a temporary halt in the characterisation of the absolute CMB energy spectrum. Most the CMB community effort in these years was devoted to the characterization of the spatial anisotropies (e.g.,  
\cite{planck2020}). 
Currently, there are some interesting space-based proposals, including  PIXIE \cite{kogut2011}, PRISM  \cite{andre2014} or the filter bank spectrometer devised for the ESA Voyage 2050 mission, \cite{Delabrouille2019}. However, results from space missions, if approved,  will not be available until at least by the mid-2030s. In the meantime,  new ground-based technological solutions can be implemented to pave the way, and to  observe at lower frequencies, allowing us to complete the spectral information of the CMB over a wider spectral range and providing a more robust description of the Galactic foregrounds. A first step in this direction is the Tenerife Microwave Spectrometer (TMS) \cite{spie:rubino2020}, a novel instrument to measure the sky spectra with great sensitivity in the unexplored band between 10--20\,GHz from the Teide Observatory.

The absolute temperature measurements to be performed by TMS require very high sensitivity and immunity to systematic errors. These features are provided by a receiver with a pseudo-correlation architecture, similar to that of the Low Frequency Instrument \cite{bersanelli2010} on board of the PLANCK satellite, \cite{tauber2010}. The pseudo-correlator continuously compares the signal received from the sky and a highly stable reference signal. In order to ensure this stability, for the TMS experiment we use an internal source mimicking the sky emission including the CMB component. Therefore, its brightness temperature must follow a blackbody curve depending solely on its physical temperature. In this way, the difference between the sky and the calibrator signals at the output of the radiometer shall be nearly zero and only the deviations from the blackbody curve are measured. 

The aim of the TMS calibrator, the 4\,K Cold Load (hereafter 4KCL), is to emit as an approximate blackbody in the frequency range of operation between 10--20\,GHz. The 4KCL will provide a stable signal that shall fill the Field Of View (FOV) of the antenna in the reference arm of the radiometer. In this paper, we present the process of design, characterisation, test and verification in the laboratory of the TMS 4KCL. This paper is organized as follows:  in Section\,\ref{sec:overview} we briefly describe the TMS instrument, with  special focus on the radiometer design and the cryogenic system. Section\,\ref{sec:prestudy} is devoted to the preliminary study of the calibrator subsystem, including the requirements, an analysis of the TMS feedhorn and the justification on the choice of materials and geometry. We divide the design formulation and simulation results under two headings depending on their nature: Section\,\ref{sec:rfdesign}, and \ref{sec:thsimulation} for radiofrequency (RF) and thermal simulations, respectively. Consequently, sections\,\ref{sec:EEM} and \ref{sec:measurements} gather the final characterisation of the load, derived from an equivalent emissivity model and the test campaign. This work finishes with the conclusions included in section\,\ref{sec:conclusions}.

\section{General overview of the instrument}\label{sec:overview}

\subsection{From scientific goals to technical requirements}\label{sec:sciencegoals}

The scientific goals of TMS have been already presented in \cite{spie:rubino2020}. TMS is intended mainly to measure the absolute sky spectrum in the 10--20\,GHz range, searching for possible deviations from a pure blackbody spectrum at the level of 10\,Jy/sr, which corresponds to the expected level of distortion signals due to reionisation and structure formation in the Universe, \cite{chluba2014}. Absolute measurements of the sky brightness temperature are badly affected by systematic effects. They require high stability, which can be achieved  by the continuous comparison of the sky signal with a well-known source in differential radiometer architectures. Artificial blackbodies --- very convenient because their brightness temperature and effective temperature match --- fill the reference feedhorn aperture and provide this calibration signal.
Thus, the TMS cold load shall be a representative blackbody between 10--20\,GHz with $\geq$0.999 emissivity, so that the measured temperature corresponds to 99.9\,\% of the physical temperature of the calibrator. With a $\geq$0.999 emissivity, we ensure an ideal absolute measurement capability of  $0.6$\,mK/6\,K or approximately 1\,mK, while the precision of a relative measurement of any deviation with respect to a perfect blackbody across the band will be at the level of few $\mu$K. Both absolute and relative measurements are of scientific interest, but the absolute measurement is by far the most difficult due to the local atmospheric and interference signals that will appear in the data. The  4KCL shall be held at a temperature between 5--10\,K (nominal  temperature of $\mathrm{T_{load}}=$6\,K). This value for the effective temperature corresponds to the expected temperature brightness of the sky in the TMS operating band, calculated as the sum of the CMB and atmospheric temperature contributions, 
\begin{equation}
    T_{\rm sky}(\nu) \approx T_{\rm CMB} (\nu) + T_{\rm atm}(\nu),
\end{equation}
and omitting the contribution of the galactic emission ($\mathrm{T_{\rm gal}(\nu)}$). The contribution of the atmospheric emission in the zenith $\mathrm{T_{\rm atm}(\nu)}$ has been generated using Paine's tool for atmospheric modelling (the \emph{}{am} model, \cite{paine_scott_2019}). It includes the expected $\mathrm{O_3}$ content over the Teide observatory, inferred from the $\mathrm{O_3}$ profiles measured in different observatories. We also assumed a value of 3.5\,mm of Precipitable Water Vapour (PWV), which is the median value for the Izaña Observatory, \cite{Otarola2018}.

In order to achieve its scientific objectives, TMS will allow two modes of operation to conduct two types of surveys, providing a QUIJOTE-like coverage of the sky \cite{QUIJOTE_SPIE2012}. We plan to conduct both wide surveys, with observations at constant elevation (typically 60\textdegree)  and deep surveys, through dedicated scan observations to explore certain sky areas. Observing campaigns will have a duration of 1 month, during which the load temperature shall be held within $\pm$100\,mK. Its absolute temperature shall be held stable within $\pm$1\,mK for observation intervals of 1 hour. In addition, an appropriate temperature measurement scheme shall be used to know the absolute temperature of this calibrator to $\pm$15\,mK.

\subsection{Description of the radiometer design}\label{sec:radiometer}

\begin{figure*}[ht]
\centering
\includegraphics[width=0.95\textwidth]{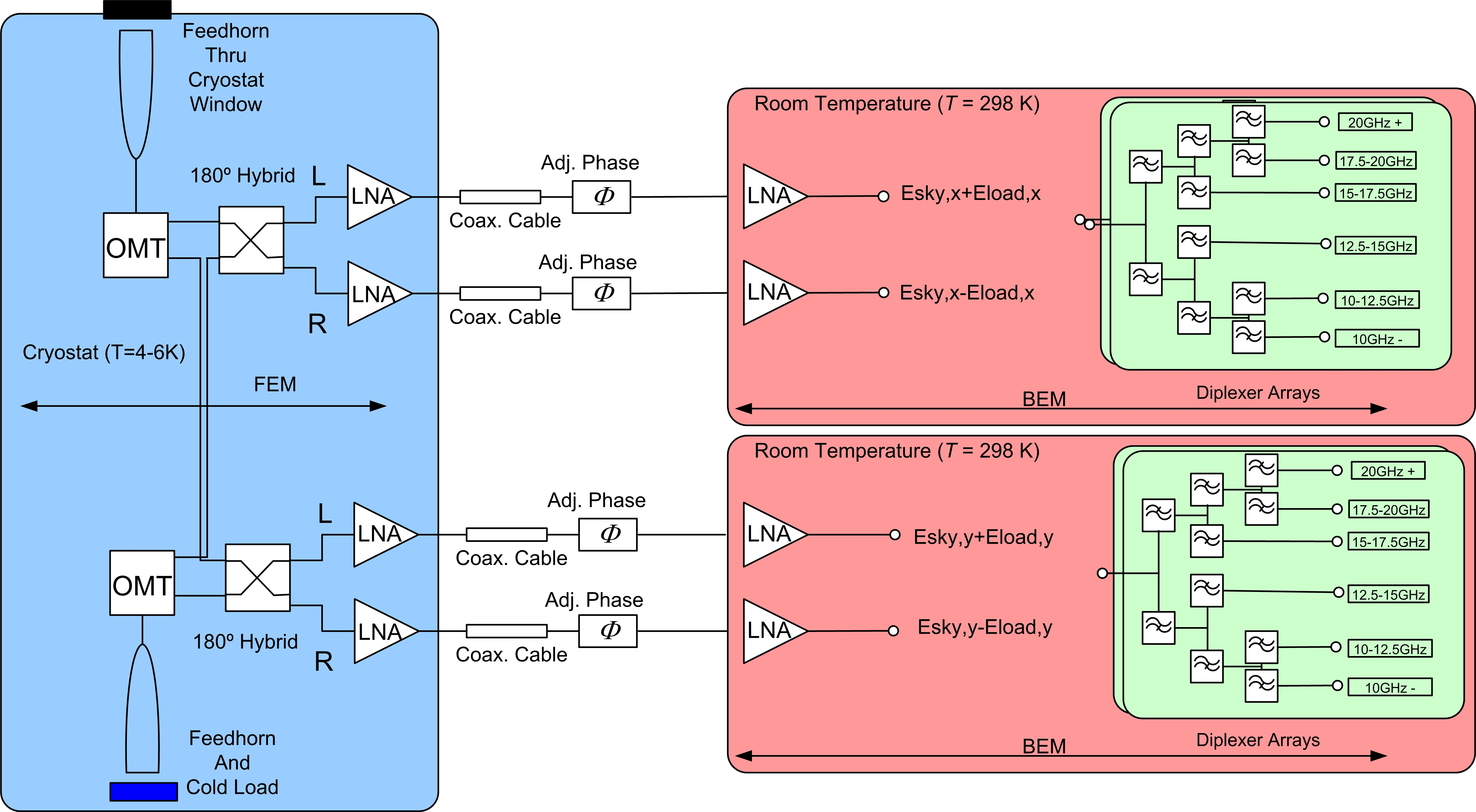}
\caption{The pseudo-correlation approach used for the design of the TMS radiometer. In \emph{blue} background, we represent the first stage, cooled to 4--10\,K, with the Front-End Modules. In \emph{pink}, we represent the Back-End Modules in the second stage, kept at room temperature. The down-conversion stage is not included in this representation. Figure taken from \cite{spie:rubino2020}.}
\label{fig:tms}
\end{figure*}

The  receiver is a pseudo-correlated microwave radiometer based on the Low Frequency Instrument \cite{bersanelli2010} on board of the PLANCK satellite \cite{tauber2010}, and followed by a spectrometer block to measure CMB spectra between 10--20\,GHz. In order to implement the spectrometer block, we use a fast data acquisition system,  discarding some microwave components used in the pseudo-correlation architecture, resulting in a simpler design. In the pseudo-correlation scheme, two wide-band corrugated feedhorns, one looking at a cold blackbody calibrator and the other out of the cryostat through a microwave-transparent window, capture the reference signal and the sky emission, respectively. Both sky and load signals are fed to either correlator arm after isolating two orthogonal polarisations with a broadband Orthogonal Mode Transducer (OMT). Thus two independent linear polarisations are measured, maximising the sensitivity of the instrument. The polar signals of both sources are correlated and decorrelated through broadband hybrid couplers before being fed to the spectrometer block. Within the correlator chain, the signals undergo several stages of amplification, as well as a filter and phase switch that ensures the needed robustness against 1/f gain variations. In the TMS, the switching scheme has been replaced by a digital acquisition using high speed ADCs and FPGAs, making the receiver robust against gain drifts once the signal has been digitised. In addition, the second 90\textdegree\,hybrid coupler- which is part of the correlation strategy --- is digitally implemented in the FPGA module. In order to digitise the coupled signals from both correlator arms, the full band has to be divided into subbands of $\leq$2.5\,GHz and then down-converted to baseband. Figure\,\ref{fig:tms} shows the schematic diagram of the radiometer design for the TMS instrument. 

Regarding stability, the performance of the instrument depends to a large extent on the load and sky temperatures and the symmetry of the design. Theoretically, a pseudo-correlation scheme is completely insensitive to 1/f noise if it presents a perfect balance between the levels of the sky and reference signal; in practice, the design will always present a slight offset between them. Although \cite{seiffert2002} demonstrate that by introducing a proper gain modulation factor the sensitivity becomes independent from the reference signal level, in general minimising the imbalance between sky and reference temperatures is a critical instrument requirement for measuring absolute temperature.

\subsection{Description of the cryostat}\label{sec:cryostat}

\begin{figure}[ht]
\centering
\subfloat{\includegraphics[width=0.5\textwidth]{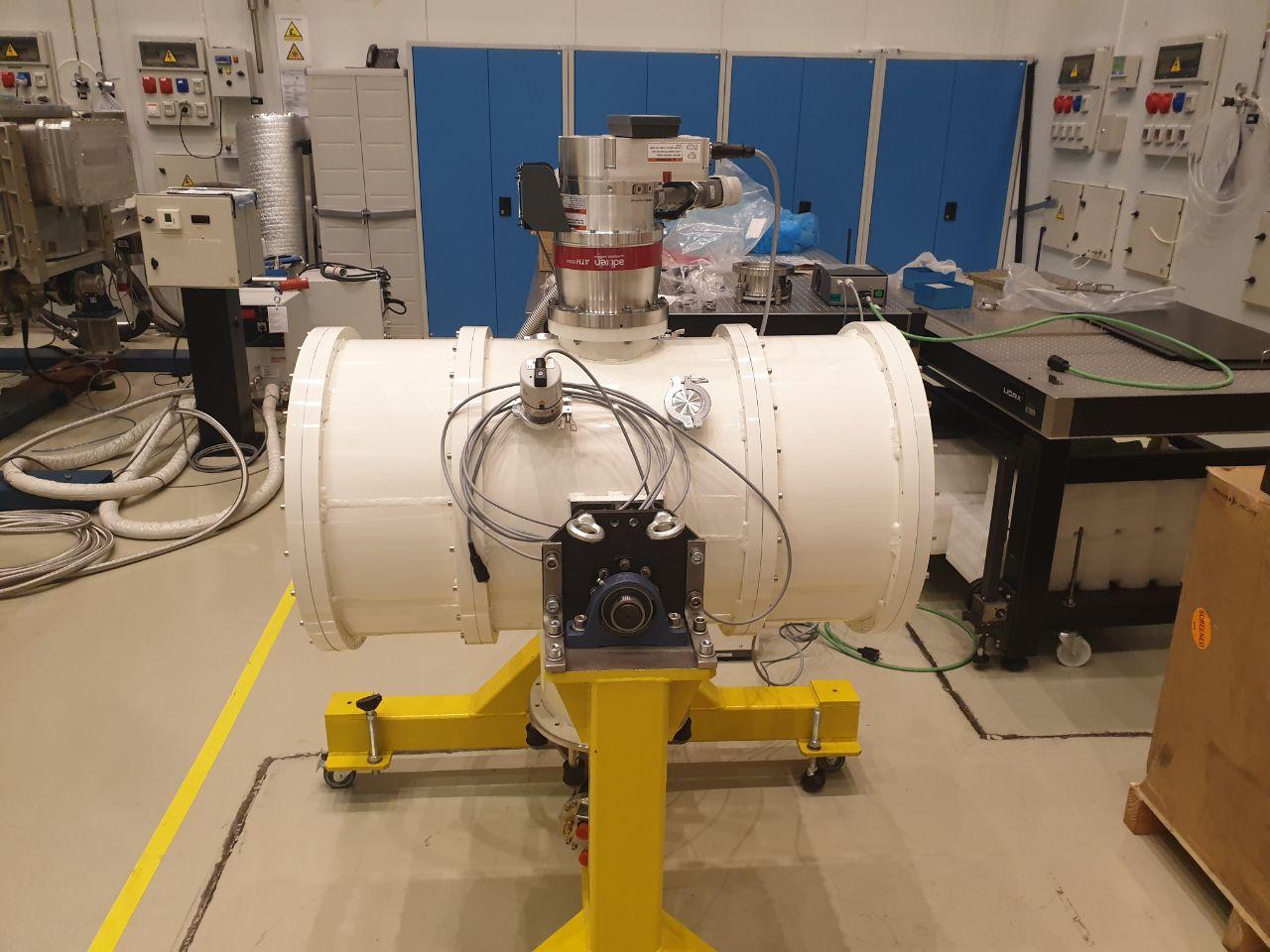}}
\subfloat{\includegraphics[width=0.4\textwidth]{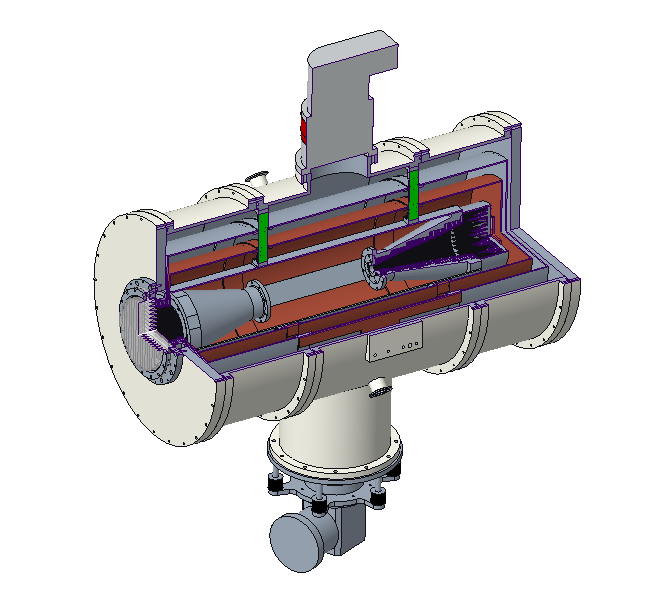}}
\caption{\emph{Left:} The cryostat subsystem of the TMS. The tests on pumping and thermal performance have already been performed at the IAC facilities.   An  ultimate  pressure  of  $\mathrm{5\times 10^{-7}\,mbar}$  is  reached  after approx.  100 min. Image taken from \cite{alonsoarias2020}. \emph{Right:} CAD 3D representation of the TMS cryostat including the window, the sky feedhorn and the reference feedhorn mated to the 4KCL.}
\label{fig:tms-cryostat}
\end{figure}

In order to achieve the sensitivity needed to reach the level of $\mathrm{10^{-25}\,W\,m^{-2}\,Hz^{-1}\,sr^{-1}}$ in subbands of 1\,GHz, \cite{spie:rubino2020}, the radiometer is cooled by a closed cycle helium cryostat between 4--10\,K. The spectrometer is mounted on a simple telescope mount capable of keeping the cryostat at a given declination angle while scanning through the azimuthal axis. Figure\,\ref{fig:tms-cryostat} shows the assembled cryostat, currently available at the IAC facilities.

\begin{figure}
    \centering
    \includegraphics[width=0.8\textwidth]{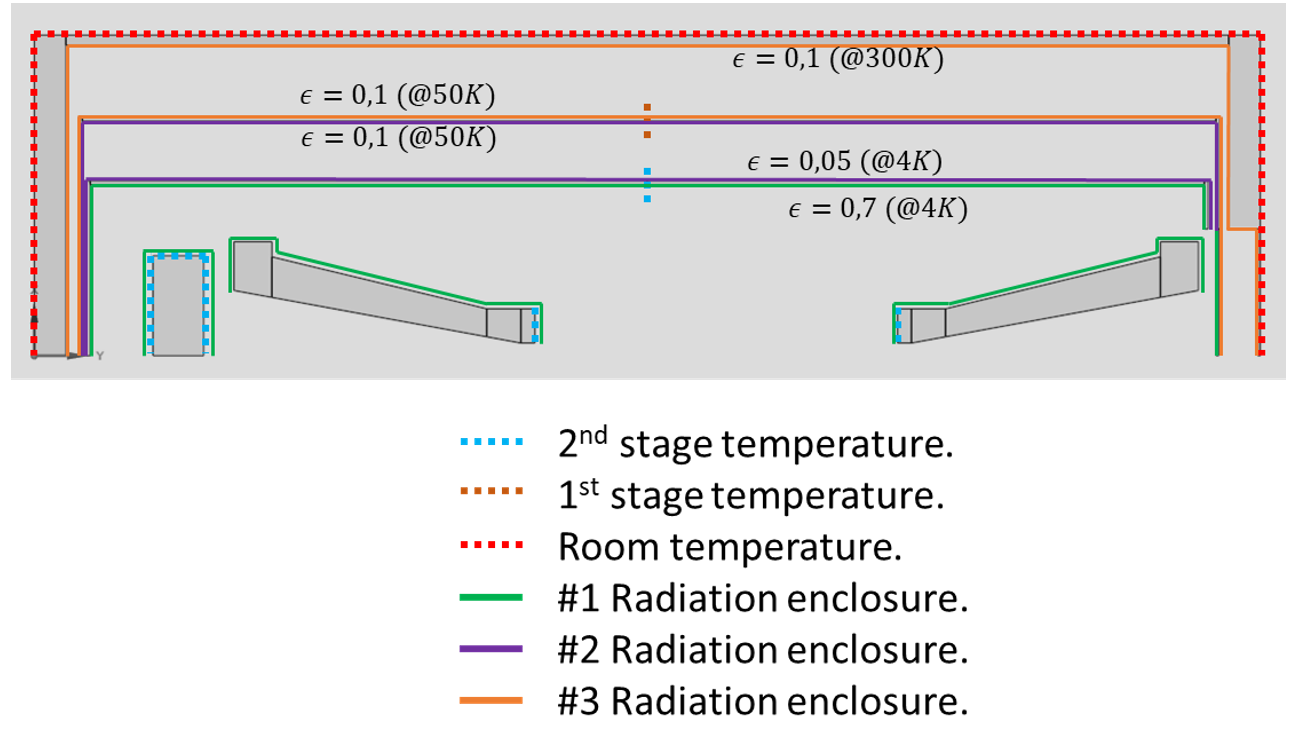}
    \caption{Schematic representation of the internal architecture of the TMS cryostat. The different temperature stages have been highlighted with different colours. }
    \label{fig:cryo-architecture}
\end{figure}

The TMS instrument is integrated in a cylindrical axis symmetric vacuum chamber made of Aluminium AA6061-T6 of 970\,mm in length by 550\,mm in diameter and a maximum weight of 150\,kg, where pressures of the order of $\mathrm{10^{-7}\,mbar}$  are reached in cryogenic conditions. The mechanical structure is divided in two cylindrical stages in order to minimise radiation loads. Figure\,\ref{fig:cryo-architecture} shows the internal architecture of the cryostat, including a breakdown of the different temperature stages.  The first stage is formed by an aluminium shield with a Multilayer Insulator (MLI) foil in order to reduce the thermal load from 32\,W to 12\,W. The second shield, connected to the second stage of the cryocooler, is manufactured in copper OFHC (IR emissivity $>$0.7), and designed to improve the thermal homogeneity inside the second stage cavity. The MLI is used to enhance the RF cavity effect around the mechanical gap, and get a better control of the temperature of the leaking radiation at microwave wavelengths. In this way, the environmental radiation leaking  through the gap can be assumed at the same thermodynamic temperature (4\,K in nominal case) of the load (and baffle).

The cryostat is coupled by a DN 160 CF flange, an air-cooled Pfeiffer ATH-500M turbomolecular repressurisation pump with a nominal flow of 550\,$\mathrm{l/s}$ and a final pressure of $\mathrm{5\times 10^{-7}\,mbar}$. The vacuum chamber has an external leakage rate in blind cover configuration of \mbox{3.54$\mathrm{\times 10^{-9}}$\,mbar $\mathrm{l/s}$.}  The cryostat subsystems provide thermal insulation and cryogenic volume for the receivers. The cryogenic hardware consists of a Helium compressor and a two-stage combined cycle cold head (Sumitomo RDK-415D) providing nominal operating temperatures of 50\,K and 4.2\,K for the first and second stage respectively.

Regarding the cryostat window, its design is based on a panel of Ultra-High-Molecular-Weight Polyethylene (UHMWPE) with a diameter of 176\,mm and 2\,mm thickness.  The surface of both sides have a pattern of triangular grooves  perpendicular to one another to reduce reflection losses and cross-polarisation. The dimensions of these grooves, as optimised with CST Studio Suite, are 10.4\,mm and 19.4\,mm for the width and height, respectively.  In order to guarantee sufficient stiffness of the window, its outer perimeter thickness is kept equal to the triangle height on the air side. Both simulation results and RF testing of the window prototype in the IAC facilities confirm return losses below $-$25\,dB. A full characterisation of the optical subsystem, including the effects of the cryostat window can be found in \cite{AlonsoArias2021}.

\section{Preliminary Study of the Cold Load}\label{sec:prestudy}

\subsection{Technical specifications}\label{sec:techrequirements}

We  established in section\,\ref{sec:radiometer} the need of a stable source as a reference to continuously compare with the sky emissions. The 4KCL subsystem must fulfill two general requirements, which often play against each other. First, it must be a quasi-perfect blackbody to provide a temperature between 5--10\,K. This value corresponds to the expected sum of the CMB and atmospheric temperature contributions, calculated using Paine's tool for atmospheric modelling, \cite{paine_scott_2019}. In this way, a similar spectra to that of the sky above the Teide Observatory, in the operational bandwidth between 10--20\,GHz shall fill the field of view of the reference antenna. As was pointed out in the previous section, minimising the imbalance between sky and calibrator temperatures mitigates the 1/f noise effects. Secondly, the load must be radiometrically stable within $\pm$1\,mK/0.5\,h in order to not produce spurious anisotropies or other spectral errors.

Blackbodies are difficult to design, manufacture and measure. Both the modelling and the measuring strategies are based on the reflection instead of the emission mechanisms of the calibrator, drawing on the principle of reciprocity in thermodynamic equilibrium. In addition, a very good knowledge of the thermal distribution is required, as the overall performance results from the convolution of radiometric and thermal performance.

The emissivity or emission coefficient $e(\nu)$ of an opaque body is defined as the ratio of its emitted spectral power, per unit area at frequency $\nu$, to that of a blackbody emitting at the same physical temperature $T$. In a closed system in thermodynamic equilibrium, the emitted power equals the absorbed power and thus, 

\begin{equation}
\label{eq:emissivity}
\centering
e(\nu)=1-r(\nu),
\end{equation}

with $r(\nu)$ being the reflection coefficient of said object. Kirchoff's law implies that the brightness temperature $T_B$, in thermal equilibrium, is given by 

\begin{equation}\label{eq:Tb}
    T_B(\nu)=e(\nu)T=\left(1-r\left(\nu\right)\right)T
\end{equation}

where $T$ is the physical temperature of the opaque body, expressed in Kelvins. This brightness temperature $T_B$ is defined as the temperature that would give the same intensity as a blackbody at long wavelengths, in the Rayleigh Jeans (RJ) regime. 

An ideal blackbody exhibits an emissivity $e(\nu)$ equal to unity and its temperature distribution is homogeneous, and thus, the brightness temperature fully corresponds to its physical temperature. Non-idealities in either of these two parameters produce differences between the physical and brightness temperature, diminishing the accuracy of the calibrations. Therefore, requirements on emissivity and temperature gradients are the most demanding. 

\begin{figure}[ht]
\centering
\includegraphics[width=0.65\textwidth]{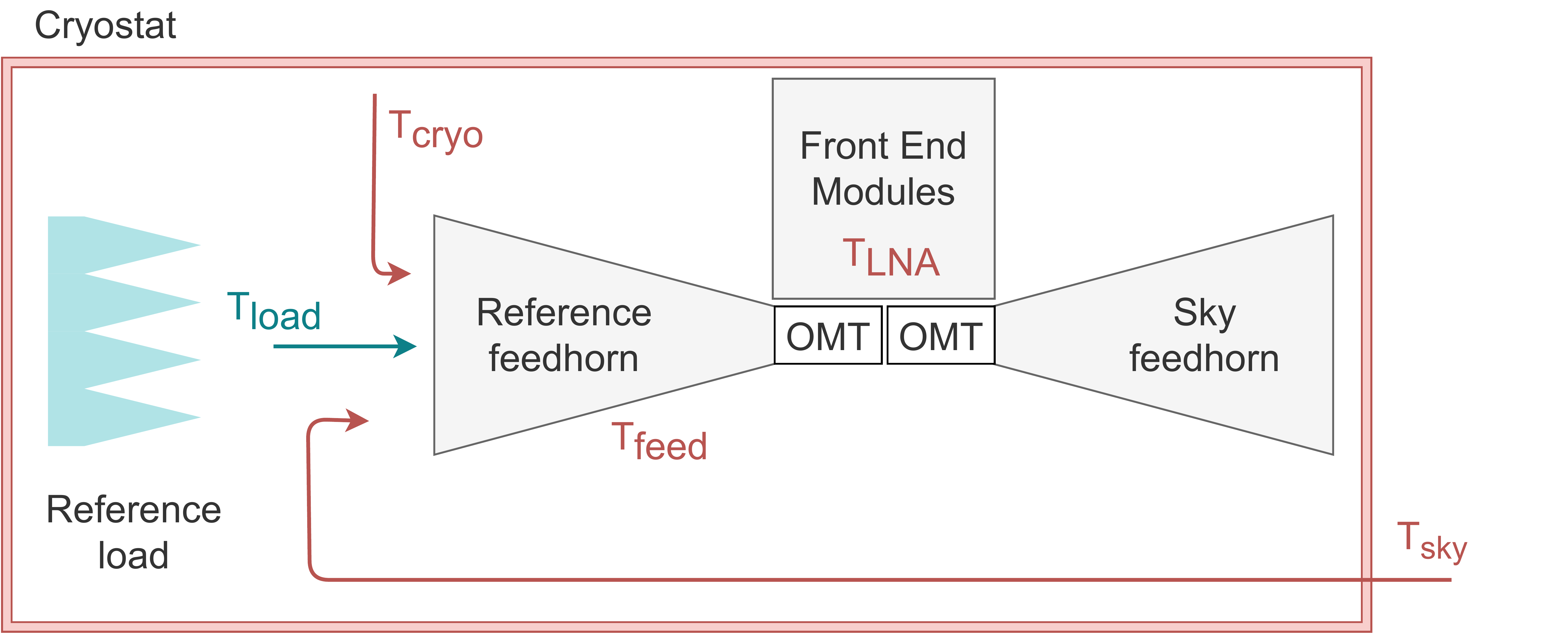}
\caption{Schematic of the temperature contributions in the reference arm feed.}
\label{fig:Tempcontrb}
\end{figure}

Invoking the principle of reciprocity, according to eq.\,\ref{eq:emissivity}, we can measure the emissivity by illuminating the load with a feed and measuring the return losses over a sphere. The goal set in section\,\ref{sec:sciencegoals} for the emissivity of $e\geq$0.999 translates hence into a goal for maximum return loss $r_{\rm tot}\leq-$30\,dB. However, also other effects contribute to the power collected by the reference feedhorn: for example, the radiation leaking from the  cryostat chamber through the mechanical gap between the load and the feedhorn. A simplified sketch of the main temperature contributions is shown in figure\,\ref{fig:Tempcontrb}. In fact, the antenna power in the reference arm of the radiometer $P_{\rm ref}$ can be expressed in general as
\begin{equation}
    P_{\rm ref}\propto [T^{b}_{\rm load}+T^b_{\rm env}]\cdot (1-RL_{\rm feed})+T^b_{\rm inst},
\end{equation}

where $T^b_{\rm load}$ and $T^b_{\rm inst}$ are the brightness temperature of the load and the instrument itself, respectively; and $T^b_{\rm env}$ is the brightness temperature of the environment surrounding the feedhorn and load system. $RL_{\rm feed}$ represents the return losses of the feedhorn in free space. Invoking the RJ approximation, the following equivalences can be derived:
\begin{align}
    \nonumber T^b_{\rm load}=\epsilon_{\rm load}\cdot T_{\rm load}\cdot(1-\eta_{\rm spill,int}) \\
    \nonumber T^b_{\rm env}=\epsilon_{\rm cryo}\cdot T_{\rm cryo}\cdot\eta_{\rm spill,int}+T_{\rm sky}\cdot \eta_{spill,ext} \\
    T^b_{\rm inst}= IL_{\rm feed}\cdot T_{\rm feed}+T_{n,\rm LNAs}\cdot (1-\epsilon_{\rm load})\cdot (1-\eta_{\rm spill,int})
\end{align}

where $T_{\rm load}$, $T_{\rm feed}$ and $T_{\rm cryo}$ are the physical temperatures of the load, feed and cryostat chamber; $T_{\rm sky}$ is the temperature of the sky radiation; and $T_{n,\rm LNAs}$ is the noise temperature of the LNA modules. These temperature terms are weighted by the corresponding terms of emissivity from both the load and the cryostat chamber, $\epsilon_{\rm load}$ and $\epsilon_{\rm cryo}$; the insertion and return losses of the TMS horn, $IL_{\rm feed}$ and $RL_{\rm feed}$; and the spillover ratio of the whole load-feedhorn system, $\upeta_{\rm spill,int}$. We draw a distinction between the external and internal spillover ratios to account for the external signals entering in the chamber or through the horn-target gap. Going into more detail, the spillover should also be separated in two terms: direct spillover (radiation entering the horn straight through the horn-load gap) and indirect spillover (radiation entering the horn bounced through the non ideal load). Since the second term must be weighed by $(1-\epsilon_{\rm load})$, in order to simplify things we can consider only the direct spillover. The resulting brightness temperature at the reference arm can be expressed by the final equation
\begin{align}
\label{eq:pref}
    T^b_{\rm ref}\approx \epsilon_{\rm load}\cdot T_{\rm load}\cdot\left(1-RL_{\rm feed}-\eta_{\rm spill,int}\right)+IL_{\rm feed}\cdot T_{\rm feed}+\\ \nonumber
    T_{n,\rm LNAs}\cdot(1-\epsilon_{\rm load}-\eta_{\rm spill,int})+ \eta_{\rm spill,ext}\cdot T_{\rm sky}+ \eta_{\rm spill,int}\cdot\left  (\epsilon_{\rm cryo}\cdot T_{\rm cryo}\right)
\end{align}

where all temperature terms weighted by $\eta_{\rm spill}\cdot RL_{\rm feed}$ coefficients have been eliminated to simplify the equation, as they are virtually negligible.

In summary, the main non-ideal features affecting the quality of the calibration are: the mismatching between the reference load and the receiver, including the feedhorn and the cascaded subsystems; and the spillover, that measures the leakage of the external signals outside the load, inside or outside the cryostat.

Equation\,\ref{eq:pref} represents a simplified signal model of the TMS receiver, describing only the 4K Load arm. The different contributions that add up in this model are a worst case: some of them are significantly reduced when the receiver pseudo-correlation scheme is considered \cite{bersanelli2010},\cite{cuttaia2004}. For a detailed description of the TMS pseudo-correlation receiver refer to \cite{PazThesis}. Based on this information, it can be inferred that the knowledge of the load physical temperature $T_{\rm load}$ is crucial to estimate the reference temperature as it is the main contribution. Nevertheless, it is to be considered that the spillover \textbf{could} introduce non-negligible uncertainties and that the feedhorn return loss affects the measurement as well. These parameters contributing to eq.\,\ref{eq:pref} were derived starting from top level requirements. The most critical contributions are the sky brightness temperature and the cryostat temperature, comparable to the reference load nominal temperature. Both will be damped at levels below $\sim$mK, so we have set $\mathrm{\eta_{\rm spill}<-}$40\,dB. The feedhorn return loss $RL_{\rm feed}$<$-$30\,dB, which has been set to obtain a given accuracy in the radiometric measurement, is equally important. This is a stringent requirement and can only be met in the majority of the band. Near the cutoff frequency (<10.5\,GHz) the $RL_{\rm feed}$>$-$30\,dB, and it is addressed in \cite{demiguel2021:metamaterial}.

To meet the thermal requirements, we have favoured materials with high values of thermal conductivity, ensuring higher temperature uniformity and faster responses to temperature change (short transient regimes), although the planned observation strategy of the instrument does not include the application of large temperature changes. The requirement for the thermal gradients is 0.1\,K.

In parallel, the reference temperature is required to be stable and not fluctuate. As it shall be the main contribution, temperature fluctuations in the load translate directly as errors at the correlator outputs. The operating temperature should be of the order of 6$\pm$2\,K, and be stable to $\pm$1\,mK for at least 30\,minutes. This stability requirement ensures the fulfillment of the scientific goals for the TMS regarding absolute temperature measurements, as explained in section\,\ref{sec:sciencegoals}.

The above requirements shall be met taking into account the space available inside the cryostat: the available baseline room for the calibrator is 7\,cm, setting a loose constraint on the reference load dimensions at the beginning of the modelling.

\subsection{Preliminary Design}\label{sec:predesign}

\subsubsection{Selection of Materials}\label{sec:materials}

The choice of the materials responds to both RF and thermal requirements. It must ensure high emissivity and a good thermal conductivity, better than 0.1\,W$\mathrm{m^{-1}K^{-1}}$ at cryogenic temperatures. These two properties are virtually incompatible with each other since they are typical of absorbing and metallic materials, respectively. However, a compromise can be obtained by combining a broadband absorbing material with a metallic backing and the adequate geometry, able to taper the radiation matching and provide a strong absorption while preventing thermal inhomogeneity across the radial and pointing axis direction.

Taking into account above statements, the ECCOSORB\texttrademark~ materials of the CR/MF-series\footnote{\url{https://www.eccosorb.eu}}, which are widely used in astrophysics for cryogenic applications, were chosen. In particular, the requirement of absorbing capability in layers of a few mm (critical to prevent thermal inhomogeneities) led to the selection of a material of CR/MF\,114 or higher. CR/MF\,114, CR/MF\,117 and  CR/MF\,124 were all available, but we discarded  CR/MF\,124 due to its too large intrinsical reflectivity, leaving CR/MF\,114 and CR/MF\,117 which are also space qualified materials in terms of outgassing.\footnote{\url{https://dokumen.tips/documents/outgassing-data-for-emerson-cuming-microwaves-material-tml-cvcm-wvr-nasa-data.html}} CR/MF\,114 and CR/MF\,117 are widely characterised down to  4\,K in several applications, \cite{valenziano2009}. Concerning the metallic core, Aluminium 6082 was considered suitable to provide the required thermal performance. Additionally, it has optimal mechanical properties for a reliable manufacturing process.

\subsubsection{Selection of the Geometry} \label{sec:geometry}

The most popular structures for calibration targets are cavities \cite{Bedford:75} or protruding tapered geometries \cite{Mather_1999}, such as cones \cite{Rostem2013}, wedges \cite{Gaidis1999} or pyramids \cite{Fixen2006}. In projects with limited room or space, such as satellite instruments, pyramidal absorbers prove to be an optimum solution, especially at the highest frequencies (from 30 GHz) because of the smaller size. The reference load unit on board the Planck-LFI, \cite{valenziano2009}, was designed with a geometry which combined both small cavities and pyramidal absorbers, while the ground calibrators for the Planck-LFI \cite{Mennella2010} and HFI \cite{Pajot2010} instruments used a similar model to the TMS 4KCL, shaped as a bed of pyramids.

The design of the 4\,K Load is based on an array structure of square-based  pyramids. Each pyramid consists of a metal core covered by a coating of a material with high absorption capabilities. The array is confined by a metal shield,  shaped as a cylinder, to suppress the spillover radiation and enhance the temperature uniformity.

A Geometrical Optics (GO) approximation is not entirely valid because the wavelength is comparable to the size of the structures and we would be disregarding scattering effects (diffraction, interference) and assuming specular reflections. 
A more correct approach than this involves multiwavelength analysis. In order to improve taper-based radiation trapping, we need to optimise two key geometrical parameters, the height $H$ and width $W$ of the core metal pyramids. By maintaining the total size of the load mould, the coating thickness $Th$ will also vary along with the aforementioned parameters, and consequently, the thermal gradients and absorptivity of the load.

\subsection{Analysis of the feedhorn}\label{sec:feedhorn}

Two properties of the antenna are essential to understand the starting point for a design strategy: the return loss $RL$ or parameter $S_{11}$  and the Near Field (NF) pattern.\footnote{By definition, return loss has opposite sign to the scattering parameter $S_{11}$ when expressed in dB, taking the value $RL\mathrm{=-20\times\log{|S_{11}|}}$.} The return loss establishes a limit at which the behaviour of the load-feedhorn system will be dominated by the feedhorn. On the other hand, we need to calculate beforehand the NF patterns in order to get an initial idea of the minimum load diameter needed, in such a manner that the Field of View (FOV) will be completely covered in the whole band. 

The feedhorn was designed to ensure high performance  over an octave bandwidth between 10--20\,GHz. In particular, levels below $-$25\,dB for the return loss, and $-$35\,dB for the cross polarisation were the strictest requirements imposed. During the project, we updated the return loss requirement to  $-$30\,dB for the horn, in order to achieve a better  matching. These demands are met with a novel design based on metamaterials, fully described in \cite{demiguel2021:metamaterial}. The manufactured horn presents a total length of 240\,mm, with an aperture diameter of 103\,mm, external diameter of 180\,mm, and a 20\,mm-diameter feedhorn mouth.

\paragraph{RF and Thermal verification.} In \cite{demiguel2021:metamaterial}, the radiation properties of the TMS feedhorn are reported, including measurements of the return losses. We performed a few RF verification tests at the IAC facilities to cross check the results obtained. In addition, and due to its complex mechanical design, a thermal verification was proposed to guarantee the stability of the horn after being thermalised at cryogenic temperatures, and its repeatability. $S_{11}$ was measured over the full TMS frequency range before and after three complete thermal cycles ---performed in the TMS cryostat ---  from room ambient temperature down to 11\,K. The torque of the screws tightening the platelets was checked as well. RF and mechanical results confirmed that the behaviour remained the same before and after cryogenics. 
 
\begin{figure}[ht]
    \centering
\subfloat{\includegraphics[width=0.68\textwidth]{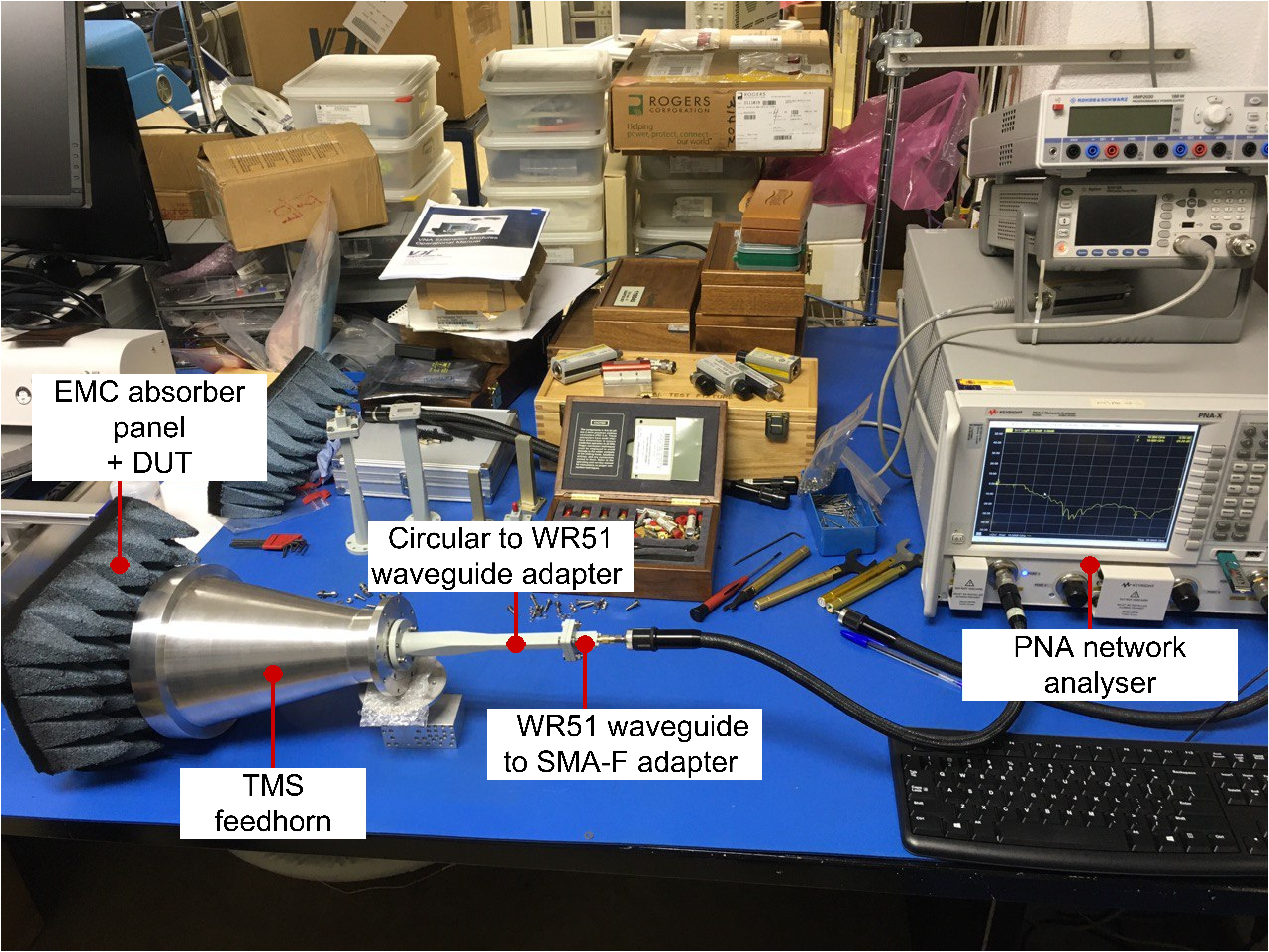}}\\
    \vspace{-0.3cm}
\subfloat{\includegraphics[height=0.45\textwidth]{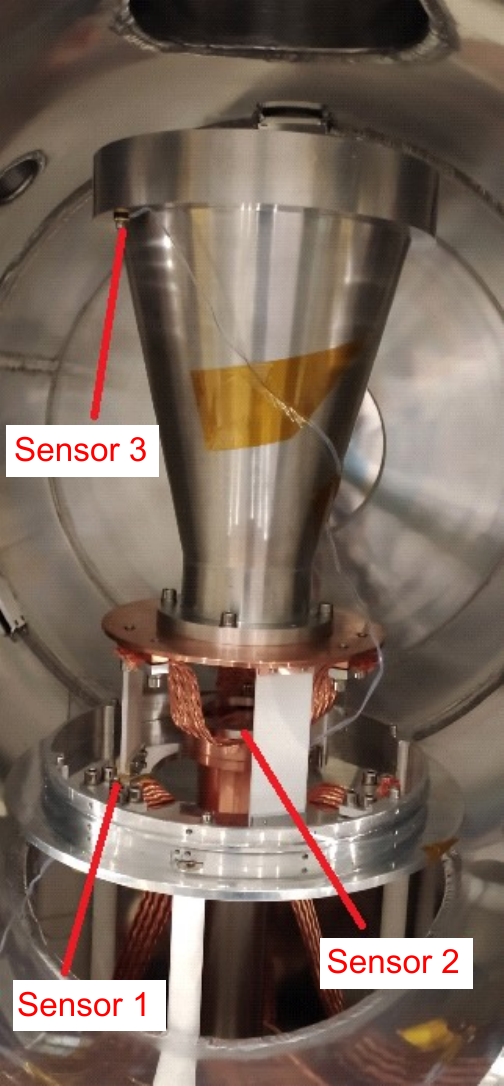}}
     \hspace{-0.8cm}
    \qquad
\subfloat{\includegraphics[height=0.45\textwidth]{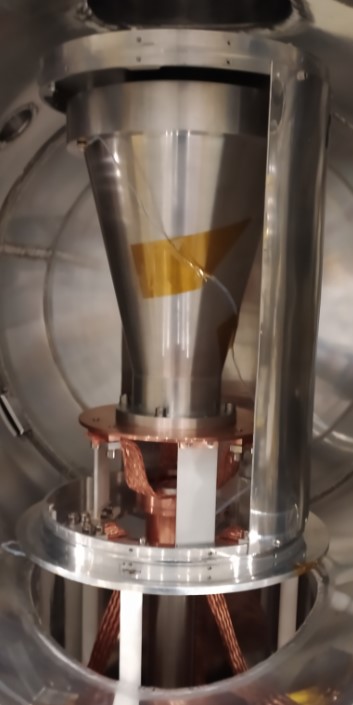}}
   \hspace{-0.8cm}
    \qquad
\subfloat{\includegraphics[height=0.45\textwidth]{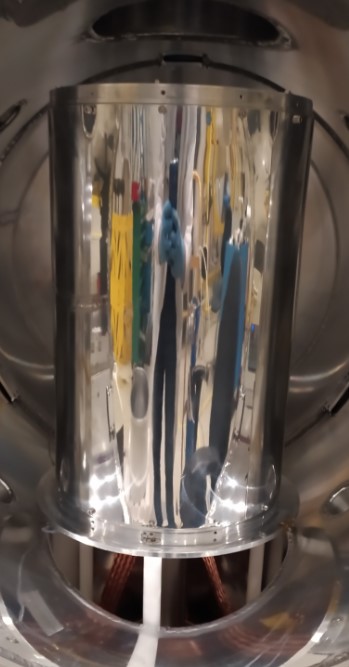}}
    \caption{Setups for the RF and thermal verification of the TMS feedhorn. \emph{Top:} RF setup, consisting of a simple setup for the measurement of S-parameters. Circular (20\,mm) to WR75 FBP120 and WR51 FBP180 adapters, WR75 and WR51 waveguide to SMA-F adapters, a PNA network analyser and EMC absorber have been used. \emph{Bottom:} Thermal test setup, which relies on the TMS cryostat and its cold head. The positions of the three thermal sensors are emphasised in \emph{red} colour.}
    \label{fig:feed_testsetups}
\end{figure}
 
In figure\,\ref{fig:feed_testsetups}, we show both setups for the RF (\emph{top}) and thermal (\emph{bottom}) tests.  The RF setup was based on a PNA network analyser: a time domain window was applied to de-embed the test-setup-related effects from the feedhorn $S_{11}$. For the thermal tests setup, the horn was connected to the second stage of the cold head through a Copper plate, surrounded by a radiation shield-connected to the first stage- to minimise the environmental temperature. The different stages are thermally isolated by means of Teflon\texttrademark\,struts to monitor the first and second stage plates temperatures. Temperature at the steady state condition reached  11\,K for sensor 3, which represent the hottest point in the horn.

\begin{figure}[ht]
\centering
\subfloat{\includegraphics[width=0.7\textwidth]{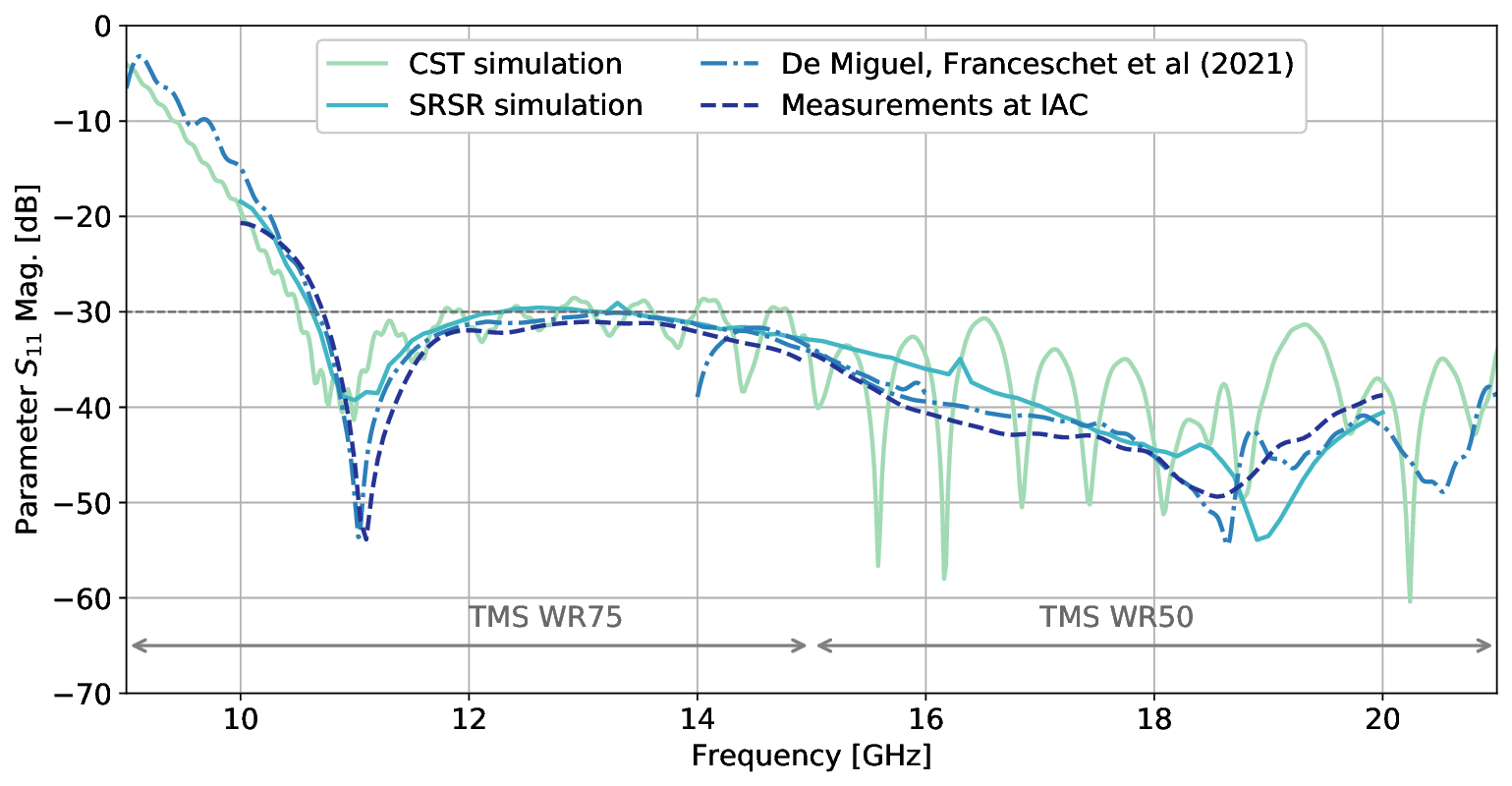}} \\
\subfloat{\includegraphics[width=0.7\textwidth]{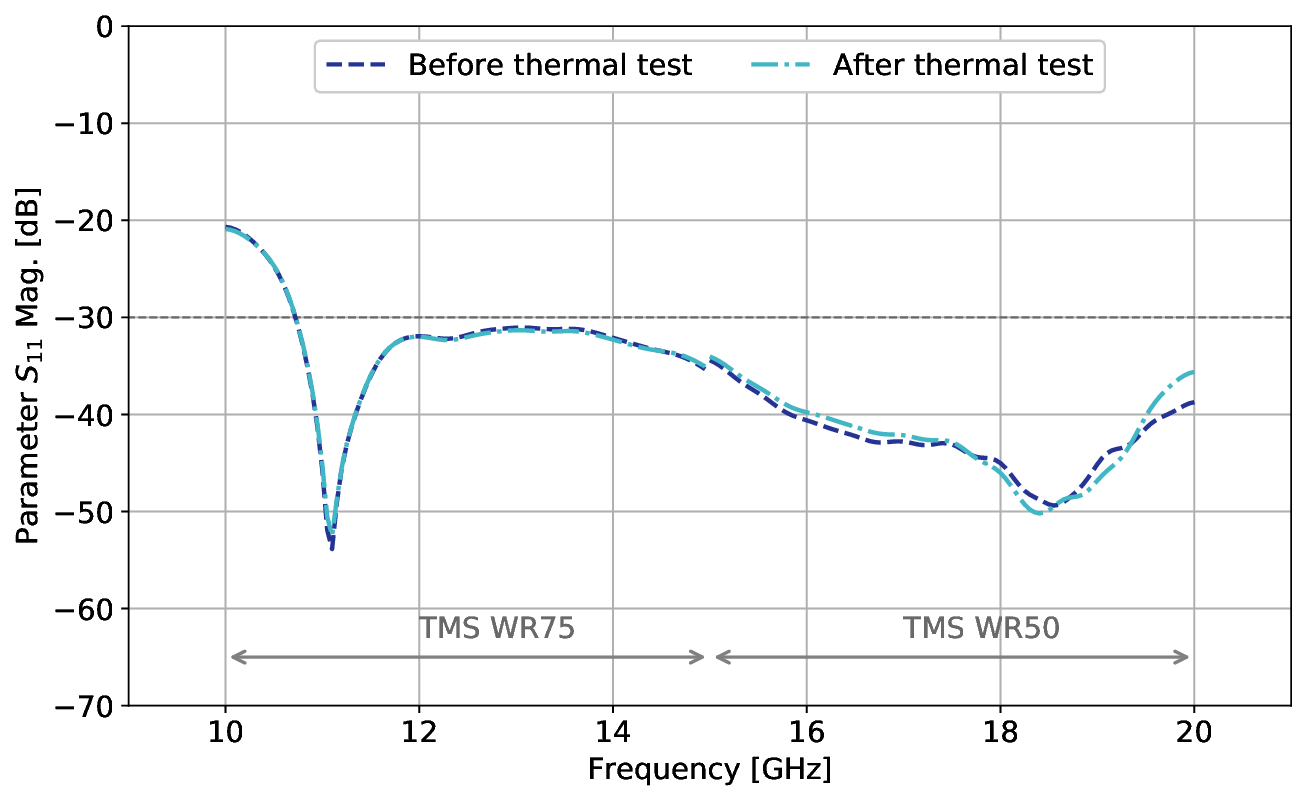}} 
\caption{Return loss ($S_{11}$) of the TMS feedhorn in the operative frequency band between 10--20\,GHz. \emph{Top}: Comparison between simulation results (\emph{solid lines} obtained with CST Studio and SRSR software and the return loss measurements (\emph{dashed lines}) reported in \cite{demiguel2021:metamaterial} and those performed in the IAC facilities. Both results are consistent with each other and with the simulations. Measurements have been obtained using two setups for the frequency range between 10--15\,GHz (WR75) and 15--20\,GHz (WR51).  \emph{Bottom}: $S_{11}$ measurements of the horn before and after the thermal cycles. No significant differences in the TMS feedhorn performance were found.}
\label{fig:feedhornS11}
\end{figure}

In figure\,\ref{fig:feedhornS11}, we compare the results obtained with two different simulators, CST\texttrademark~ and SRSR\texttrademark~, in comparison to the measurements obtained in the testing activities. In both cases, the results for the return loss show the required $S_{11}<-$30\,dB is met nearly over the whole band between 10 and 20\,GHz. This value of return loss is in fact only slightly higher than the goal of $S_{11}<-$30\,dB described in section\,\ref{sec:prestudy} and, consequently, the horn can be considered, to some extent, the limiting agent.

\paragraph{Feedhorn patterns in the Near Field region.} We analysed the Near Field (NF) patterns to establish the dependence of the FOV or footprint size on both frequency and distance from the aperture of the horn. The Electrical field was calculated over a flat sub-volume in front of the aperture at the position of a virtual load. Considering $\mathrm{z=0}$ the plane containing this aperture, we allocated for the load the maximum volume allowed inside the cryostat (7\,cm maximum length allocated for the full load, including the metal inner; and 9\,mm as the closest distance considered safe between the load tips and the feedhorn mouth plane). These constraints define the following relevant distances for the RF simulation:

\begin{itemize}
\item $\mathrm{z=9}$\,mm, corresponding to the plane containing the load's closest surface (tips) facing the horn;
\item $\mathrm{z=68}$\,mm is the distance at which the base of the load shall be located, taking into account the maximum room restriction specified in section\,\ref{sec:techrequirements}; and, 
\item $\mathrm{z=44}$\,mm, a distance chosen to represent the distance at which we expect a maximum peak of absorption (estimated at about $\mathrm{3/5}$ of the length of the pyramids).
\end{itemize}

\begin{figure}[ht]
\centering
\includegraphics[width=0.7\textwidth]{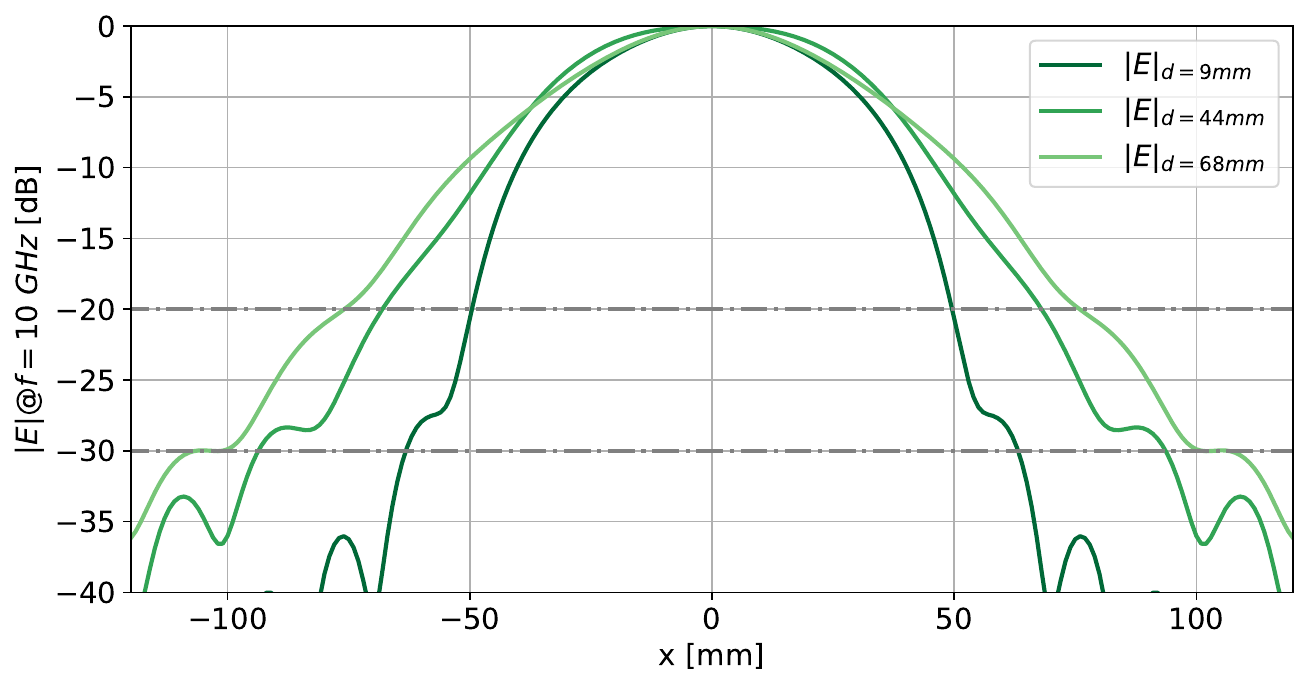}
\caption{Comparison of Near Fields at 10\,GHz at distances $d=\{9,44,68\}$\,mm. The NF power is normalised to the maximum intensity at each distance from the aperture. The symmetry of the NF patterns allows to represent a single cut, in the x-axis direction. Beamwidths at $-$30\,dB in both x and y axis are 120, 160 and 180\,mm.} 
\label{fig:NF-freq10} 
\end{figure}

\begin{figure}[ht] 
\centering
\includegraphics[width=0.6\textwidth]{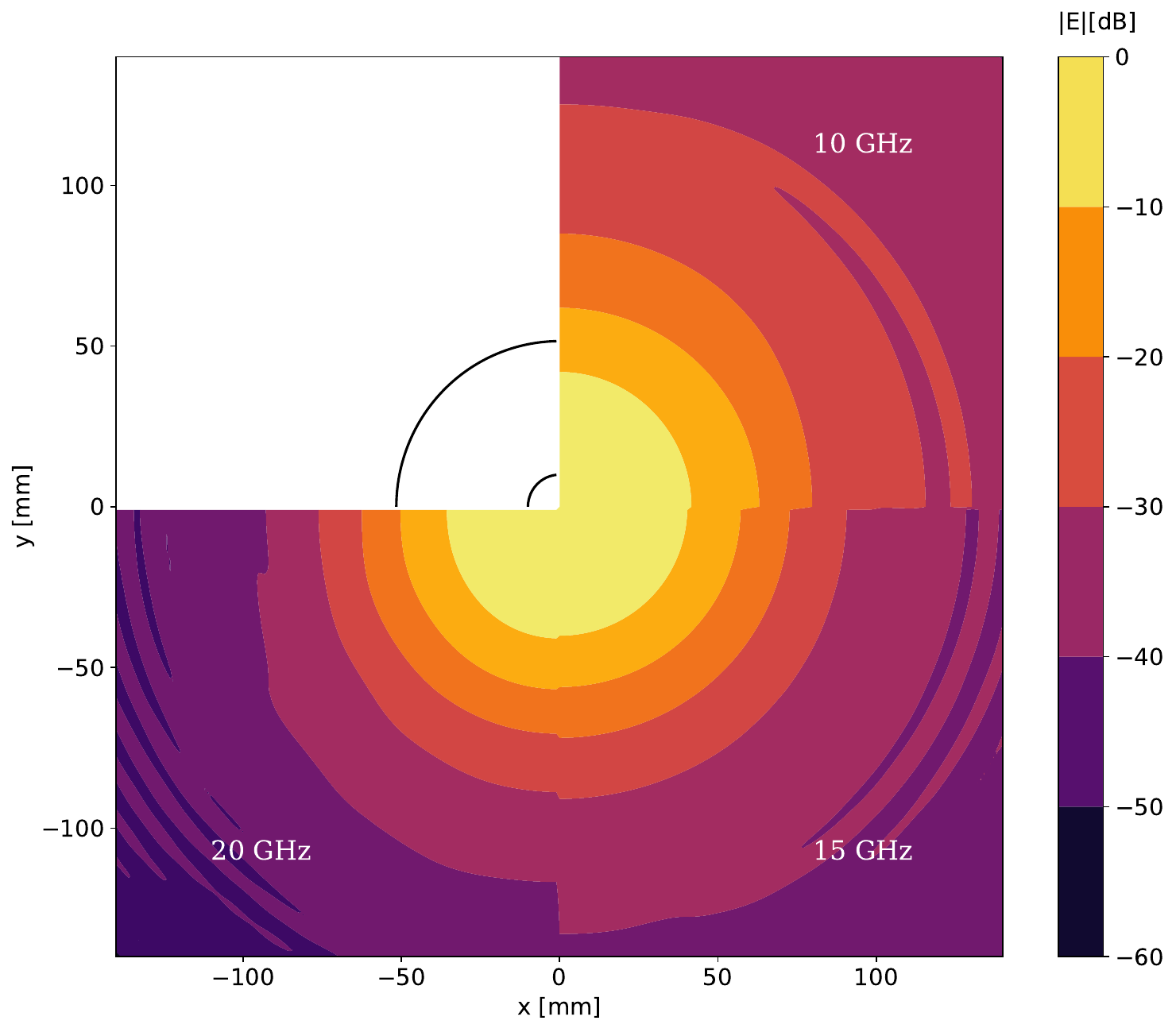}
\caption{Near Field (NF) frequency dependence at a distance of $d=$44\,mm. Starting from the upper right quadrant and moving clockwise, NF at 10, 15 and 20\,GHz, respectively.  The NF power is normalised to the maximum intensity at each frequency and distance from the aperture. The upper left quadrant shows the diameters of the feed mouth (20\,mm) and aperture (103\,mm), for reference.} 
\label{fig:NF-distance44} 
\end{figure}

Figure\,\ref{fig:NF-freq10} shows the CST simulation results at 10\,GHz for the z- planes defined above. In gray dashed lines, $-$20\,dB and $-$30\,dB levels have also been plotted. The Near Field cuts in the x-axis and y-axis barely show differences between them and the maximum beamwidth $\mathrm{BW_{-30dB}=180}$\,mm in the furthest plane ($\mathrm{z=68}$\,mm). In addition, the evolution of the  NF footprints over frequency is shown in figure\,\ref{fig:NF-distance44} for the virtual distance $\mathrm{z=44}$\,mm of maximum power absorption.

This analysis provides an indication about the level of the signal reaching a certain region of the Cold Load, here identified with respect to the radial distance from the feedhorn axis. From these results, we infer minimum dimensions for the Cold Load, knowing that there is a trade-off between maximum available space and budget, and NF footprint coverage. The information about which curve to consider will be provided by the analysis of the pyramids reported in section\,\ref{sec:rfdesign} (suggesting that $d=44$\,mm is the most representative choice).

\section{Simulation Results and Design Refinement}\label{sec:rfsimulation}

The detailed design is a result of the combination of RF and thermal modelling, as well as experimental tests performed on representative samples. Both the optimisation of the design and the estimation of the RF performance involved the use of a Finite Element Method model (FEM) of the reference unit and the load-feedhorn system, and an iterative process from the preliminary baseline and requirements for a refined design. We used both the CST Studio\texttrademark~ and ANSYS HFSS\texttrademark~ software to solve and optimise the problem. The optimisation of the Cold Load design was based on the computation of several electromagnetic (EM) parameters. 
Table\,\ref{tab:EM-design} summarises the correspondence between these EM and the design parameters themselves.  

\begin{table}[ht]
\centering
\resizebox{0.5\textwidth}{!}{ 
\begin{tabular}{|c|c|c|c|c|c|c|c|}
\hline
\textbf{EM Parameter}            & \multicolumn{7}{c|}{\textbf{Design Parameters}} \\ \hline
 & \rotatebox{90}{\textbf{4KCL shape}} &  \rotatebox{90}{\textbf{4KCL material}} &  \rotatebox{90}{\textbf{4KCL length}} &  \rotatebox{90}{\textbf{4KCL diameter}} &  \rotatebox{90}{\textbf{Gap}} &  \rotatebox{90}{\textbf{TSH coating}} &  \rotatebox{90}{\textbf{Thermal gradients}} \\ \hline
\textit{Return Loss}             & \texttimes     & \texttimes    & \texttimes    &      &      &\texttimes    &      \\ \hline
\textit{Bandwidth}               & \texttimes     & \texttimes    & \texttimes    &      &      & \texttimes    &      \\ \hline
\textit{FH Near Field}           &       &      &      & \texttimes    & \texttimes    &      &      \\ \hline
\textit{Spillover}               &       &      &      & \texttimes    & \texttimes    & \texttimes    &      \\ \hline
\textit{Power Loss density (Z)}  &       &      &      &      &      &      & \texttimes    \\ \hline
\textit{Power Loss density (XY)} &       &      &      &      &      &      & \texttimes    \\ \hline
\end{tabular} 
}
\caption{Correspondence between EM and design parameters.}
\label{tab:EM-design}
\end{table}

We report below the final design of the TMS 4KCL subsystem, as well as the electromagnetic and thermal results  obtained from the simulations and verification tests performed during the design refinement process.

\subsection{RF design}\label{sec:rfdesign}

\begin{figure}[ht] 
\centering
\includegraphics[width=1.\textwidth]{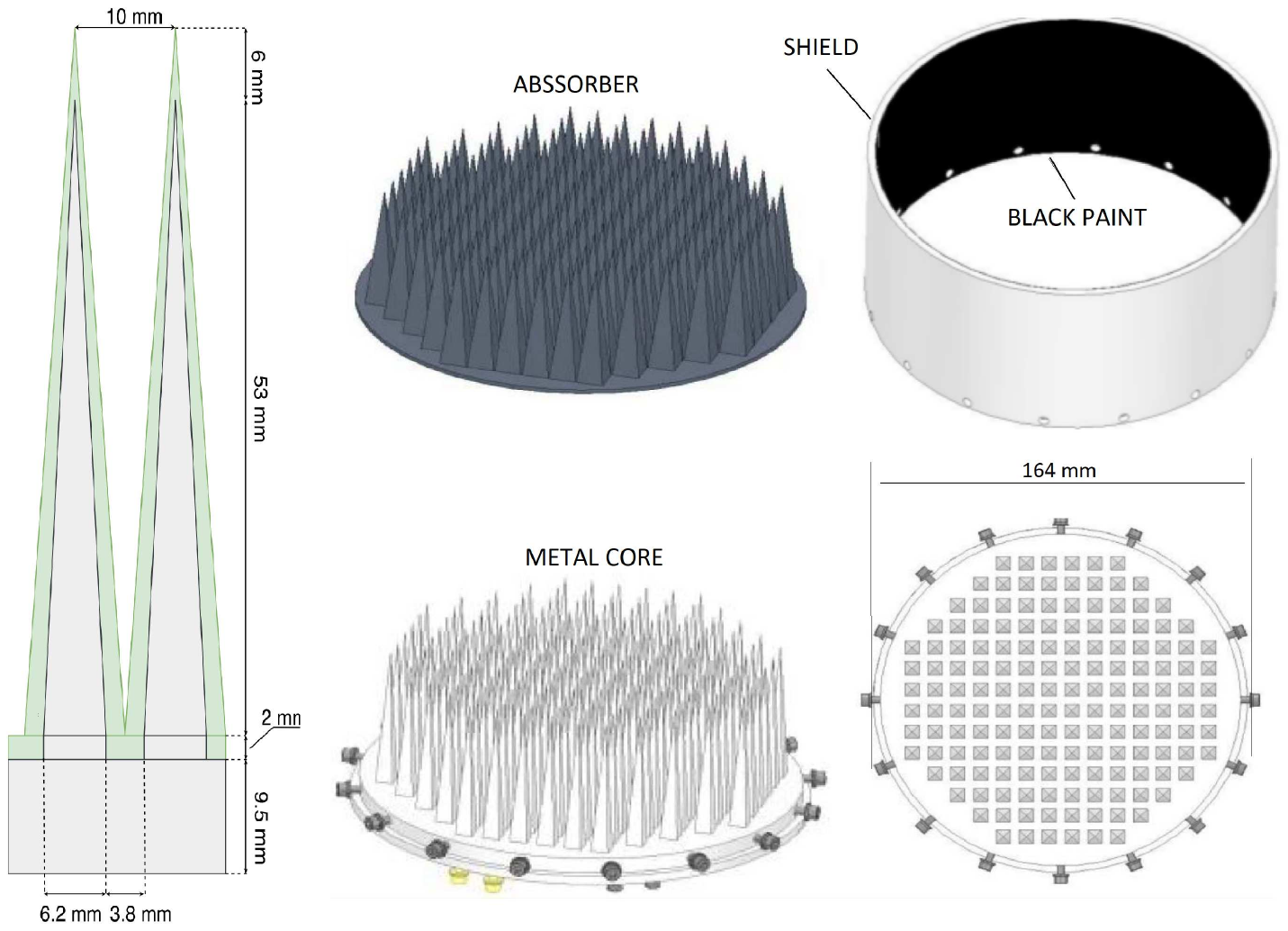}
\caption{Final geometrical configuration of the 4KCL model. \emph{Left}: optimised dimensions of the pyramid cores are 
$W=$6.2\,mm and $H=$53\,mm, coated with $Th=$1.9\,mm of absorber material. \emph{Right}: The 4KCL includes the metal core (baseplate and array of pyramids) and the absorber coating and a metallic shield. The array configuration presents an octagonal arrangement with a diameter of 140\,mm. } 
\label{fig:pyr-dim} 
\end{figure}

The final design of the 4\,K load is shown in figure\,\ref{fig:pyr-dim}, including the projection from above and the dimensions of the pyramidal elements. The load  includes a metal core and the absorber coating, as well as a metallic shield. The metal core is divided in two sub-units, namely the baseplate and the core shaped as a bed of pyramids. Both are manufactured  with Aluminium AI6082 in a single piece to improve thermal conduction.

The baseplate is a circular plate with  a diameter  164\,mm and 9.5\,mm  thickness, coated with 2\,mm of absorber material. The back side is designed to host several thermal sensors. In addition, it presents a circular step to fix the shield with screws and spring washers. On the other hand, the optimised base width and height of the metal core pyramids are $W=$6.2\,mm and $H=$53\,mm. The absorber thickness at the base of the pyramids is $Th=$1.9\,mm, so the total size of the base of each absorber element is $10\times 10\mathrm{\,mm^2}$, as shown in detail in figure\,\ref{fig:pyr-dim}. We opted for a final array configuration inside the metal shield with a diameter of 140\,mm in an octagonal arrangement, a maximum of 14 pyramids, a trade-off between mechanical and manufacturing requirements and the footprints resulting from the feedhorn NF  shown in section\,\ref{sec:feedhorn}. 

The shield is a simple cylinder concentric to the baseplate and made of Aluminium too. The  main function of the shield is the suppression of the spillover radiation  while enhancing the  emissivity and temperature homogeneity. It is a piece of 72\,mm in height, presenting an external and internal diameters of 164\,mm and 158\,mm, respectively. The external surface is polished, while the internal face is finished with an absorptive black paint (Chemglaze Z306\footnote{\url{http://esmat.esa.int/Services/Preferred_Lists/Materials_Lists/a12_3.htm}}) to improve the thermal homogeneity. The shield is prepared also to host one thermal sensor.

\begin{figure}[ht] 
\centering
\includegraphics[width=0.50\textwidth]{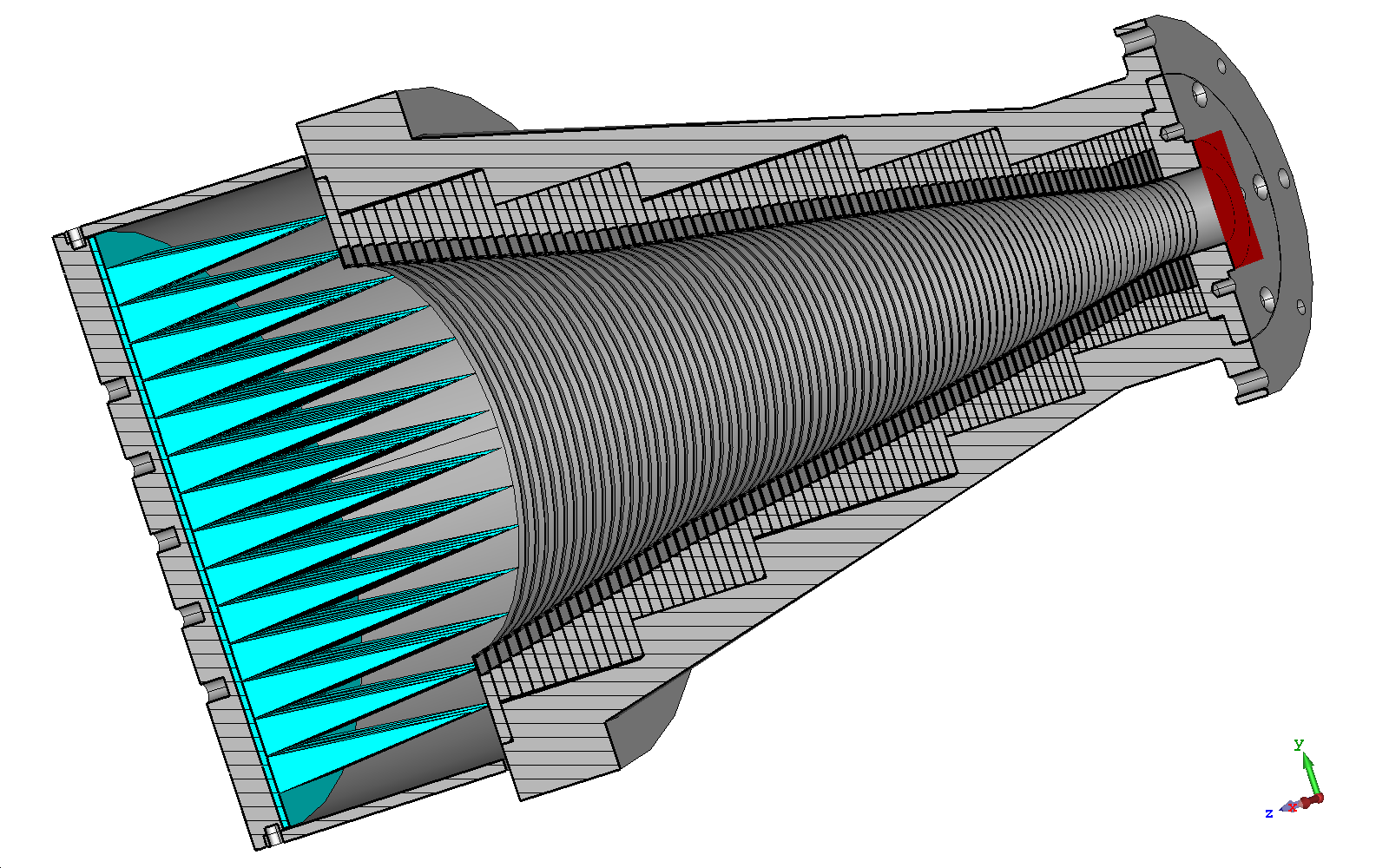}
\caption{Model of the horn -  reference load system used for the FEM analysis. Between the load shield and the feedhorn there is a 1\,mm gap. In order to reduce computation time and resources, in the first design iterations the feedhorn and the metallic part of the 4KCL were modelled as perfect electrical conductors (PEC), instead of Aluminium.} 
\label{fig:load-feedhorn} 
\end{figure}

\begin{figure}[ht] 
\centering
\includegraphics[width=0.8\textwidth]{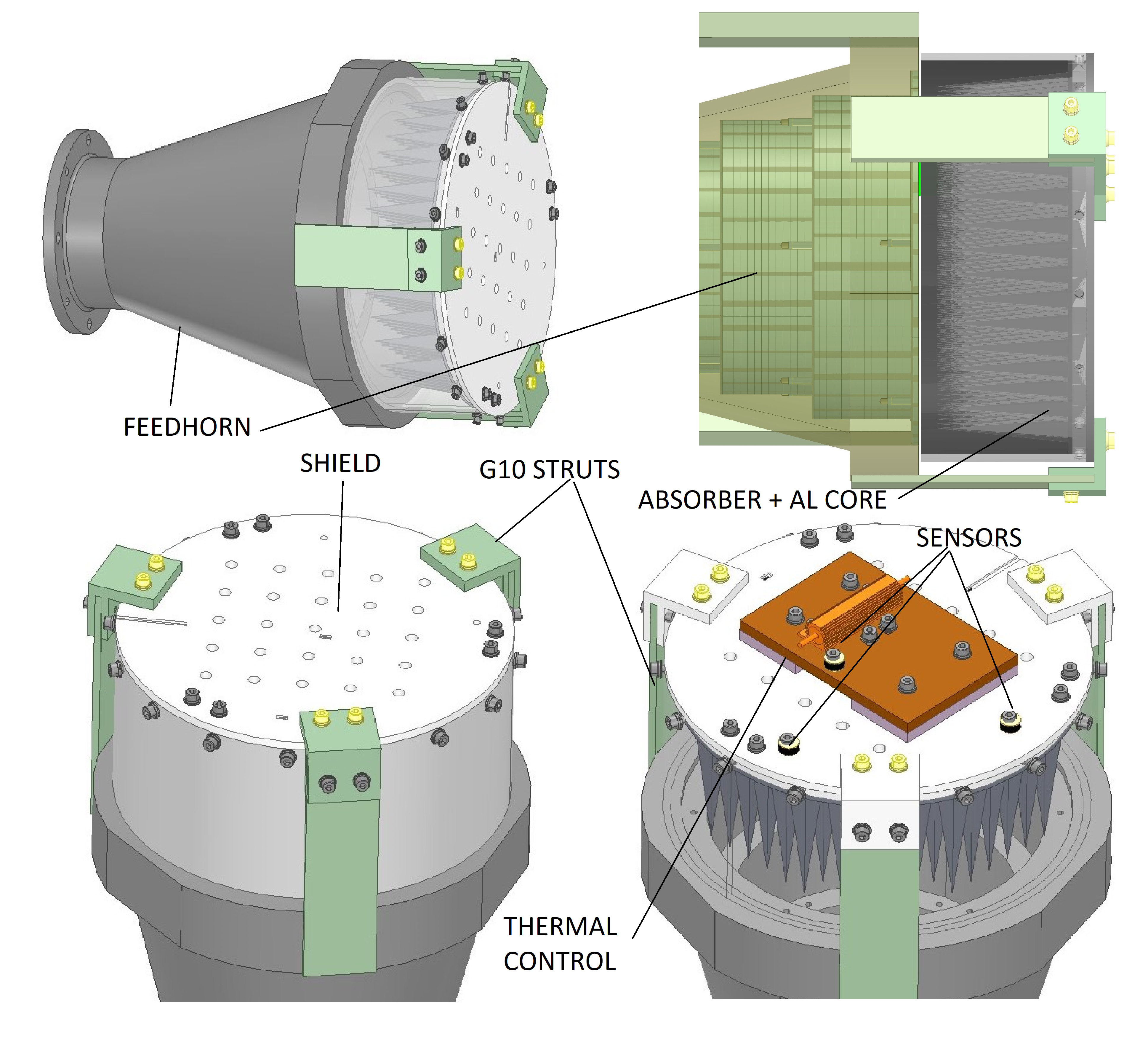}
\caption{Feedhorn-reference load mechanical assembly. All the relevant parts are identified; the 1\,mm gap is visible, together with the G10 thermal struts. Thermal interfaces and sensors are shown in the bottom right panel} 
\label{fig:4KCL_assembly} 
\end{figure}

\begin{figure}[ht] 
\centering
\includegraphics[width=0.70\textwidth]{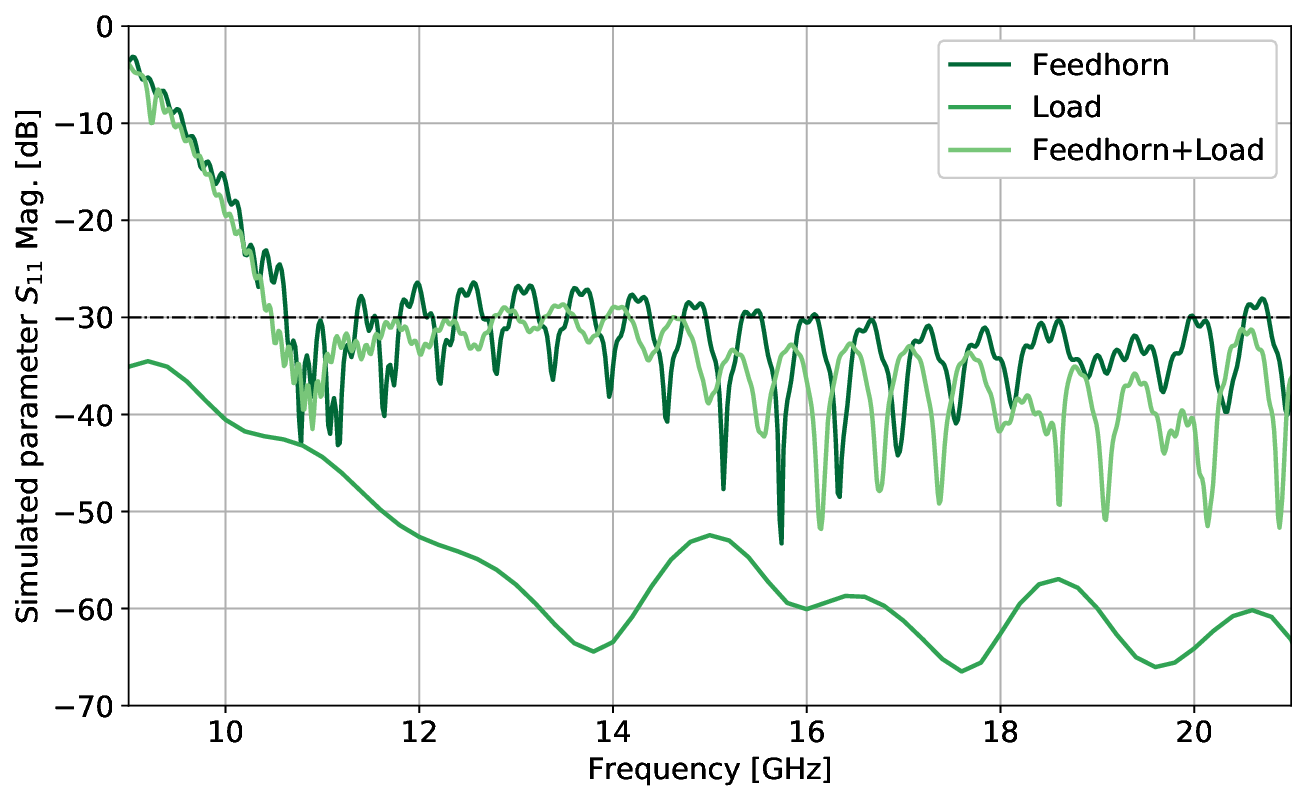}
\caption{Simulated return loss ($S_{11}$) of the cold load, feedhorn and their combined operation over the TMS bandwidth between 10--20\,GHz.
At the lower end of the spectral range, the $S_{11}$ of the assembly exceeds the requirement of $-$30\,dB due to the feedhorn performance.The $S_{11}$ of the 4KCL never exceeds $-$40\,dB. 
} 
\label{fig:load-S11} 
\end{figure}

The EM design was verified by combining two different models able to separately estimate both the total reflectivity, $S_{11}$, due to the pyramids array (used a plane wave as excitation) and the coupling between the feedhorn and the load as arranged inside the cryostat (figure\,\ref{fig:load-feedhorn}), so as to estimate the power intercepted by the load and the spillover generated by the mechanical gap between feedhorn and shield (1\,mm, due to thermal needs and the manufacturing process of the meta-horn described in \cite{demiguel2019:hornmanufacture}, \cite{demiguel2019:hornfund}. Although the simulation of the complete model is dominated by the $S_{11}$ of the feedhorn alone (especially at low frequency), however, it is powerful to evaluate the spillover component. Table\,\ref{tab:load-meritfigures} summarises two figures of merit of the calibrator performance, the return loss and spillover, at three relevant operational frequencies. Additionally, $\mathrm{S_{11}}$ is represented over the full range in figure\,\ref{fig:load-S11}.

\begin{table}[ht]
\centering
\resizebox{0.4\textwidth}{!}{ 
\begin{tabular}{ccc}
\hline
\textbf{Frequency (GHz)} & $\mathbf{S_{11}\,(dB)}$ & $\mathbf{\upeta_{spill}\,(dB)}$ \\ \hline
10 GHz & -40.1 dB & -47.1 dB \\ 
15 GHz & -52.5 dB & -48.9 dB \\ 
20 GHz & -64.2 dB & -49.5 dB \\ \hline
\end{tabular} 
}
\caption{Figures of merit for the cold load performance at $10,\,15,\,20\,\mathrm{GHz}$.}
\label{tab:load-meritfigures}
\end{table}

Results correspond to the ideal positioning of the load, which is completely coaxial with the feedhorn: G10 struts (figure\,\ref{fig:4KCL_assembly}) were used to couple the feedhorn and load, preserving the thermal break between them. A tolerance study was performed to assess how much coaxial displacement and tilt angle is acceptable.

The limited computational resources allowed us to perform only an approximated analysis, replacing the TMS feedhorn by its equivalent electric field generated at the aperture. $S_{11}$ is calculated by means of the reflection coefficient $\rho_{\rm R}$. An E-field probe is placed between the E-field source and the load to calculate $\rho_{\rm R}=E_{\rm ref}/E_{\rm inc}$. The reflection coefficient $\rho_{\rm R}$ has also been calculated for a perfect conductive (PEC) sheet placed in front of the load, in order to provide a reference. While this method is approximate, it significantly reduces computational time and offers valuable insights into the relative results during the design phase. Absolute results have been extensively verified by laboratory tests, as discussed in section\,\ref{sec:measurements}.

\begin{figure}[ht] 
\centering
\includegraphics[width=0.70\textwidth]{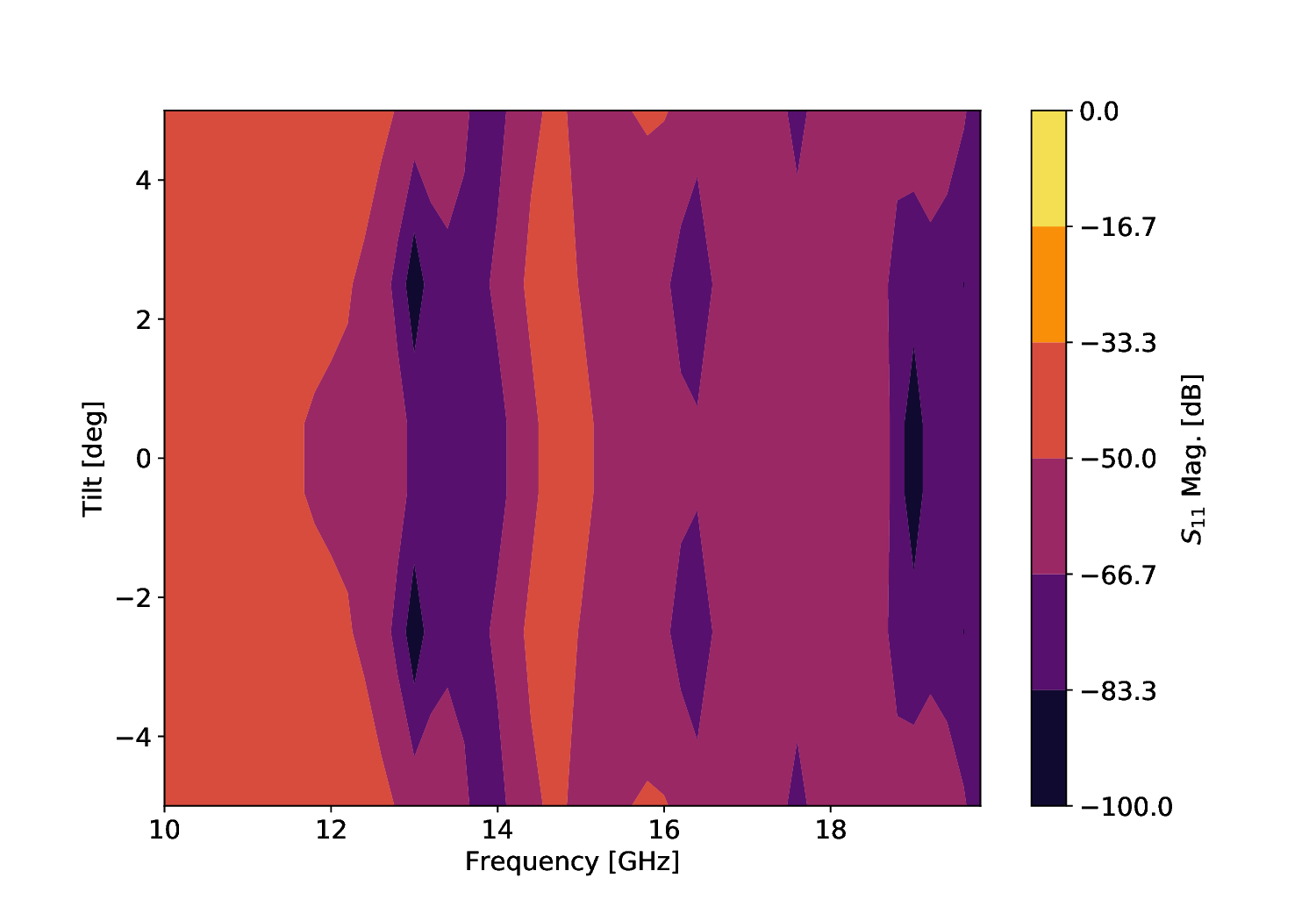}
\caption{Spectrogram with the simulation results for different tilt angles. $S_{11}$ over frequency for different angular displacements between -5\textdegree\, and 5\textdegree. RL remains under $-30$\,dB (meeting technical requirement), and under $-$40\,dB outside the range 11--12\,GHz, at angular displacements of up to 5\textdegree.} 
\label{fig:load-offset} 
\end{figure}

Parametric simulations were performed to obtain limiting values for angular and linear displacements of the load during the design phase. Even though the results, given the simplified model described above, are largely affected by the feedhorn performance at frequencies below 12 GHz,  we can set a conservative acceptance criterion of the order of $<$5.0\textdegree\,tilt for the angular tolerance and 2.5\,mm for the linear displacement along X-Y. Larger deviations would reduce the operating bandwidth and peformance in general. Figure\,\ref{fig:load-offset} shows the relative results obtained for the angular displacement.  

\begin{figure}[ht] 
\centering
\includegraphics[width=0.70\textwidth]{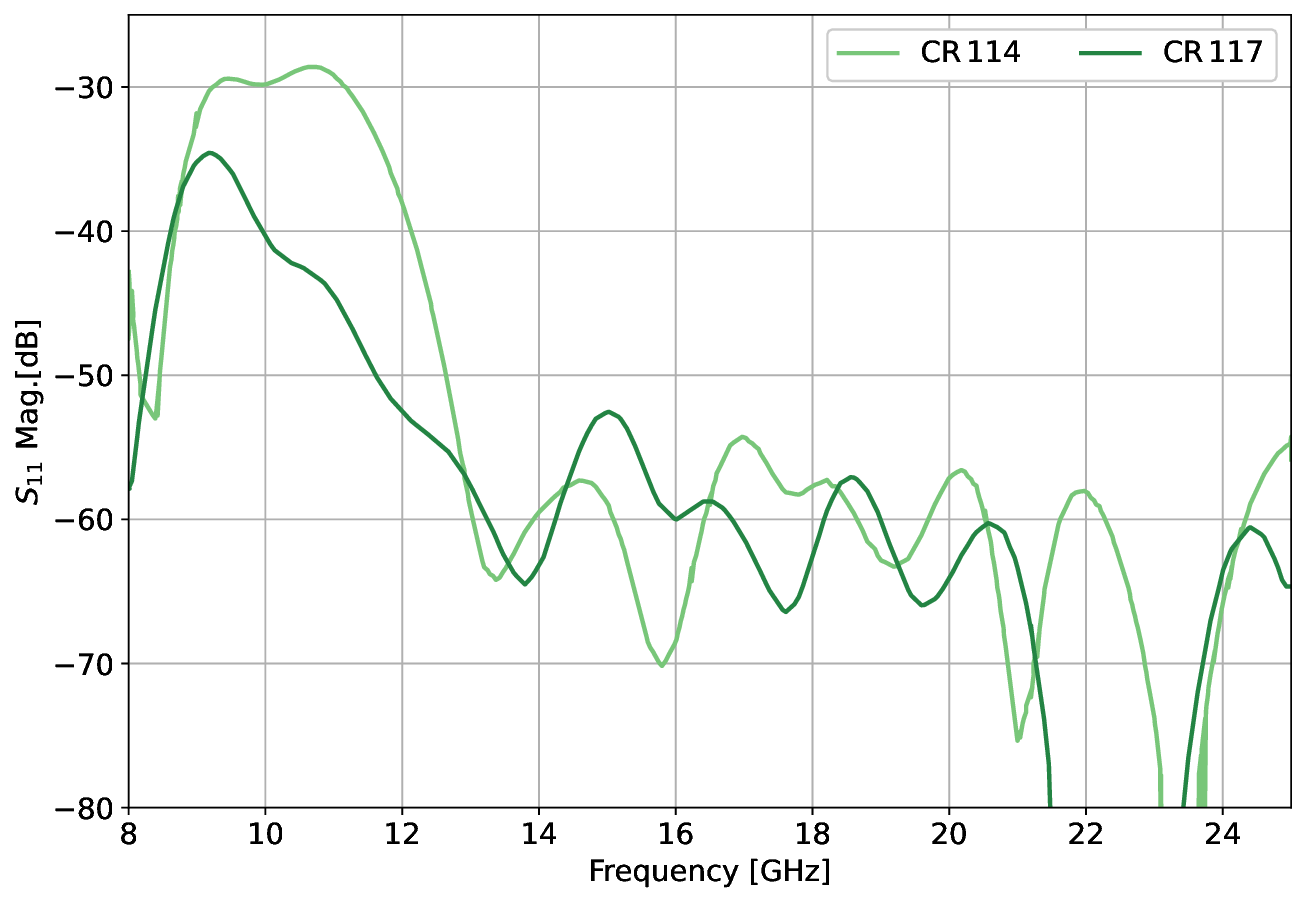}
\caption{Evaluation of the single layer design and choice of coating materials  based on $S_{11}$ over frequency. We compared single layer designs with CR\,114 and CR\,117. With the  CR\,117 based design, RF requirements are satisfied  ($\mathrm{S_{11}}<-$30\,dB).} 
\label{fig:load-S11-materials} 
\end{figure}

In addition, we ran different simulations to select the best option for the coating material. We assessed the RF properties of single-layer ($\mathrm{Th_{MAX}=1.9}$\,mm) designs with CR\,114 and CR\,117, and also a double-layer design based on a combination of both. The last option, even if more promising in terms of performance, was nevertheless abandoned due to the higher costs and risks connected to the complexity of the manufacturing. It will be in case reconsidered in case of future upgrades.  We present the results for the return loss up to 25 GHz in figure\,\ref{fig:load-S11-materials}. We finally opted for a single layer coating made of Eccosorb CR\,117, showing simultaneously better $\mathrm{S_{11}}$ and larger thermal conductivity.  

The analysis of the distribution of the dissipated power completes the RF characterisation of the load.  This study provides the input data for the intrinsic emissivity model of the target discussed in section\,\ref{sec:EEM}. In order to fully characterise the distribution of dissipated power in the load, we distinguished two different cases: the power loss density variation along the the load radius (the radial distribution) and along the pyramid height (axial distribution).

\begin{figure}[ht] 
\centering
\includegraphics[width=0.75\textwidth]{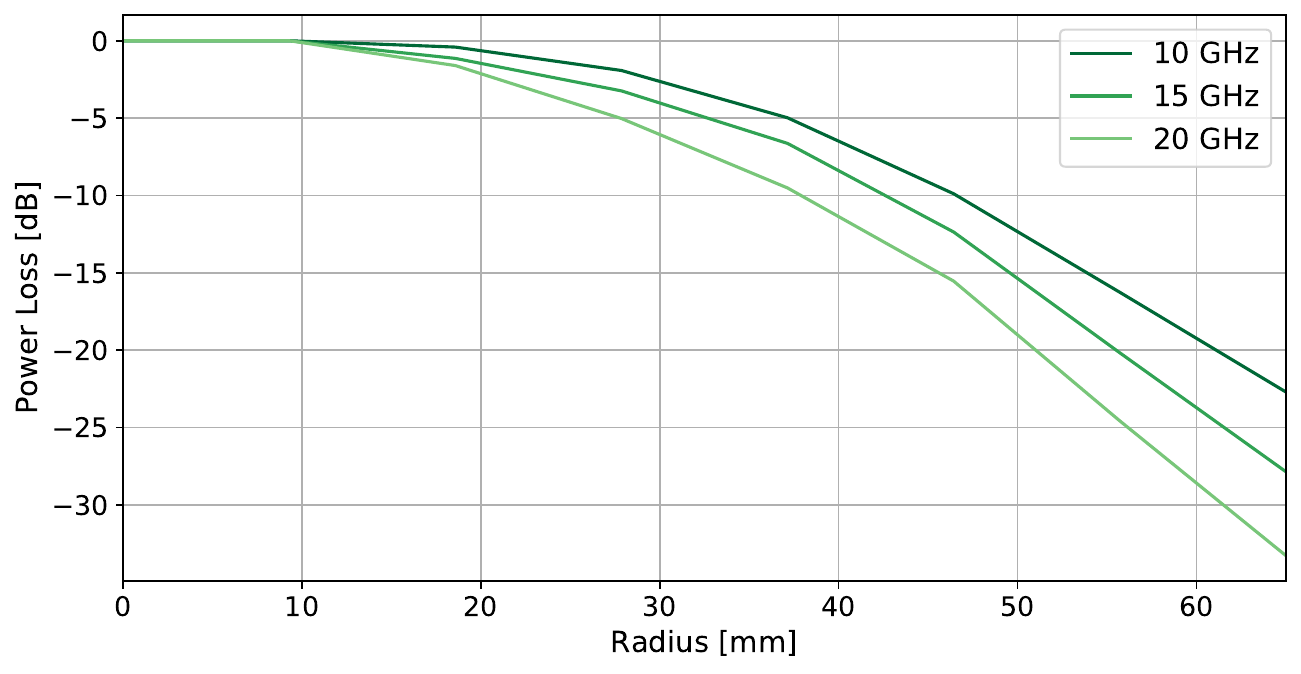}
\caption{Dissipated power radial distribution at 10, 15 and 20\,GHz. The power loss density was computed for each single pyramid along the maximum radial direction. The dissipated power is normalised to the maximum intensity at each frequency.} 
\label{fig:load-powerprofile-radial} 
\end{figure}

First, the power loss density was computed for each single pyramid along a radial direction. The symmetry of the design in both x- and y- directions results in a symmetry in the power loss distribution and allows a simple computation to obtain the radial variation. The normalised power loss is shown as a function of the radius, at three relevant frequencies, in figure\,\ref{fig:load-powerprofile-radial}.

The radial distribution of the power loss reflects a similar dependence with the distance from the centre and the frequency of operation as that seen in the NF results of the feedhorn (see section\,\ref{sec:feedhorn}). In terms of emissivity and total brightness, this feature plays a crucial role as the contribution of the central pyramids can be $\mathrm{10^3}$ higher than the contribution of the outlying ones at 20\,GHz. This feature comes to our help since it strongly relaxes the thermal homogeneity requirement (see section \ref{sec:techrequirements}).

\begin{figure}[ht] 
\centering
\includegraphics[width=0.8\textwidth]{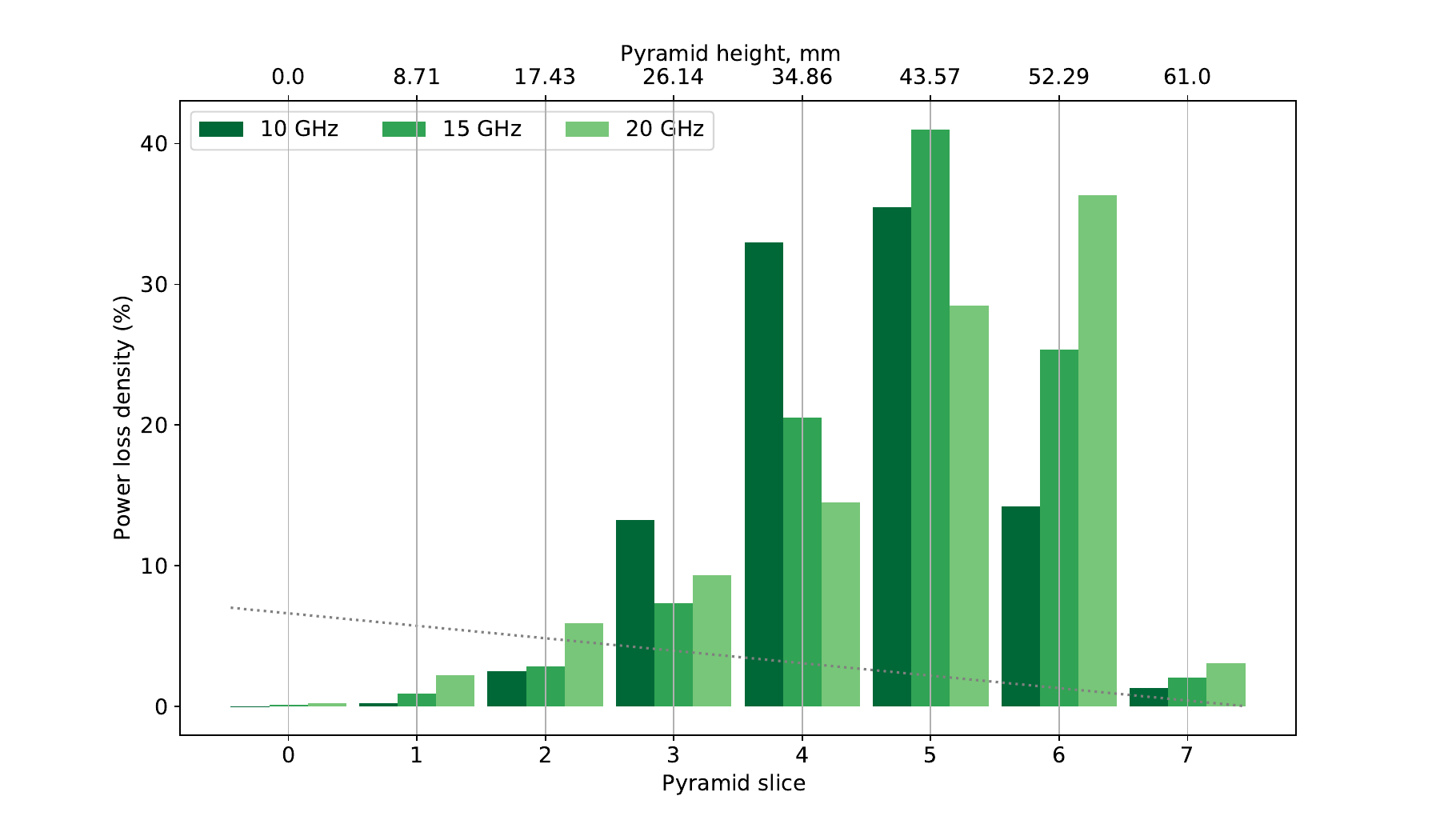}
\caption{Dissipated power profile at 10, 15 and 20\,GHz. Seven equal-height parts where considered to calculate the power loss axial distribution. The mechanism of dissipation is more relevant between 35--50\,mm from the base, moving towards the tip at shorter wavelengths.} 
\label{fig:load-powerprofile-axial}
\end{figure}

The axial distribution of the dissipated power was obtained by dividing  each pyramidal element of the target in seven equal-height slices and calculating the power loss density in each slice and then stacking and averaging. As a result, we obtained the fraction of the total dissipated power in each section, or depth, of a single pyramidal element. Results in figure\,\ref{fig:load-powerprofile-axial} show the height at which the mechanism of power dissipation is more relevant, between 35 and 50\,mm from the baseplate. At shorter wavelengths, the peak of maximum absorption moves towards the tip of the pyramids. 

\begin{figure}[ht] 
\centering
\includegraphics[width=1\textwidth]{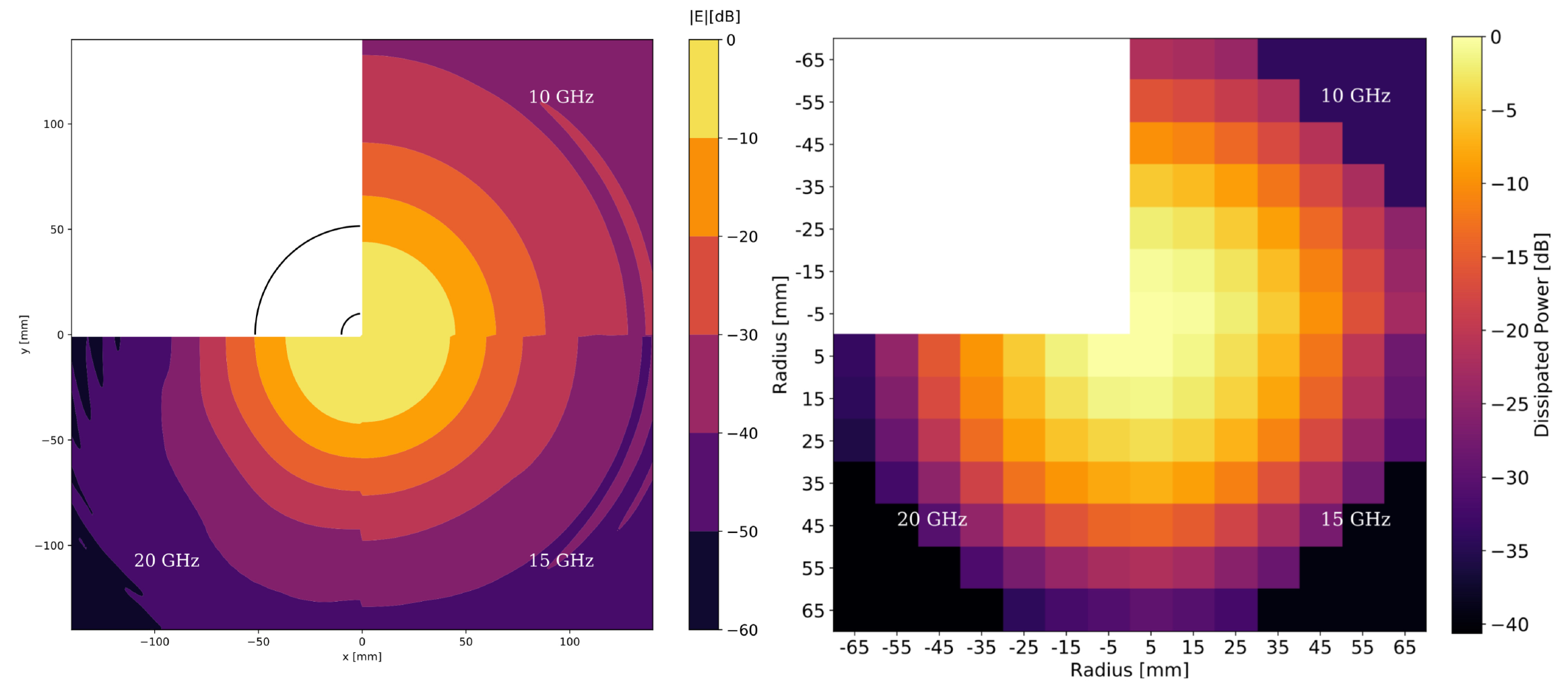}
\caption{\emph{Left:} Near Field frequency dependence at a distance of $d=$53\,mm.  The NF power is normalised to the maximum intensity at each frequency. \emph{Right:} Dissipated power distribution at 10, 15 and 20\,GHz.  The dissipated power has been normalised to the maximum intensity, corresponding at the central pyramids,  at each frequency.} 
\label{fig:NF-distance52} 
\end{figure}

The total emissivity and thus, the total brightness temperature, is due largely to this middle section, and to a lesser degree to the tips and base section of the load. In particular, slices 4 to 6  account for more than 80$\,\%$ of the total dissipated power, a factor to keep in mind when calculating the thermal gradients. Going back to the study in section\,\ref{sec:feedhorn}, it is interesting to obtain a representation consistent with the final result, considering the power distribution at the actual distance of maximum power dissipation. We considered the average of the height at which the peak of power dissipation is reached at each studied frequency, added to the 9\,mm of additional distance between the feedhorn aperture and the tips of the pyramids, being this the real location of the load.  We therefore recalculated the electric field at this distance of 53\,mm, obtaining the distribution shown in figure\,\ref{fig:NF-distance52}.

\subsection{Thermal simulations}\label{sec:thsimulation}

A thermal model of the load has been developed in ESATAN-TMS software package and a set of simulations has been performed to verify the final design compliance to requirements, in terms of thermal gradients and stability. The model includes both conductive and radiative heat transfers and their impact is evaluated at different temperatures of the feedhorn. In each case, the feedhorn is assumed to be a boundary surface, with a consistent and uniform temperature.

Results of the simulations confirm that the metal substrate allows a good thermal homogeneity, keeping gradients within desired levels when the load radiative environment, consisting of the corresponding feedhorn, stays within the expected temperature range. 

The steady state of the load has been simulated considering the cold finger at 4.2\,K, and a constant power of 0.2\,W on the active control flange, dimensioned in such a way that the load base become stable at a temperature of about 6\,K. A parametric study of the whole gradient across the load, both vertical along the pyramids axis and radial across the base, has been studied by varying the temperature of the feedhorn facing the load. Results are given in the table\,\ref{tab:th-steady}.

\begin{table}[ht]
\centering
\resizebox{0.55\textwidth}{!}{ 
\begin{tabular}{c c c}
\hline
\textbf{Feedhorn T (K)} & \textbf{4K Load Mean T (K)} & \textbf{4K Load gradient (K)} \\ \hline
5.0 & 6.197 & 0.001 \\
10.0 & 6.211 & 0.004 \\
50.0 & 6.369 & 0.072 \\
80.0 & 6.722 & 0.413 \\ \hline
\end{tabular} 
}
\caption{Estimated 4\,K load temperature distribution while exposed to different radiative environment temperatures. The gradient reported is the difference between the maximum and minimum temperatures evaluated across the whole load.}
\label{tab:th-steady}
\end{table}

\begin{figure}[ht] 
\centering
\includegraphics[width=0.45\textwidth]{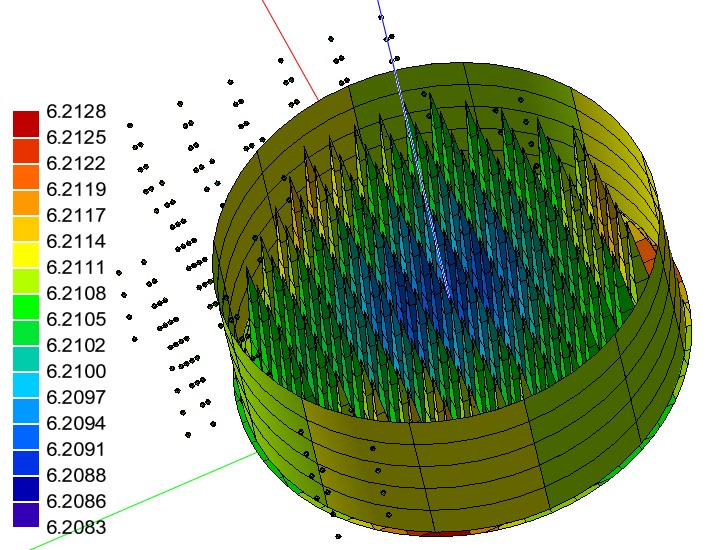}
\includegraphics[width=0.47\textwidth]{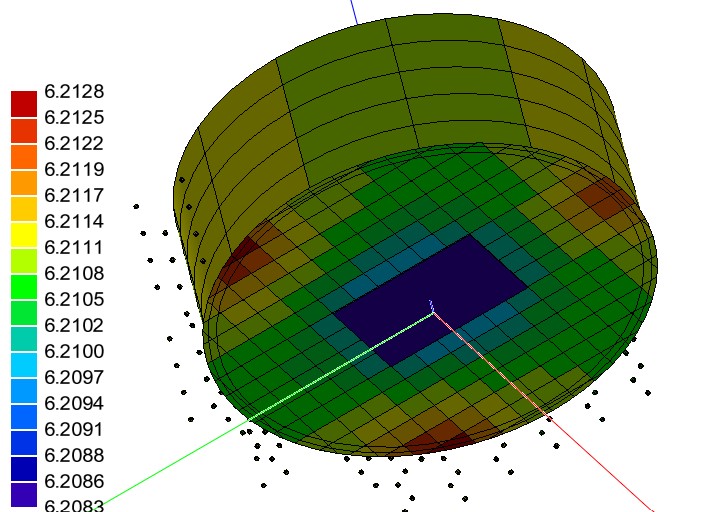}
\caption{Temperature distribution of the load with a 10\,K feedhorn temperature. \emph{Left}: View from the horn side of the load temperature distribution. \emph{Right}: View from the base side of the load temperature distribution. The cold spot where the cold flange is attached can clearly be seen in the centre of the load. Three symmetrical hot spots are seen around the load perimeter which correspond with the G10 supports to the feedhorn.} 
\label{fig:load-Tmap} 
\end{figure}

As is evident in  figure\,\ref{fig:load-Tmap}, the active control flange, connected to the cold head flange, is the load cold spot, while three hot spots can be seen at the interface of the G10 support struts coming from the horn assembly. The gradient is very limited also in this case; more relevant thermal inhomogeneity is found with higher feedhorn temperature, outside the expected operational range: these cases were indeed simulated in order to verify which level of temperature imbalance has to be created in the verification test phase in order to get reliable measurements of the 4\,K load temperature differences.

The temperature stability estimate is harder to be evaluated for two main limitations at an early phase of the project:
\begin{itemize}
    \item there are no measured temperature variation curves of the real system. We then performed simulations taking into account temperature curves of the same cold head model, expected to operate in the TMS cryostat, measured during tests in a different cryostat and setup;
    \item the active thermal control through a PID algorithm has to be carefully tested and tuned on the real systems to reach the required level of stability. The software routine  used for simulation can only give a rough assessment of the feasibility of control,  and an order of magnitude of the power needed. 
\end{itemize}

\begin{figure}[ht] 
\centering
\includegraphics[width=0.4\textwidth]{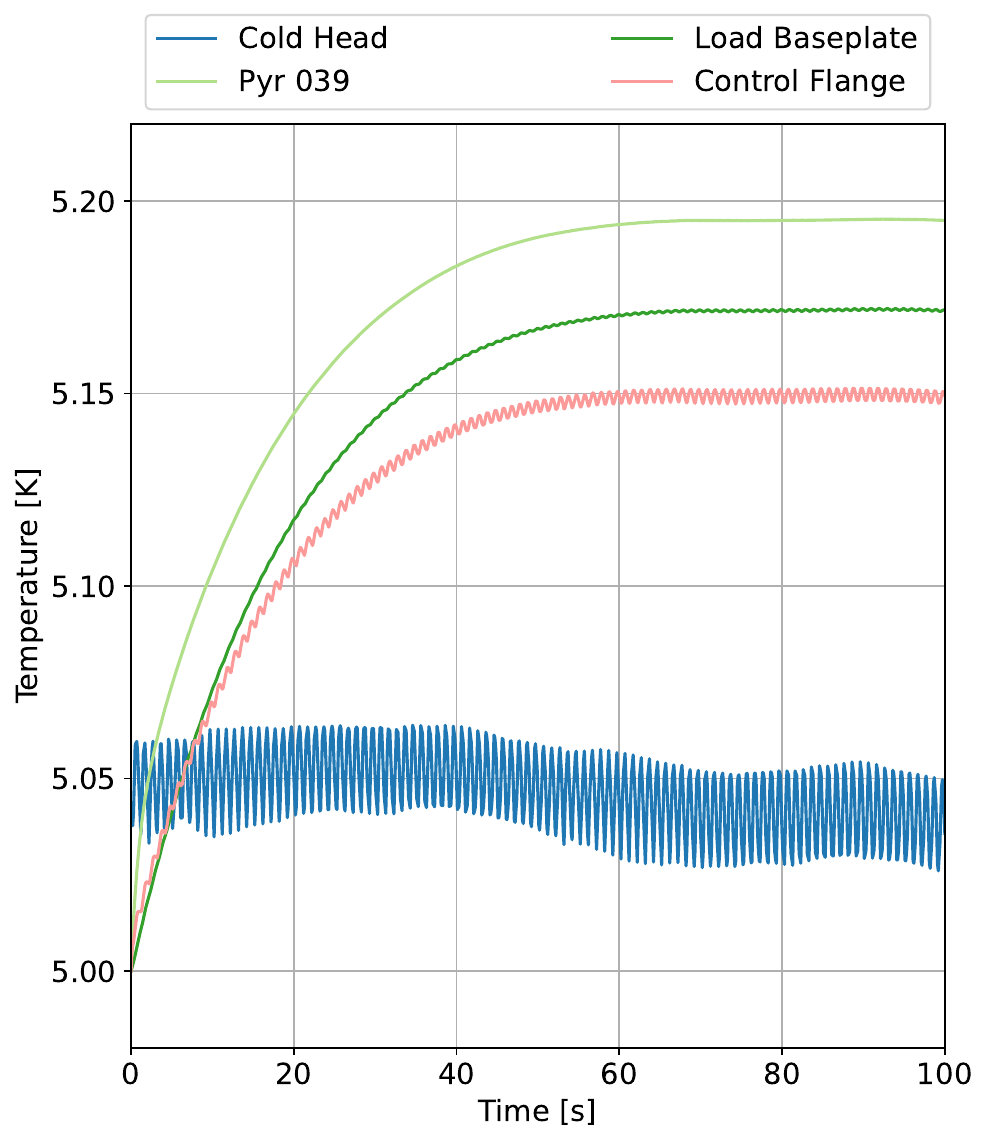}
\includegraphics[width=0.4\textwidth]{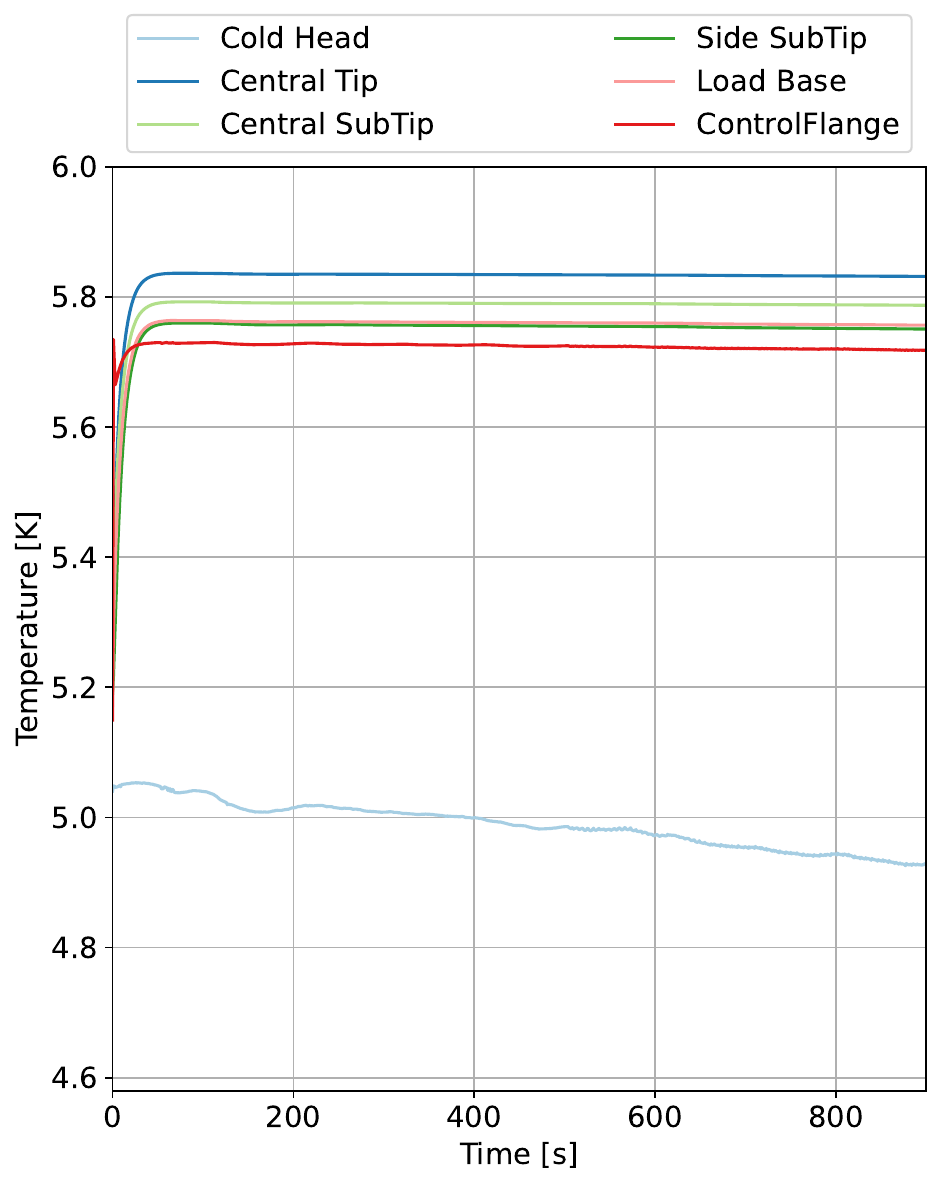}
\caption{\emph{Left}: Temperature curve evaluated at different locations of the load by the model with input oscillation from the cold head; no active control is applied. \emph{Right}: Temperature curve evaluated at different locations of the load by the model with input oscillation from the cold head; an active control based on a PID algorithm is simulated at the dedicated flange level.} 
\label{fig:transient-Tcurve} 
\end{figure}

Keeping in mind these limitations, we can conclude that results of transient simulations (shown in figure\,\ref{fig:transient-Tcurve}) give a good margin for achieving the desired stability.

\section{Performance verification and test results}\label{sec:measurements}

\subsection{RF Room Temperature Verification}\label{sec:RTtest}

We fully characterised the emissivity of the load using a feed transmitting towards the target load and performing a reflection measurement. Monostatic and a bistatic setup were implemented to obtain two different measurements, the specular return loss and the  diffusive return loss (spillover).

\subsubsection{Setup} \label{sec:RFsetup}

\begin{figure}[t] 
\centering
\subfloat{\includegraphics[width=0.60\textwidth]{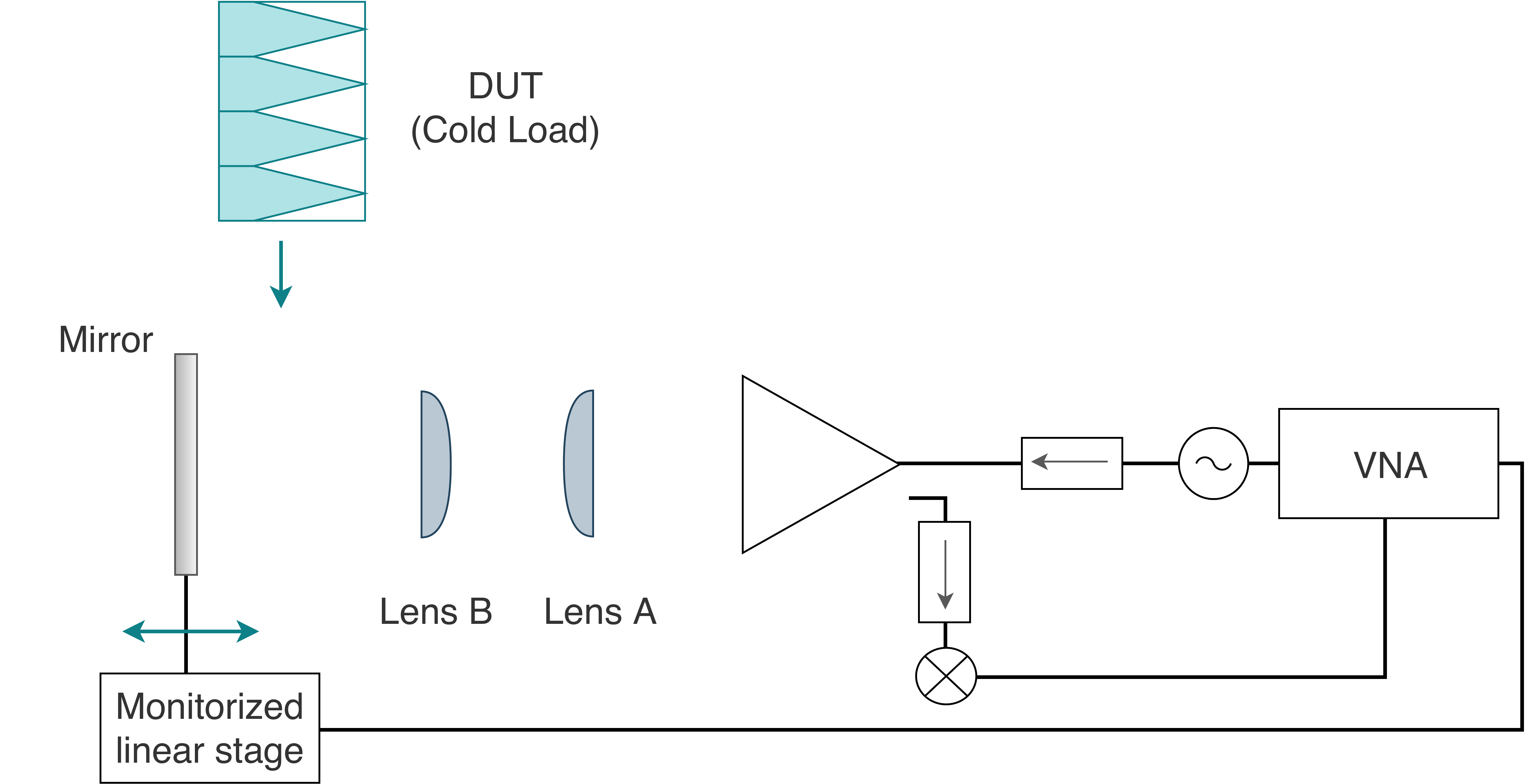}}
\qquad
\subfloat{\includegraphics[width=0.3\textwidth]{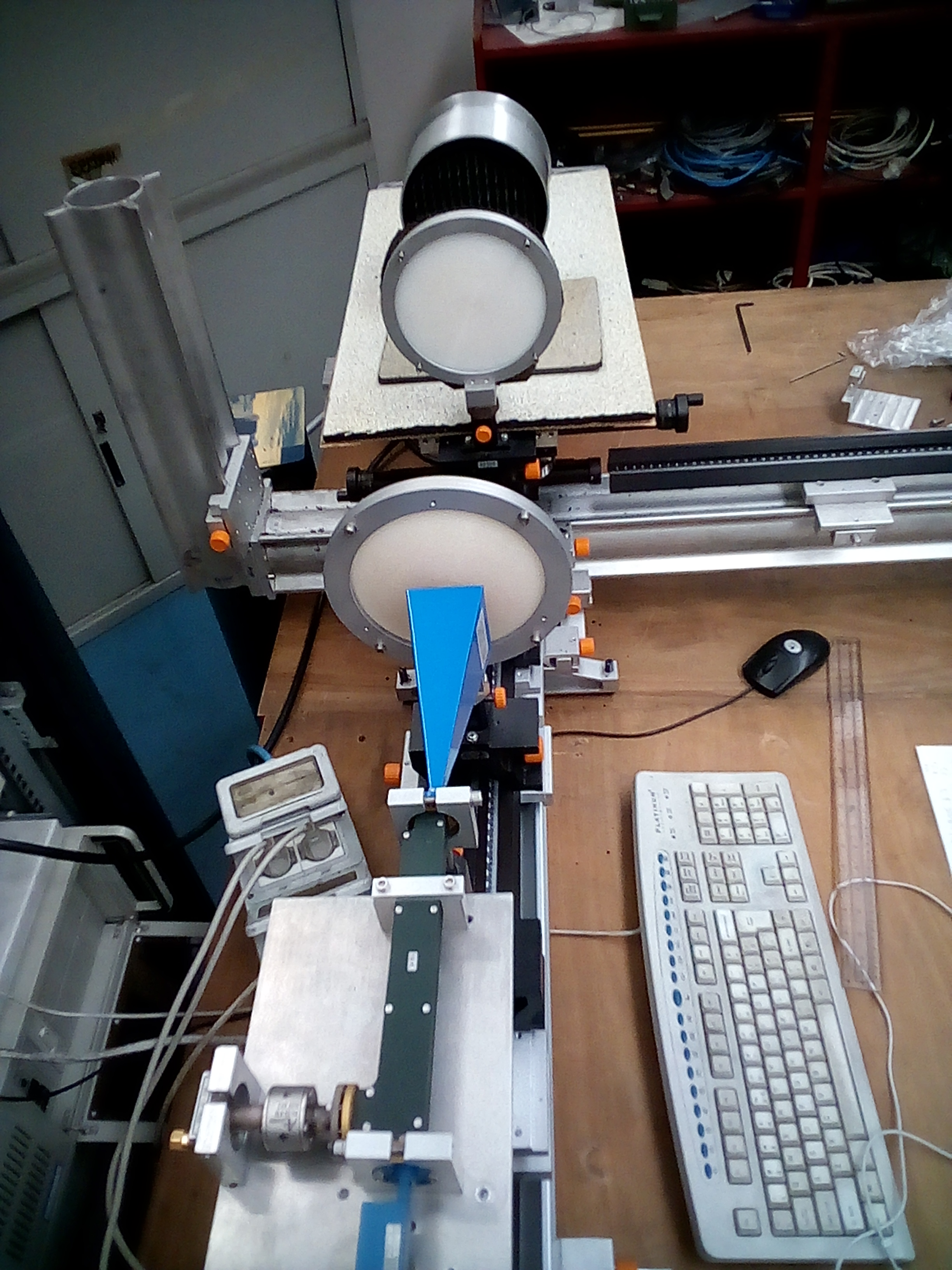}}
\caption{Monostatic setup to perform free-space measurements of low-reflectivity
targets. \emph{Left:} Illustration of the monostatic setup used to measure the on-axis or specular Return Loss. \emph{Right: } Photograph of the load integrated in the RF bench.} 
\label{fig:setup-monostatic} 
\end{figure}

The monostatic setup is shown in figure\,\ref{fig:setup-monostatic}; the optical bench allows the Device Under Test (DUT) to be displaced both longitudinally and transversally with great precision, through a dedicated mechanical support equipment. 
The VNA transmit and receive heads are connected to a directional coupler.
Ku and K band heads were used over the 9.6--18.6\,GHz and 17.4--20\,GHz bands respectively.
Standard gain horns, which have reasonably good coupling to astigmatic Gaussian beams, were used as antennas.
The setup is the same as \cite{Simonetto2021MillimeterwaveRT} and it is only briefly described here.
The two lenses are arranged as a confocal telescope, focusing the beam onto the load reference plane (its front aperture).

The (computed) size of the incident beam at the reference plane varies for Ku band between 40.1$\mathrm{\times}$36.7\,mm at 9.6\,GHz and 31.1$\mathrm{\times}$29.9\,mm at 18.6\,GHz, and for K band between 26.4$\mathrm{\times}$24.9\,mm at 17.4\,GHz and 24.5$\mathrm{\times}$23.7\,mm at 20\,GHz.

The setup was formerly envisaged for higher frequencies, so the size of the available lenses was sub-optimal: the diameter of the lens close to the antenna is 0.9 and 1.1 times the H and E plane diameter of the astigmatic beam at 9.6\,GHz, so truncation effects can not be considered negligible.
Already at 13\,GHz the lens-to-beam diameter ratio increases to much safer values of 1.25 and 1.2 in the two planes.

The DUT can be moved with high precision along the axis of the incident beam with a motorised linear stage. When the DUT is moved over approximately a half wavelength, assuming that all the other components of the test chain keep stationary, the measured return signal describes a circle in the complex plane, with a radius proportional to the DUT reflectivity and its centre mostly set by systematics (non-ideal components, environment reflections).
Calibration is made by comparison with a thin dedicated mirror (assumed perfectly reflecting) mounted in front of the Cold Load just 2\,mm above the tips of the pyramids. 
The ratio of the circle radii measured with the Cold Load and the mirror provides thus the reflectivity of the Cold Load. 
The extent of the linear movement is set as a half wavelength at the lowest frequency in the band, and circles are fitted in the complex plane. The standard deviation of the fit is used as the error bar.
Measurements were made in this configuration for both vertical and horizontal beam polarisation and for 0\textdegree\, and 45\textdegree\, azimuthal orientation of the Cold Load. 
A manual linear stage allows us to displace the DUT across the axis of the incident beam. The DUT mount allows us to rotate it in 45\textdegree\, steps around the axis of the incident beam (azimuth). We have verified the susceptibility of the measurement during operations to eventual misalignments between the horn and load. For this purpose, we have dedicated specific tests to measure the RL for horizontal displacements of 2.5, 5.0, 7.5, 10.0, 12.5\,mm for horizontal and vertical polarisation, and for azimuth orientation of the load at 0 and 45\textdegree. The largest displacement is greater than the side length of a pyramid. Both measurements (linear and angular displacement) were designed to verify the solidity of the design and the possible susceptibility of the performance on the fine feedhorn-load alignment, especially at cryogenic temperature.
In addition, a manual rotation stage allows us to tilt the Cold  Load around the vertical axis (elevation).
Monostatic measurements with off-axis incidence were made for $\pm$5, $\pm$2, $\pm$1\textdegree\, for the four combinations of beam polarisation and azimuth.

\begin{figure}[t] 
\centering
\subfloat{\includegraphics[width=0.60\textwidth]{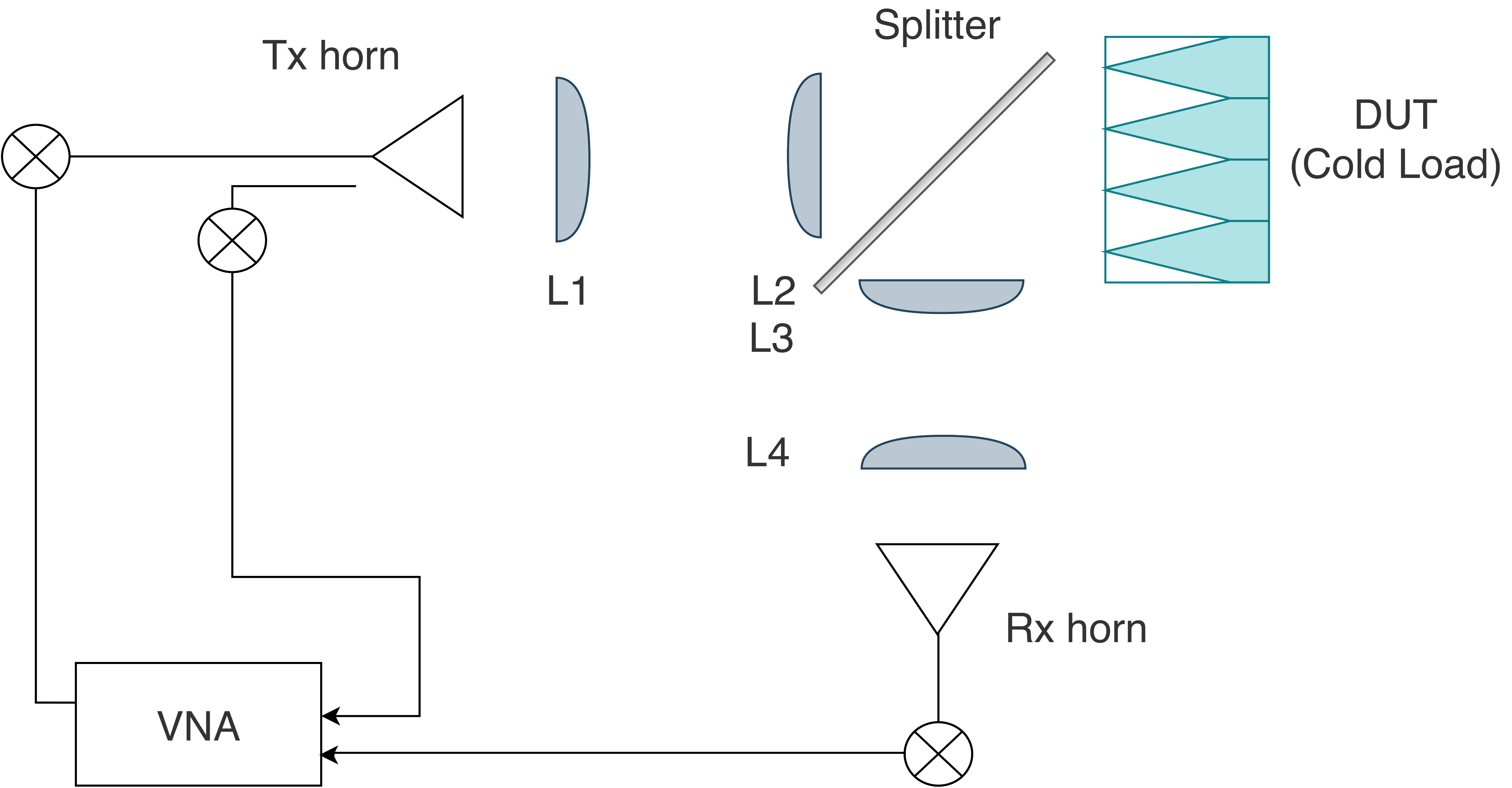}}
\qquad
\subfloat{\includegraphics[width=0.3\textwidth]{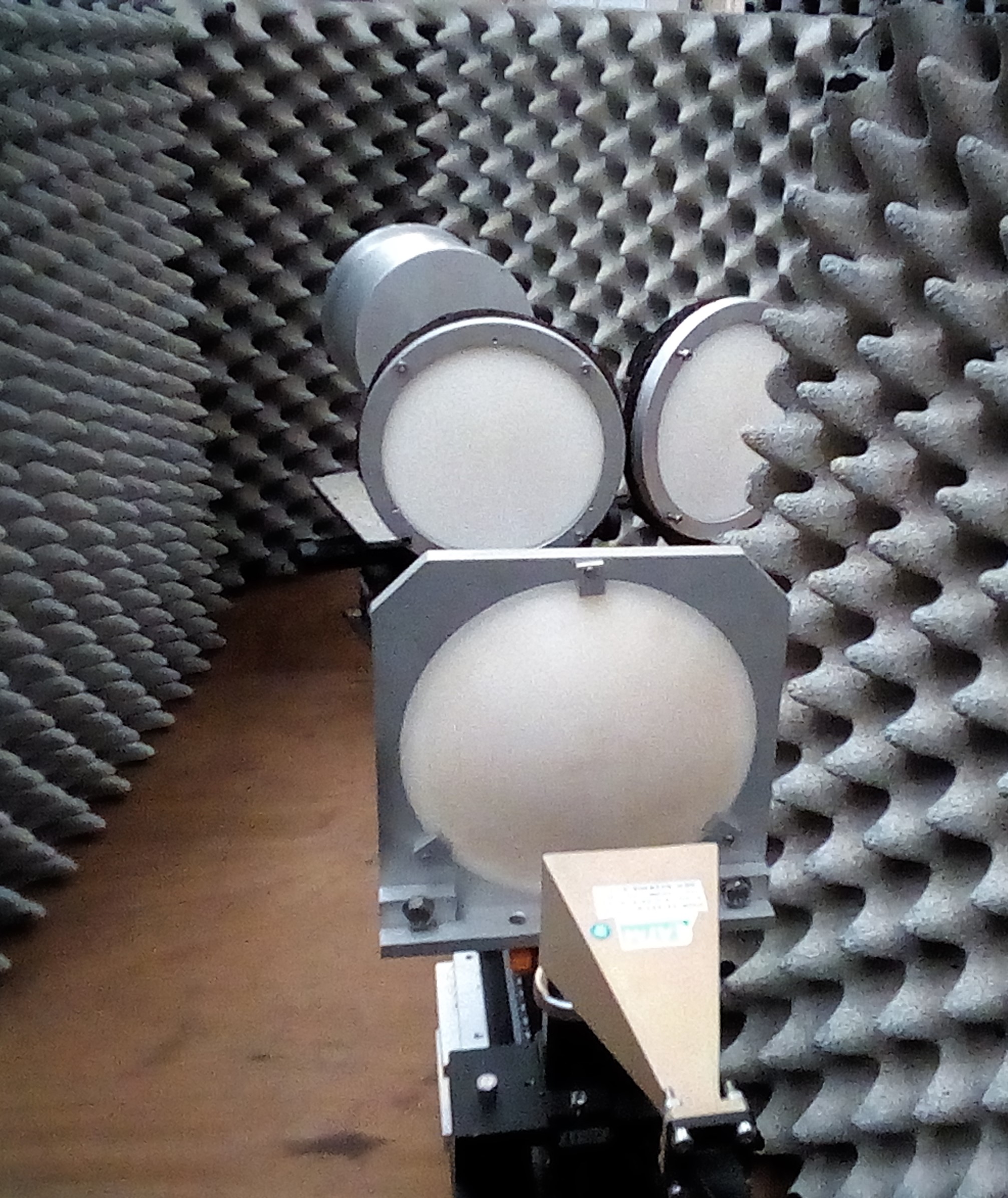}}
\caption{Bi-static setup to perform free-space measurements of low-reflectivity targets. Measurement of the off-axis diffuse Return Loss.} 
\label{fig:setup-bistatic} 
\end{figure}

To have an estimation of the spillover - intended as the radiation leaking into the feedhorn from the environment surrounding the load, through the small gap between the shield and the feedhorn- measurements were also made in a bistatic configuration, using the same setup as in \cite{Simonetto2021MillimeterwaveRT}; the scheme adopted is shown in figure\,\ref{fig:setup-bistatic}.
Two rails with identical confocal telescopes are hinged on a bench, the DUT being mounted above the fixed point where the two beams cross. The DUT can be rotated in the plane of the bench. The VNA transmitting (Tx) and receiving (Rx) heads are connected to the two arms. Calibration is made simply by correction of the response: the reflectivity is measured by the ratio of the measurement obtained with the DUT facing the incident beam and the calibration run, made with the mirror mounted on the Cold Load oriented for specular reflection.
Systematics are not as dominant as in the monostatic case, so that this simple calibration is adequate, as long as the measurement is made in a semi-anechoic environment: for this purpose, absorbing panels were interposed between the TX and RX arms to prevent crosstalk through antenna sidelobes.

\begin{figure}[t] 
\centering
\includegraphics[width=0.70\textwidth]{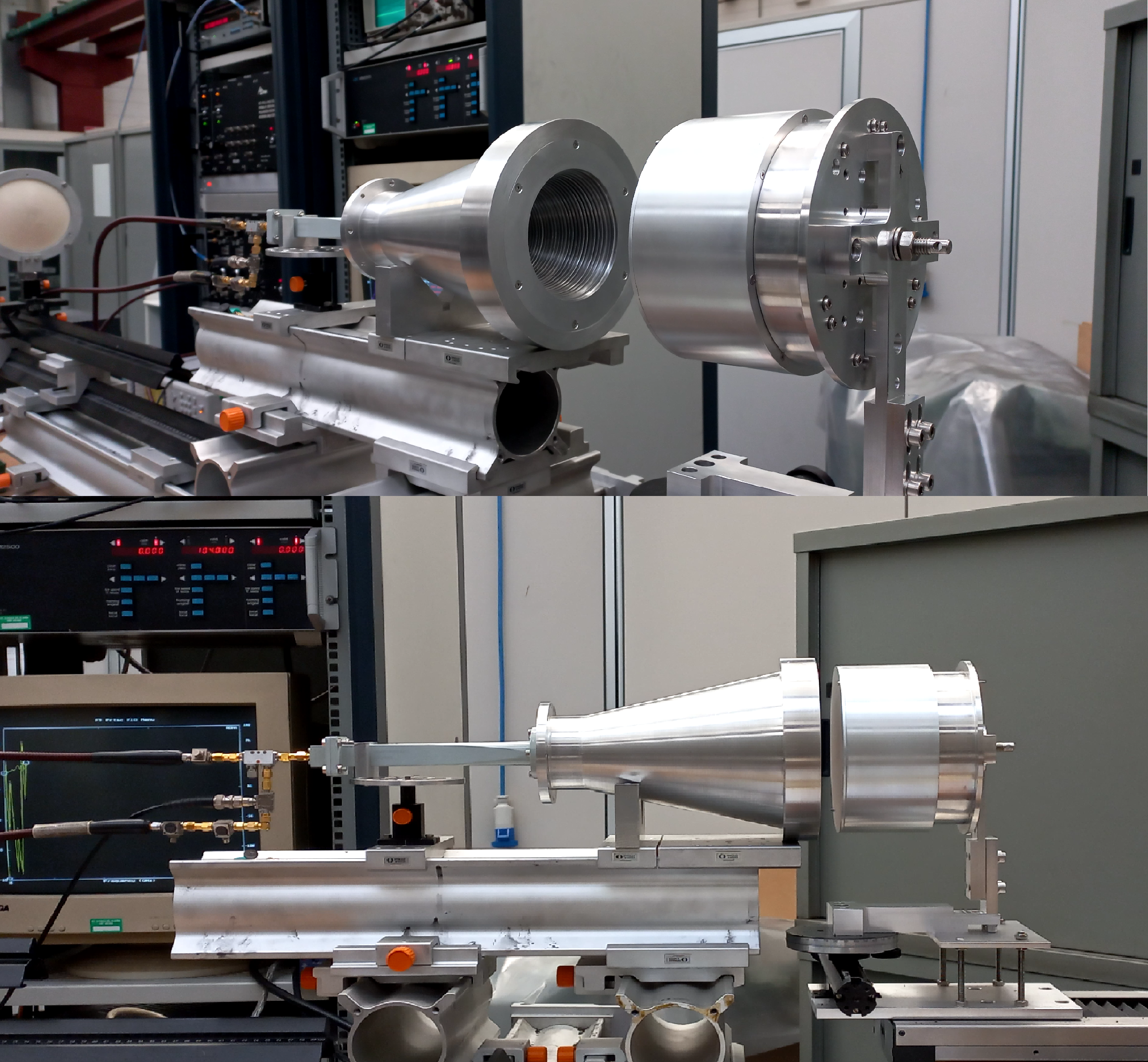}
\caption{Monostatic setup based on the TMS feedhorn. Measurement of the specular Return Loss.} 
\label{fig:TMS_FH_RL_setup} 
\end{figure}

Monostatic measurements were also repeated with the same calibration technique (i.e., moving the Cold Load), but using the TMS feedhorn instead of the confocal telescope coupled to standard horns. The purpose was to characterise the load reflectivity in the same experimental conditions in which the TMS instrument operates (figure\,\ref{fig:TMS_FH_RL_setup}).

 The 4\,K Load design optimisation was based on the EM input data (dielectric and magnetic loss tangent, permittivity , permeability) available in literature at these frequencies. Unfortunately, there are available data only at 10\,GHz and 18\,GHz in the range relevant for TMS. The same data, at cryogenic temperatures, are missing and can only be extrapolated by model. Based on \cite{almamemo}, we can infer that the CR\,117 (Eccosorb CR series having the RT properties  closest to MF\,116) transmission loss at 5\,K, could be about  90\,\%, of the value measured at 300\,K; even if this behaviour is mapped at higher frequencies, basing on the model it should be frequency independent. In addition, we see that the change with temperature decrease passing from MF\,112 (having the lower transmission loss) to MF\,116 (having larger transmission loss). If we base instead on measurement performed at 40\,GHz on CR\,117 (having transmission loss about double that of MF\,116, see \cite{eccosorb}), within the Planck LFI 4\,K reference Load development program \cite{CuttaiaThesis1}, \cite{CuttaiaThesis2}, the CR\,117 transmission Loss is expected not change much from 300\,K to 5\,K; this behavior is all the more evident the greater the metal component present in the Eccosorb CR series. The uncertainty in the behaviour with frequency does not look as large as to compromise the validity of the overall design. For all these reasons, the possible degradation of reflection loss expected at the operating temperature (4\,K--10\,K) compared to the value at the verification temperature (300\,K) is such as to allow the requirement emissivity $\mathrm{e\geq 0.999}$ to be met with a large margin. Nevertheless, dedicated measurements are envisaged to measure the EM parameters of CR\,117, in the bandwidth 10--20\,GHz, at temperature down to 7\,K and to possibly characterise the reflection loss of the 4\,K Cold Load at cryogenic temperatures.

\subsubsection{Test results} \label{sec:TESTresults}
\begin{figure}[ht]
\centering
\includegraphics[width=0.7\textwidth]{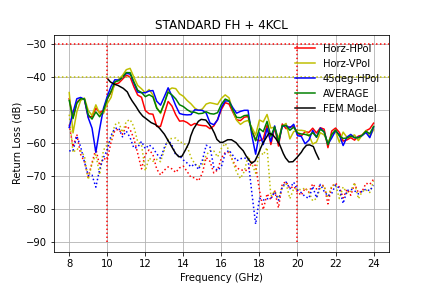}
\caption{Cold Load reflectivity (\emph{solid lines}) and precision (\emph{dotted lines}) [dB] for normal incidence,  horizontal and 45\textdegree\, azimuth, for the vertical and horizontal polarisations. The averaged values are shown in green. Two horizontal dotted lines highlight the $-$30\,dB RL requirement (\emph{{red}}) and the $-$40\,dB RL goal (\emph{yellow}). RL was measured in an extended frequency range, and the nominal operational frequency range, 10--20\,GHz, is highlighted by the two vertical dotted red lines. The experimental data are compared to the FEM Model. Even if the curves differ,  they are very close to the average return loss ($-$51.0 and $-$50.2\,dB in the range 10--20\,GHz) and the local values at\,10 and 18\,GHz, which are the only frequencies at which the electromagnetic parameters are known from experimental data (the input parameters for the FEM simulations have been linearly interpolated in frequency).}
\label{fig:NormalIncidenceAllPolPhi}
\end{figure}

\begin{figure}[ht]
\centering
\includegraphics[width=.7\textwidth]{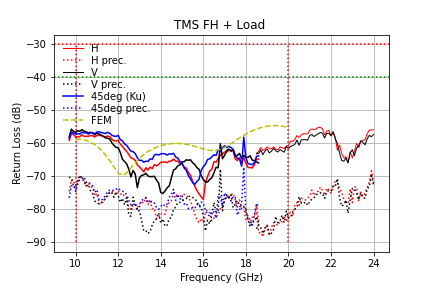}
\caption{Specular reflectivity of the TMS-feedhorn + Cold Load system.  The RL was measured with the load positioned horizontally and at 45\textdegree, for both vertical (V, \emph{black}) and horizontal (H, \emph{red}) polarisations. The $-$30\,dB RL requirement and $-$40\,dB RL goal are highlighted with \emph{red} and \emph{green, dotted, lines}, respectively. The RL was also measured outside the nominal operational frequency range (10--20\,GHz,  highlighted by the two vertical \emph{dotted red lines}). The FEM model (\emph{yellow dashed line}) is superimposed in the nominal frequency range (10--20\,GHz).}
\label{fig:TMS_FH_Load_all}
\end{figure}

The Cold Load reflectivity measured at normal incidence with the confocal telescope shows a general agreement with the free-space simulations of figure\,\ref{fig:load-S11},  
showing the largest deviations at low frequency, where the measurement setup is sub-optimal due to the mentioned undersized confocal lenses.
The specular reflectivity was tested over a frequency range much wider than the nominal, between 8--24\,GHz. Return loss is much better than the $-$30\,dB goal in this extended range, and also better than the $-$40\,dB design goal for most of the band (figure\,\ref{fig:NormalIncidenceAllPolPhi}). The same measurement made with the TMS feedhorn (figure\,\ref{fig:TMS_FH_Load_all}), gives an appreciably lower reflectivity, as discussed below.

\begin{figure}[t] 
\centering
\subfloat{\includegraphics[width=0.8\textwidth]{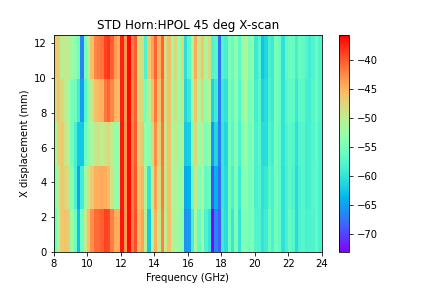}}
\quad
\subfloat{\includegraphics[width=0.8\textwidth]{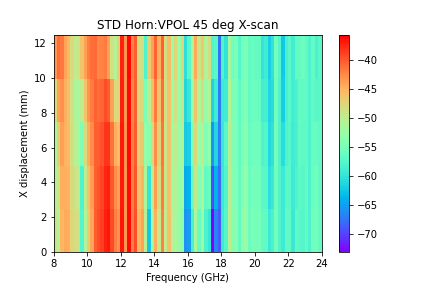}}
\caption{The two spectrograms show the Cold Load reflectivity for a displacement across the incident beam up to 12.5\,mm.
Normal incidence, horizontal and vertical polarisation, azimuthal orientation 45\textdegree. RL is always better than $-$35\,dB over the full extended range and better than $-$40\,dB outside the range 11--13\,GHz. Results are similar for H and V polarisation.}
\label{fig:45LoadVPolXscan}
\end{figure}

\begin{figure}[t] 
\centering
\subfloat{\includegraphics[width=0.8\textwidth]{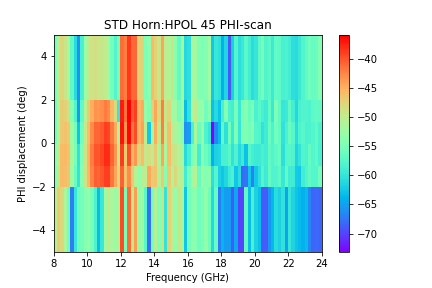}}
\quad
\subfloat{\includegraphics[width=0.8\textwidth]{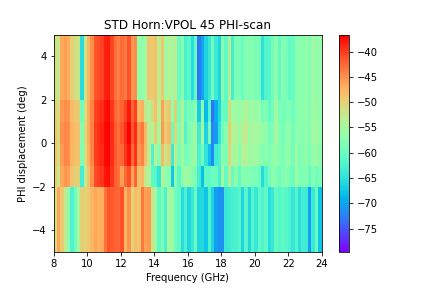}}
\caption{Spectrograms showing the 4KCL RL for normal incidence, elevation of $\pm$5, $\pm$2, $\pm$1\textdegree and azimuthal orientation of 45\textdegree. RL results are similar for the horizontal (H, \emph{top}) and vertical (V, \emph{bottom}) polarisations. RL remains lower than $-$30\,dB over the full extended range, and better than $-$40\,dB outside the range 12--13\,GHz.}
\label{fig:45dgVPolOffNormal}
\end{figure}

\begin{figure}[t] 
\centering
\subfloat{\includegraphics[width=0.8\textwidth]{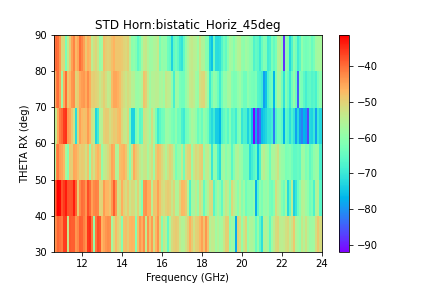}}
\quad
\subfloat{\includegraphics[width=0.8\textwidth]{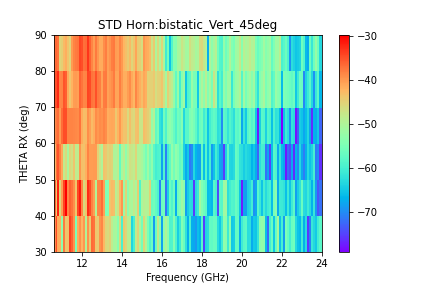}}
\caption{Spectrograms showing the 4KCL RL measured with the bistatic configuration for scattering angles from 30 to 90\textdegree (azimuthal orientation of 45\textdegree). RL results are similar for the horizontal (H, \emph{top}) and vertical (V, \emph{bottom}) polarisations. RL remains lower than $-$30\,dB over the full extended range, and better than $-$40\,dB outside the range 11--12\,GHz, and at any frequency for angles greater than 50\textdegree.}
\label{fig:STD_FH_bistatic}
\end{figure}

The susceptibility to translational displacements along the horizontal axis (parallel to the RF test bench) from 0\,mm (feedhorn aligned with the load) and 12.5\,mm, is shown in figure\,\ref{fig:45LoadVPolXscan}. The susceptibility to small angular displacements of the load around the vertical axis perpendicular to the RF test bench (elevation of $\pm$5, $\pm$2, $\pm$1\textdegree) is shown in figure\,\ref{fig:45dgVPolOffNormal}. The diffusive reflectivity was tested in the bistatic setup shown in figure\,\ref{fig:setup-bistatic}. Measurements were made at 10\textdegree~angular steps, for angles between the TX and RX antenna in the range 30\textdegree--90\textdegree. The specific setup used did not allow RL measurements at angles of less than 30\textdegree. Results are shown in the two spectrograms in figure\,\ref{fig:STD_FH_bistatic}.

Figure\,\ref{fig:TMS_FH_Load_all} shows the results obtained for the setup sketched in figure\,\ref{fig:TMS_FH_RL_setup}, where the load is directly coupled to the TMS feedhorn, and the confocal telescope is not used. Reflectivity improved compared to that measured with the confocal setup. The confocal telescope launches and receives a beam with a far-field divergence varying between 9 and 12\textdegree~in Ku band, and between 11 and 12\textdegree~in K band. On the other hand, the beam launched by the feedhorn is essentially a pure $\mathrm{HE_{11}}$ in an aperture of 103\,mm, which is best coupled with a gaussian beam with waist of 33.1\,mm and far-field divergence of 16.1\textdegree at 10\,GHz and 8.2 at 20\,GHz.

A plausible explanation lies in the different selectivity of the antennas. The standard gain horns of the confocal telescope have 88\,\%\, power coupling with the astigmatic gaussian beam, whereas the feedhorn has a 98\,\%\,coupling with a gaussian beam. 
The Cold Load is necessarily scattering reflected power into a large number of modes, which are coupled much more easily (12\%\,against 2\%) into the telescope's antenna than in the TMS feedhorn. 
Whilst the data are very noisy, the average difference in the levels is compatible with this model. 
Results from Finite Element Model simulations are superimposed to the experimental results. Simulation is based on a unitary cell including only one  pyramid, a Floquet port as RF source, Master and Slave symmetries to indefinitely replicate the pyramidal geometry over X and Y axis. Eccosorb CR117 electromagnetic parameters are the same as those used in the previous models. Measures span the range between\,8 and 24\,GHz. Even though the nominal setup (TMS horn + Cold Load) provides much better results, also the measurements made with the standard horn --- affected by the mentioned features causing a degradation in the measured Return Loss --- suggest performance far better than the RF design requirement.

Concerning the diffusive reflectivity measurements, spillover is calculated as the  value integrated over the full sphere. The resulting value  must be strictly considered as a property of the load (fully immersed in a 4\begin{math}\pi\end{math} sphere). However, in the TMS  architecture, the mechanical gap between the feedhorn and the load is negligible (less than 1\,mm). Taking this into account, the direct effective spillover must be considered just a small fraction  of the already very low value measured using the two confocal telescopes setup, and hence negligible. The indirect diffusive radiation (coming from angles different from those subtended by the circular gap) is instead further dumped by the multi-reflections on the load.

\subsection{Cryogenic verification} \label{sec:Cryotest}
From the thermal point of view, the verification consisted in 
\begin{itemize}
    \item twenty thermal vacuum cycles on a prototype sample of $\mathrm{6\times6}$ pyramids (figure\,\ref{fig:thcycles}), sharing the same design of the final load: it was a set of 20 thermal cycles between room temperature down to 4\,K and back to room;
    \item a thermal vacuum verification on the actual load, consisting of 
        \subitem -three cycles: the load was cycled between room temperature down to 4\,K and back.
        \subitem -a thermal balance test aimed at validating the load cryogenic design and at correlating to thermal model predictions.
\end{itemize}

\subsubsection{Thermal cycling (aging/stress)}  \label{sec:ThCycling}

\begin{figure}[ht] 
\centering
\includegraphics[width=0.25\textwidth]{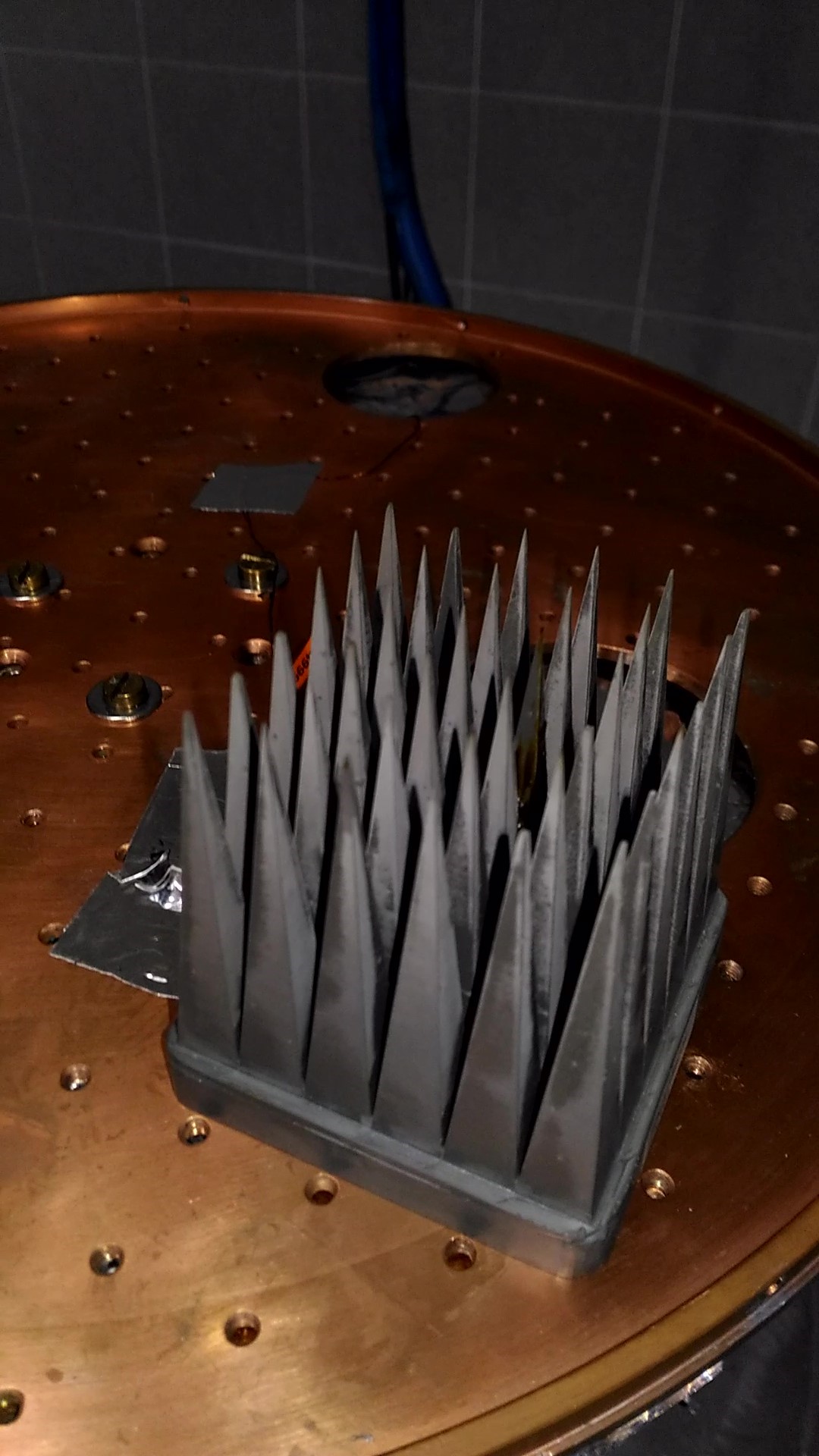}
\includegraphics[width=0.65\textwidth]{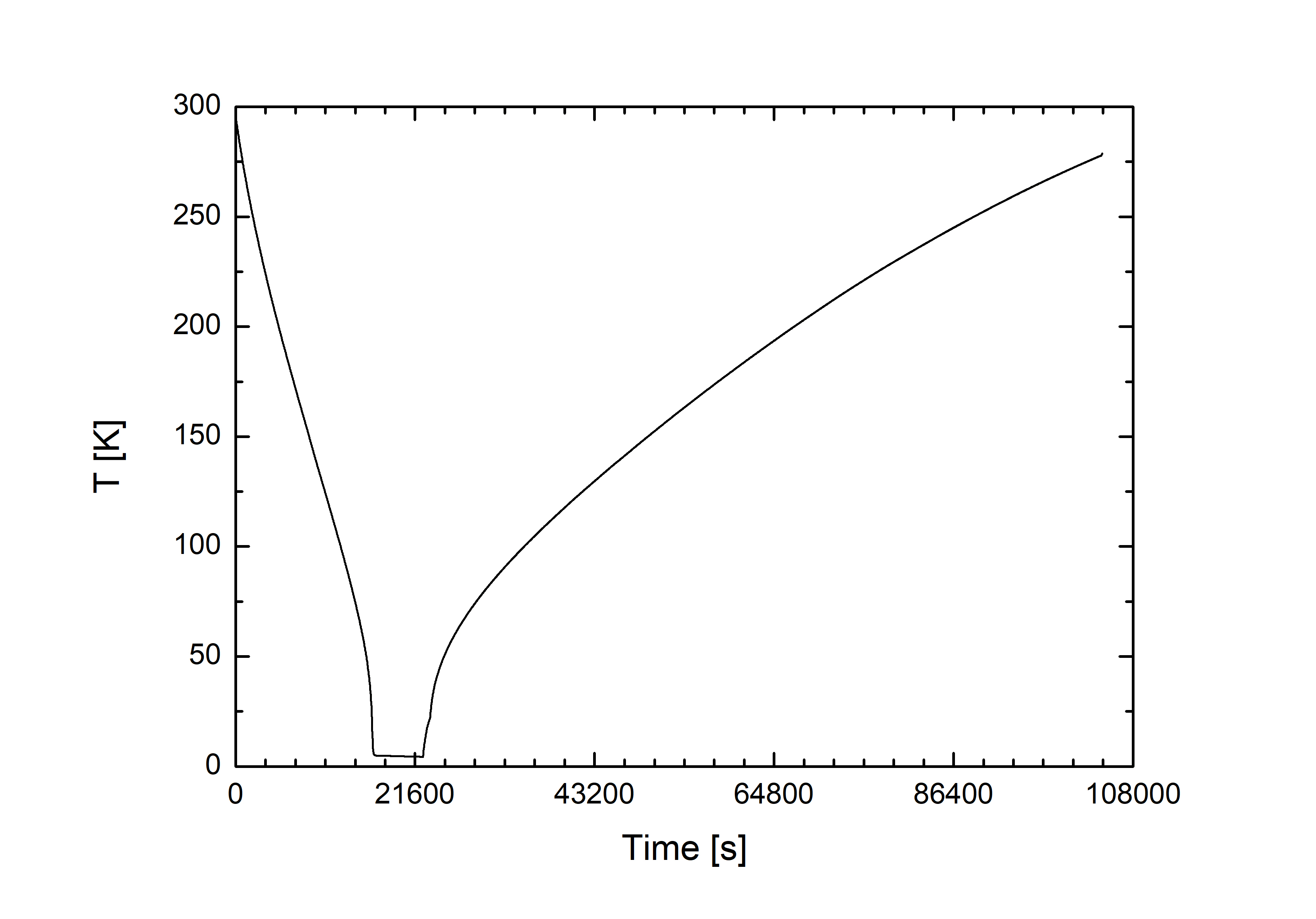}
\caption{\emph{Left}: Setup of the load sample cryogenic cycles; the device is screwed directly on the cold flange of a 4\,K cryogenic facility. \emph{Right}: Cold flange temperature curve during one of the cycles (cycle 17).} 
\label{fig:thcycles} 
\end{figure}

A sample with the same design of the load but limited to a square section of $\mathrm{6\times6}$ pyramids was integrated in the cold stage flange (figure\, \ref{fig:thcycles}, \emph{left}) of one of the cryogenic chamber of the \emph{Cryowaves} laboratory at INAF OAS Bologna. The chamber is supplied with a Sumitomo Heavy Industries two-stage cold head model RDK-415D, providing 1.5\,W heat lift at the 4\,K cold stage.

The sample experienced 20 cycles from room temperature down to below 5\,K  following the typical profile shown in figure\,\ref{fig:thcycles}, \emph{right}); the minimum dwell time was 30 minutes and the cooldown and warmup rate was less than 10\,K/min. The visual inspection of the load during and after the cycling process showed no effect. 

The nominal TMS 4K load (figure\,\ref{fig:thsetup}) underwent three cycles between room temperature and 3K during the thermal verification campaign. The load did not show any anomaly on visual inspection performed at the end of the third run.

\subsubsection{Thermal balance}  \label{sec:ThBalance}

In order to validate the load cryogenic design a thermal balance test was planned. The test consisted of a set of steady state and transient verification tests. 
The steady state tests were designed in order to verify the load performance with respect to the thermal gradient and stability at the operating cryogenic conditions, in a facility provided with the same cooler cold head type as the TMS cryostat. The environment surrounding the target consisted of an Aluminium shield in order to be representative of the feedhorn facing the load during nominal operation of the instrument.

\begin{table}[ht]
\centering
\resizebox{0.9\textwidth}{!}{ 
\begin{tabular}{l c c l}
\hline
\textbf{Sensor Id}	& \textbf{Type – Package}	& \textbf{Mounting}	& \textbf{Location}\\ \hline
TMS-0 & Si Diode – CU (std curve) & Screw & 4K CL control flange\\
TMS-1 &	Cernox – SD (cal) & Permanent glue & 4K CL base side hole\\
TMS-2 & Cernox – SD (cal) & Permanent glue & 4K CL base central hole\\
TMS-3 & Cernox – CU (cal) & Al tape	& 4K CL baffle side\\
TMS-4 &	Cernox – CU (cal) & Screw	& 4K CL base (opposite to TMS-1)\\
TBT-0 &	Si Diode – CU (std curve) & Screw & Cryostat 4K Flange\\
TBT-1 &	Cernox – SD (cal) & Heat shrink & 4K CL side pyramid (over TMS-1)\\
TBT-2 &	Cernox – SD (cal) & Heat shrink & 4K CL central pyr (over TMS-2)\\
TBT-3 & Si Diode – CU (std curve) & Screw & Radiative shield top\\
TBT-4 & Si Diode – CU (std curve) & Screw & Radiative shield Base\\ \hline
\end{tabular} 
}
\caption{List of temperature sensors used during the tests. SD package is a small one suitable for mounting it on small or irregular surfaces, like pyramid tips in this case.}
\label{tab:Sensor_List}
\end{table}

\begin{figure}[t] 
\centering
\includegraphics[width=1.0\textwidth]{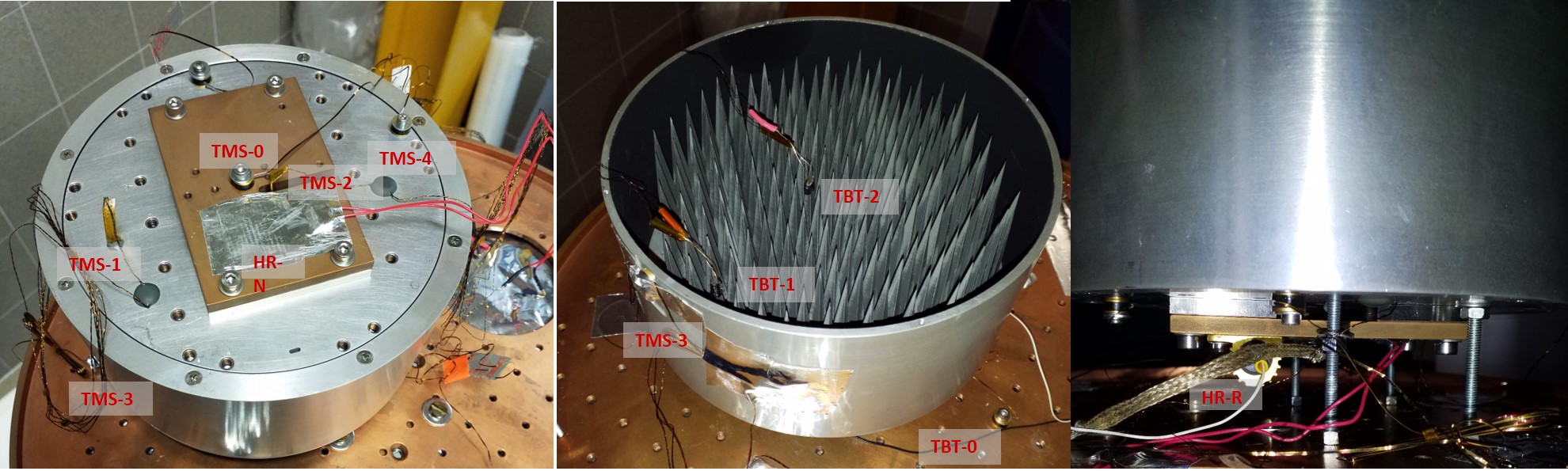}
\includegraphics[width=1.0\textwidth]{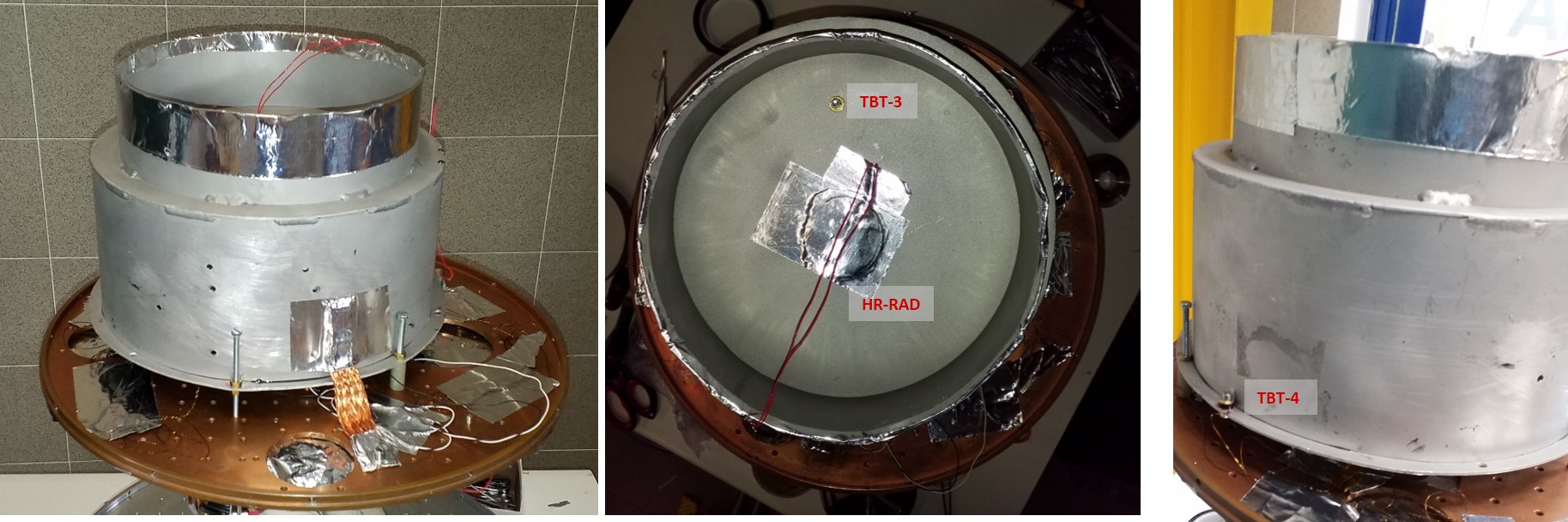}
\caption{Setup of the thermal verification test. Position of the sensors and heaters is also specified: \textit{TMS} labelled sensors are the device sensors that will be kept installed during TMS operations, while \textit{TBT} labelled sensors are sensors dedicated to the tests. \emph{Top}: Bottom, top and side view of the cold load mounted on the cryostat. \emph{Bottom}: Different views of the radiative shield, working as reference radiative environment for the load during tests.} 
\label{fig:thsetup} 
\end{figure}

Both the set of nominal TMS temperature sensors (labelled as \textit{TMS} in the corresponding figures and tables) and a dedicated set of additional temperature sensors (\textit{TBT}) were used in the tests, in order to get as complete as possible overview of the thermal behaviour of the device (figure\, \ref{fig:thsetup}).
The sensors used are Lakeshore\texttrademark~Cernox resistors and silicon diodes (table\,\ref{tab:Sensor_List}); each Cernox is provided with its calibration curve, claiming an intrinsic accuracy of about 7\,mK, taking into account additional uncertainties in the mounting and harness routing, an overall accuracy of 15\,mK is considered for this kind of sensors. Silicon diode standard curve accuracy is 0.25 K; they were actually used for monitoring and control. Sensitivity of both the sensors type is better than 1\,mK. 

A first set of steady state test consisted in setting both the cold load control flange and radiative shield setpoints at the same control temperature and measure the temperature distribution in the load. This test has provided  the confirmation that the gradient and stability of the load in operational conditions, i.e., load and feedhorn controlled at the same temperature, are meeting the requirement specifications. 

\begin{table}[ht]
\centering
\resizebox{0.7\textwidth}{!}{ 
\begin{tabular}{l c c c}
\hline
\textbf{Control T [K]} & \textbf{Gradient [mK]} & \textbf{Max $T_{rms}$ [mK]} & \textbf{Max $\Delta T_{p-p}$ [mK]} \\ \hline
6.0 & $27 \pm  30$ & 0.4 & $2.2 \pm  2$ \\
7.0 & $25 \pm  30$ & 0.3 & $3.4 \pm  2$ \\
8.0 & $26 \pm  30$ & 0.4 & $1.5 \pm  2$ \\
10.0 & $27 \pm  30$ & 0.4 & $1.9 \pm  2$ \\ \hline
\end{tabular} 
}
\caption{Results of the isothermal steady state test. The gradient is the maximum difference measured among the sensors monitored, while the stability is the maximum rms and peak-peak measured on the single sensors.}
\label{tab:isoT_test}
\end{table}

\begin{figure}[t] 
\centering
\subfloat{\includegraphics[width=0.67\textwidth]{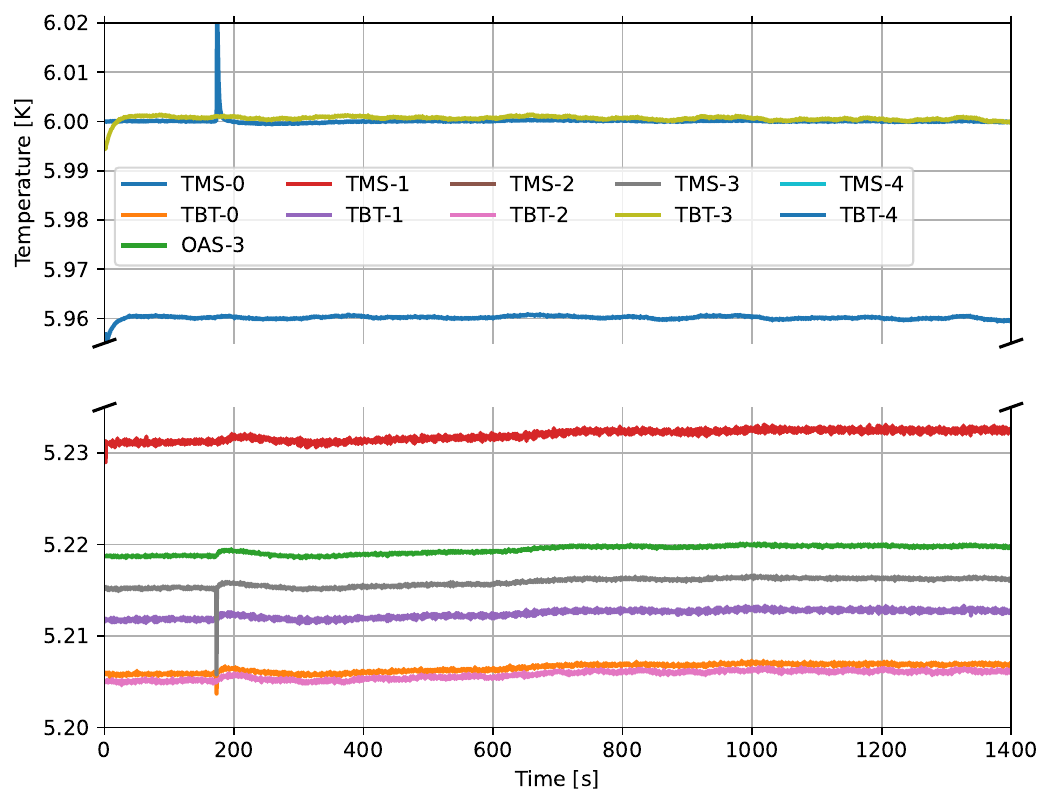}}\\\vspace{-0.45cm}
\subfloat{\includegraphics[width=0.67\textwidth]{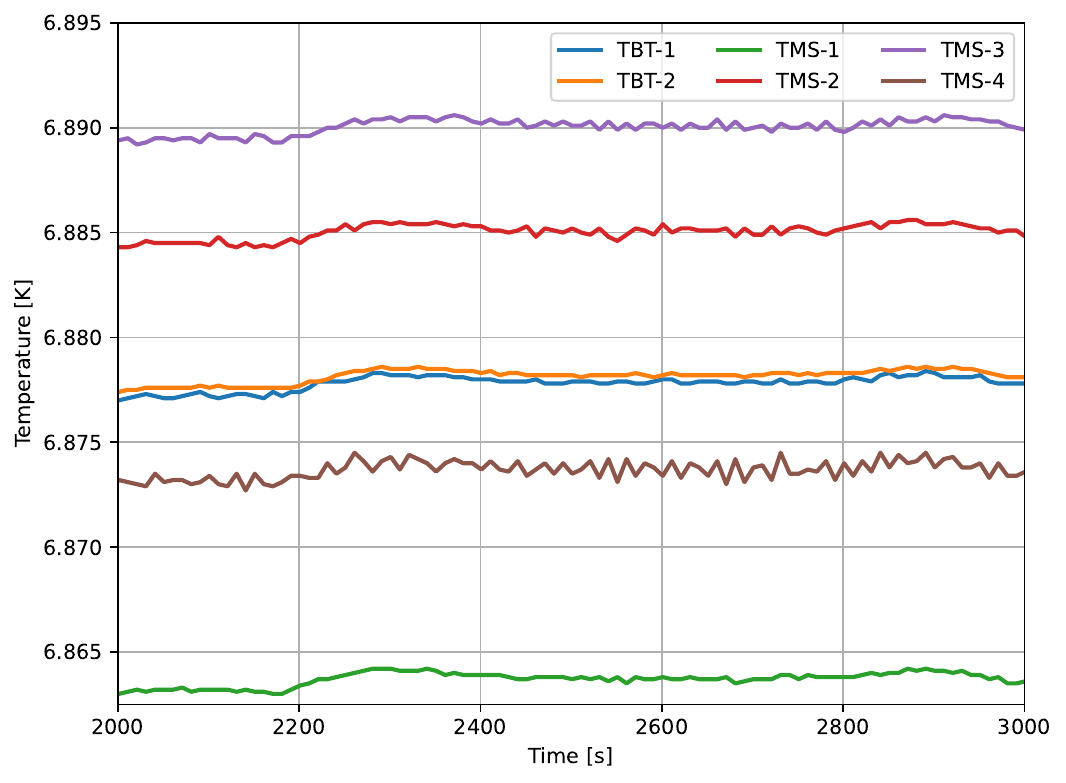}}\\\vspace{-0.45cm}
\subfloat{\includegraphics[width=0.67\textwidth]{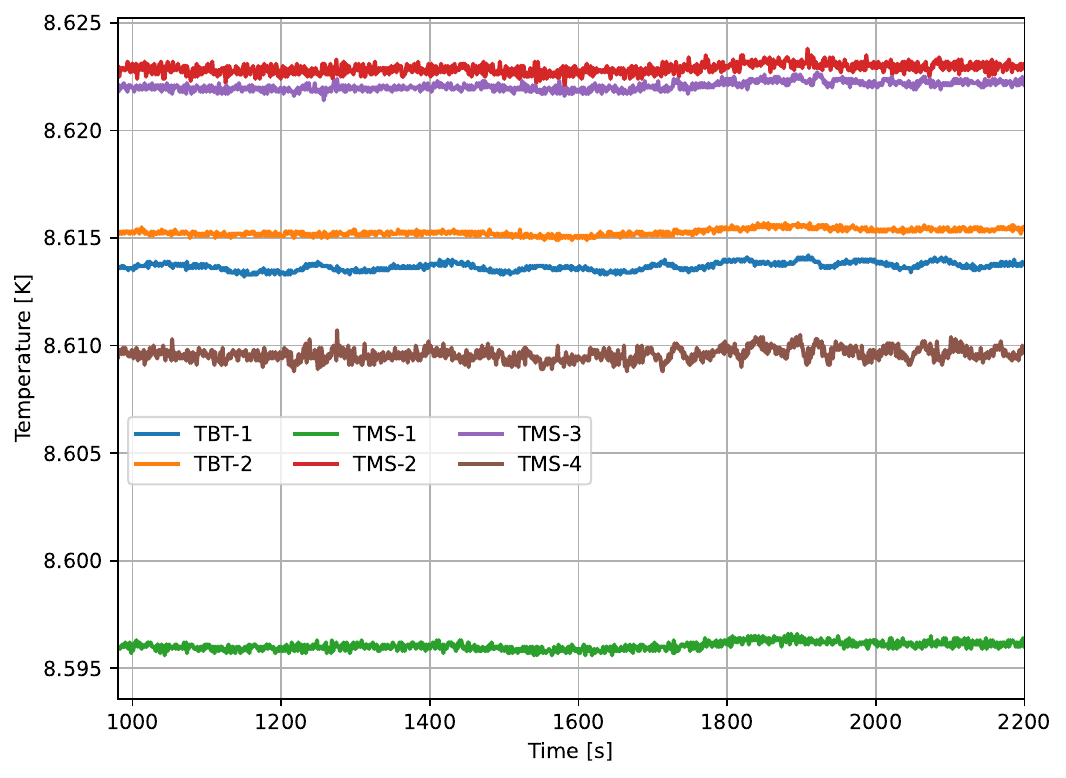}}
\caption{Relevant temperature curves during the test of isothermal states of cold load control flange and radiative shield at 6\,K (\emph{top}), 8\,K (\emph{centre}), and 10K (\emph{bottom}).} 
\label{fig:isoT} 
\end{figure}

The overall gradient (table\,\ref{tab:isoT_test}) over the load is around 25\,mK in the cases measured.
A relevant aspect which has to be noted is that the load reaches a steady temperature which is intermediate between the control flange temperature and the cryostat cold flange (figure\,\ref{fig:isoT}); this is due to the fact that the load is mounted on the cryostat cold flange through four insulating support directly linked by their base (figure\,\ref{fig:thsetup}, \emph{top right}). However the load gradient is kept homogeneous by its high internal conductivity with respect to the thermal conductance of the stainless steel spacers to both the control flange and the cryostat cold flange. 

\FloatBarrier

In order to cross-check the thermal homogeneity in the case of load and radiative environment set at the same temperature, we have exploited the 1\,mK resolution of the sensors to verify that the gradient, passing from one isothermal steady state to the other, is actually stable at the level of mK.
This additional data analysis is done by evaluating the change of the temperature difference between relevant couples of sensors among the different isothermal steady state tests. The couple of sensors chosen are:
\begin{itemize}
    \item \textit{TMS-3, TMS-1}, being the sensors showing the largest gradients in most of the cases (difference identified as $\Delta T_0$),
    \item \textit{TMS-1, TBT-1}, which monitor the gradient across one single side pyramid (difference identified as $\Delta T_1$), and
    \item \textit{TMS-2 – TBT-2}, which  monitor the gradient across one single central pyramid (difference identified as $\Delta T_2$).
\end{itemize}
Results of this double check analysis are reported in table\,\ref{tab:isoT_check}.

\begin{figure}[tb]
\centering
\includegraphics[width=.75\textwidth]{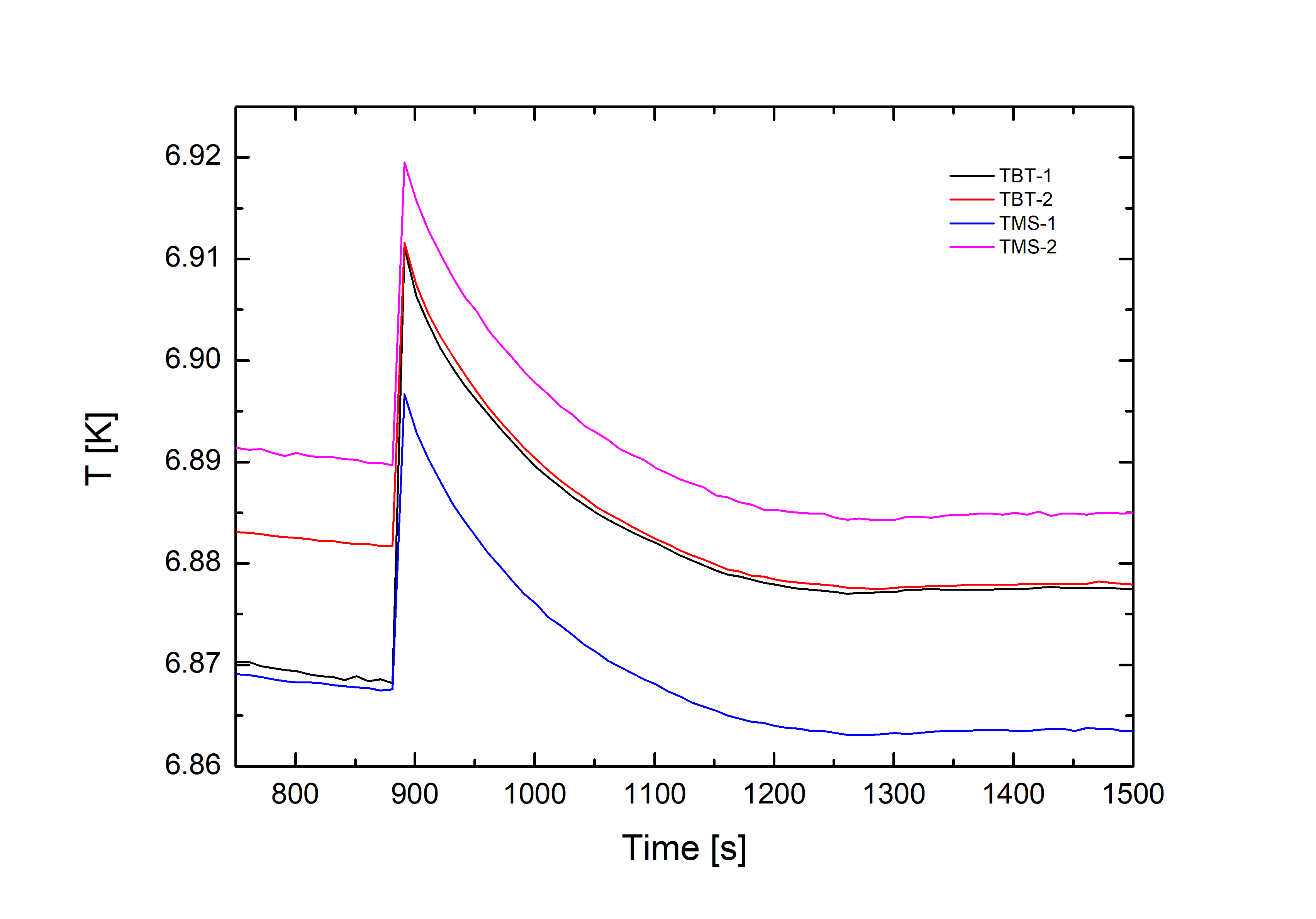}
\caption{Spike occurred during stabilisation of the isothermal test case at 8\,K, corresponding to a control power small drop, and affecting all sensors. The TBT-1 sensor ( in \emph{black} colour), running close to TMS-1 sensor (\emph{blue} colour) before the glitch, was finally running more than 10\,mK higher after the glitch, close to the TBT-2 sensor (\emph{red} colour), while differences among all other sensors were kept as before.}
\label{fig:TBT1_glitch}
\end{figure}

\begin{table}[ht]
\centering
\resizebox{0.7\textwidth}{!}{ 
\begin{tabular}{l c}
\hline
\textbf{Sensor couple and test cases}	& \textbf{Gradient variation [mK]}\\ \hline
|$\Delta T_0$\,(6K) - $\Delta T_0$\,(7K)| & 1.0 \\
|$\Delta T_0$\,(7K) - $\Delta T_0$\,(8K)| & 1.0 \\
|$\Delta T_0$\,(8K) - $\Delta T_0$\,(10K)| & 1.0 \\ \hline
|$\Delta T_1$\,(6K) - $\Delta T_1$\,(7K)| & 16.0 \\
|$\Delta T_1$\,(7K) - $\Delta T_1$(8K)| & 1.0 \\
|$\Delta T_1$\,(8K) - $\Delta T_1$\,(10K)| & 3.0 \\ \hline
|$\Delta T_2$\,(6K) - $\Delta T_2$\,(7K)| & 3.0 \\
|$\Delta T_2$\,(7K) - $\Delta T_2$\,(8K)| & 0.0 \\
|$\Delta T_2$\,(8K) - $\Delta T_2$\,(10K)| & 1.0 \\ \hline
\end{tabular} 
}
\caption{Variation of the gradient between couples of sensors through the different isothermal steady state test cases.}
\label{tab:isoT_check}
\end{table}

Results show that the variation of the gradient is at mK level, confirming that in these test cases the load temperature is homogeneous, as expected; the only outlying value is related to the difference between TBT-1 and TMS-1 in the isothermal test case at 6\,K. This is possibly due to a step change occurred in the TBT-1 sensor during the second day of measurement at cold, after the 6\,K test reference case was already performed (see figure\, \ref{fig:TBT1_glitch}). Indeed, a different value was found when we repeated the same test, in which difference between TBT-2 and TMS-2 was quite constant (see table \ref{tab:multi_6K}).

\begin{table}[ht]
\centering
\resizebox{0.6\textwidth}{!}{ 
\begin{tabular}{l c c}
\hline
\textbf{Isothermal 6\,K test date} & \textbf{$\Delta T_1$ [mK]} & \textbf{$\Delta T_2$ [mK]}\\ \hline
 2022/01/17 & -1.0 & 4.0\\
 2022/01/24 & 9.0 & 5.0\\
 2022/01/26 & 9.0 & 5.0\\
 2022/01/27 & 9.0 & 5.0 \\
 \hline
\end{tabular} 
}
\caption{Temperature difference measured between top and base of the pyramids during isothermal test case at 6\,K, repeated at different times.}
\label{tab:multi_6K}
\end{table}

A second set of steady state tests has been performed by stepping the radiative shield temperatures at 10, 20 and 40\,K, while keeping the temperature of the load flange controlled at 7\,K.
The figures of merit chosen for the test results are the global thermal gradient across the load and the gradient along the single pyramids monitored, i.e., the side pyramid monitored by the sensor couple \textit{TMS-1, TBT-1} and the central pyramid monitored by \textit{TMS-2 – TBT-2}. Due to the large height-to-base ratio ($\sim$6) in size of the pyramids, they represent the section of the load which should have larger impact from a warmer surrounding environment. 
Results of this set of measurements (table\,\ref{tab:RadSteps_test}) confirmed the hypothesis that a measurable effect of thermal imbalance between the load and its surrounding environment is obtained with a temperature difference of more than 20\,K. Then the thermal homogeneity of the cold load is conserved for a large range of temperatures of the feedhorn facing it during nominal operations.

\begin{table}[t]
\centering
\resizebox{0.7\textwidth}{!}{ 
\begin{tabular}{l c c c}
\hline
\textbf{Shield T} & \textbf{Overall} & \textbf{Side pyramid} & \textbf{Central pyramid} \\
\textbf{[K]} & \textbf{Gradient [mK]} & \textbf{Gradient [mK]} & \textbf{Gradient [mK]} \\ \hline
7.0  & 25 $\pm$ 30 & 16 $\pm$ 30 &  -6 $\pm$ 30 \\
10.0 & 23 $\pm$ 30 & 15 $\pm$ 30 & -5 $\pm$ 30 \\
20.0 & 35 $\pm$ 30 & 22 $\pm$ 30 & -8 $\pm$ 30 \\
40.0 & 66 $\pm$ 30 & 51 $\pm$ 30 & -6 $\pm$ 30 \\ \hline
\end{tabular} 
}
\caption{Summary of the results of the steady state test with radiative shield temperature steps.}
\label{tab:RadSteps_test}
\end{table}

\begin{figure}[ht] 
\centering
\subfloat{\includegraphics[width=0.65\textwidth]{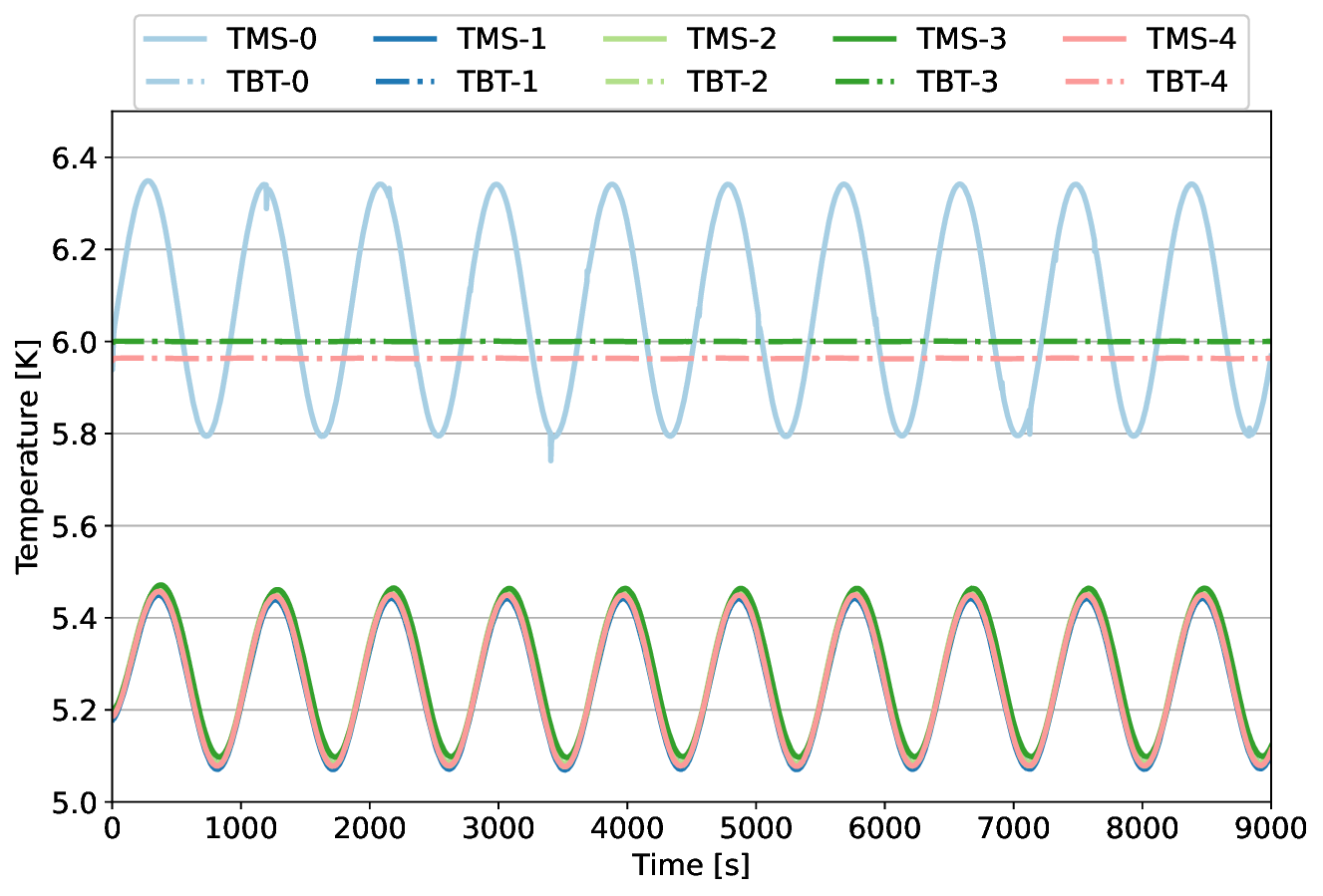}}
\vspace{-0.3cm}
\subfloat{\includegraphics[width=0.65\textwidth]{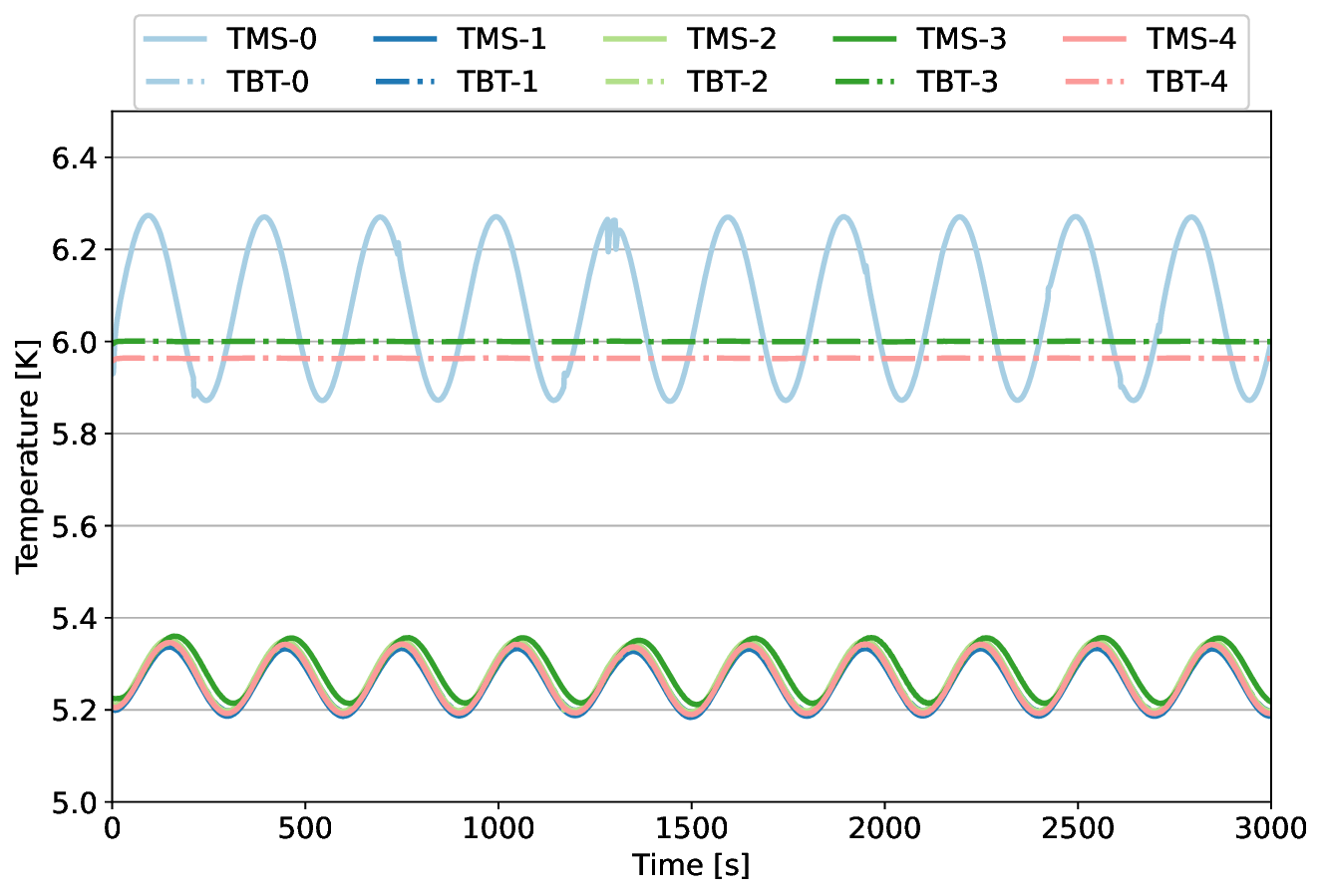}}
\vspace{-0.3cm}
\subfloat{\includegraphics[width=0.65\textwidth]{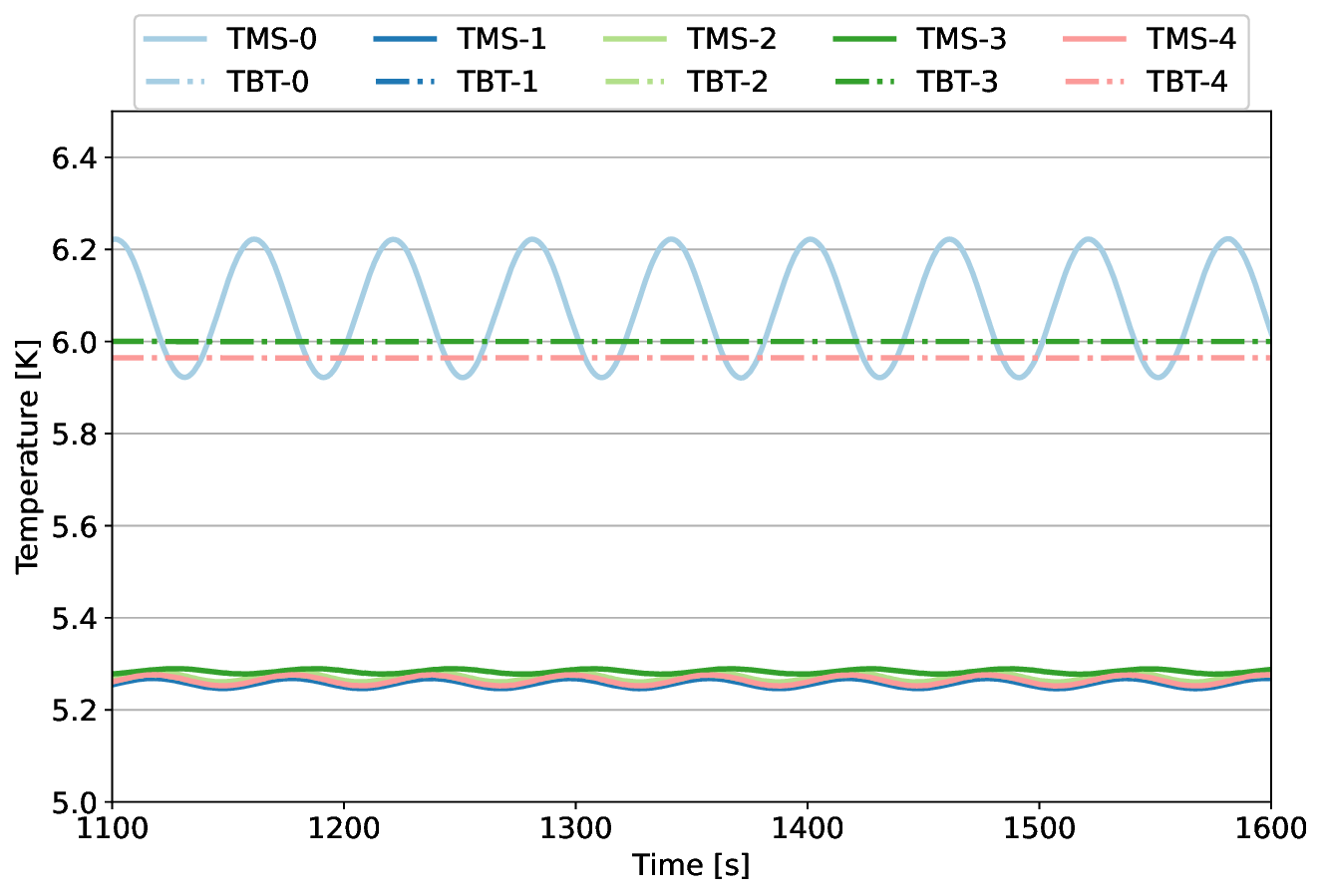}}
\caption{Relevant temperature curves during transient test with oscillations of period 900\,s (\emph{top}), 300\,s (\emph{centre}), and 60\,s (\emph{bottom}); the plot \textit{y} scale is left unchanged for all the plots in order to evidence the different damping effects of the thermo-mechanical structure on the temperature fluctuations propagation.} 
\label{fig:TransentT} 
\end{figure}

The transient test planned on the device is mainly aimed at validating the thermal model, by inducing fluctuations at different timescales on the 4KCL control flange and studying how they propagate in the system, in order to compare the measurement with model predictions. This test is then a useful tool for the thermal model fine tuning. The set of tests consisted in inducing 900, 600, 300, 100, and 60\,s periodic oscillation on the control flange with an amplitude of 0.15 to 0.5\,K, around the nominal temperature of 6.0\,K. The temperature curves at the level of different parts of the load have been fitted with the function $\mathrm{T(t)=T_0+A\sin (2\pi (t-t_0)/T)}$, where $T_0$ is the mean temperature, $A$ is the fluctuations amplitude, $T$ is the oscillation period, and $t_0$ is the phase delay. The output of the data analysis (table\,\ref{tab:Transient_60s}) is the ratio of the fluctuations amplitude at the different load sensors to the source fluctuation amplitude in the control flange $A_i/A_{\rm CTRL}$ (damping) and the corresponding phase difference  $\mathrm{\Delta\phi=2\pi(t_0^i-t_0^{CTRL})/T}$.

\begin{table}[t]
\centering
\resizebox{.8\textwidth}{!}{ 
\begin{tabular}{|c|c|c|c|c|c|c|c|} 
\hline
\textbf{Sensor}  & \multicolumn{2}{|c|}{\textbf{Amplitude}} & \multicolumn{2}{|c|}{\textbf{$t_0$}} & {\textbf{Damping}} & {\textbf{Delay}} & {\textbf{Phase}}\\\cline{2-8} 
\textbf{ID}	&	Value [K]	& Error [mK]	& Value [s]	& Error [s]	& 	& [s]	& [rad]\\\hline
\textbf{OAS-1}	& 0.01122	& 0.02 &	23.83	 &  0.01	  & 0.075	& 17.55	& 1.84\\\hline
\textbf{OAS-2}	& 0.01096	& 0.01 &	22.697	& 0.009	  & 0.073	& 16.41	& 1.72\\\hline
\textbf{TMS-1}	& 0.01100	  & 0.009 &	22.561	& 0.008	  & 0.073	& 16.28	& 1.70\\\hline
\textbf{TMS-2}	& 0.01182	& 0.01 &	22.17	& 0.01	  & 0.079	& 15.89		& 1.66\\\hline
\textbf{TMS-4}	& 0.01190	& 0.01 &	22.66	& 0.01	  & 0.079	& 16.38	& 1.72\\\hline
\textbf{TMS-3}	& 0.00580	& 0.01 &	31.86	& 0.02	  & 0.039	& 25.58	& 2.68\\\hline
\textbf{TMS-0}	& 0.15028	& 0.07 &	6.282	 &  0.005	& 1			& 0					& 0.00\\\hline
\end{tabular}
}
\caption{Summary of the transient test data analysis for the case of 60\,s oscillations.}
\label{tab:Transient_60s}
\end{table}

\begin{table}[t]
\centering
\resizebox{.83\textwidth}{!}{ 
\begin{tabular}{|c|c|c|c|c|c|c|c|c|} 
\hline
\textbf{Sensor}  & \multicolumn{2}{|c|}{\textbf{Period 900\,s}} & \multicolumn{2}{|c|}{\textbf{Period 600\,s}} & \multicolumn{2}{|c|}{\textbf{Period 300\,s}} & \multicolumn{2}{|c|}{\textbf{Period 100\,s}}\\\cline{2-9}
\textbf{ID}	&	Damping	& Phase [rad]	& Damping	& Phase [rad]	& Damping	& Phase [rad]	& Damping	& Phase [rad]\\\hline
\textbf{OAS-1}	& 0.656	& 0.58	& 0.570	& 0.82 &	0.357	& 1.18	& 0.124	& 1.64\\\hline
\textbf{OAS-2}	& 0.656	& 0.58	& 0.572	& 0.81	& 0.359	& 1.16 &	0.127 &	1.59\\\hline
\textbf{TMS-1}	& 0.654	& 0.57	& 0.570	& 0.80	& 0.359	& 1.16 &	0.126 &	1.58\\\hline
\textbf{TMS-2}	& 0.674	& 0.57	& 0.579	& 0.80	& 0.372	& 1.15 &	0.135 &	1.55\\\hline
\textbf{TMS-4}	& 0.664	& 0.57	& 0.575	& 0.80	& 0.364	& 1.16 &	0.126 &	1.59\\\hline
\textbf{TMS-3}	& 0.668	& 0.68	& 0.570	& 0.96	& 0.353	& 1.48 &	0.093 &	2.35\\\hline
\end{tabular}
}
\caption{Summary of the results of the transient test with cold load control flange temperature fluctuations.}
\label{tab:Transient_test}
\end{table}

The transient results (table \ref{tab:Transient_test}) show once again that the main thermal decoupling is between the control flange and the load itself, and its efficiency in enhancing the stability. The fit parameters measured at the different load locations are instead very close to each other, confirming the thermal homogeneity across the load absorbing section. As an exception, there is a measurable delay in the side shield with respect to the main target body, due to the additional thermal resistance at the interface. Indeed, the cylindrical shield is actually screwed to a recess in the aluminium base in eight points, as evident in figure\,\ref{fig:thsetup}, \emph{top left}. 

The thermal balance test has been performed for each of the two different test runs showing results variation within 5\,\%. Considering also a 10\,\% accuracy of the model results, we have set the same level of uncertainty in the correlation between test data and thermal model output, in particular for  the damping factor at the level of the different regions of the Cold Load. Then, by varying the following parameters of the model:

\begin{itemize}
    \item the contact resistance between the load control flange and base, 
    \item the contact resistance between the load base and side cylindrical shield,
    \item the thermal conductivity of the screw supporting the load on the cryostat cold flange,
    \item thermal conductivity and specific heat of the Aluminium, and
    \item thermal conductivity and specific heat of the Eccosorb CR\,117,
\end{itemize}  
we found a set of values that allow the model to reproduce the experimental data at the levels shown in table\,\ref{tab:Model_Corr} for each sensor and oscillation frequency. 

For shortest timescales, an equivalent absolute error in time delay produce a largest relative error in phase, so we accepted a higher relative difference between data and model for the phase values, also considering that the damping factor is the most relevant parameter for the thermal instability evaluation.

We determined the 4KCL thermal time constant by correlating the thermal balance test results with model predictions. For this, we evaluated the RC of the load (lumped capacitance as a whole, and thermal resistance with respect to the control flange). We found a value of about 120\,s, also well correlated with the stabilisation time observed in the different phases of the test campaign.

\begin{table}[t]
\centering
\resizebox{0.8\textwidth}{!}{ 
\begin{tabular}{|c|c|c|c|c|c|c|c|c|c|c|} 
\hline
\textbf{Sensor}  & \multicolumn{2}{|c|}{\textbf{900\,s}} & \multicolumn{2}{|c|}{\textbf{600\,s}} & \multicolumn{2}{|c|}{\textbf{300\,s}} & \multicolumn{2}{|c|}{\textbf{100\,s}} & \multicolumn{2}{|c|}{\textbf{60\,s}}\\\cline{2-11}  
\textbf{ID}	&	$Damping [\%]$	& $\phi [\%]$	&	$Damping [\%]$	& $\phi [\%]$	&	$Damping [\%]$	& $\phi [\%]$	&	$Damping [\%]$	& $\phi [\%]$	&	$Damping [\%]$	& $\phi [\%]$	\\\hline
\textbf{OAS-1}	& 1.2	& 1.8	& 2.5	& 0.5	& 3.7	& 3.8	&	2.3	& 10.2 &	5.3	&  13.3	\\\hline
\textbf{OAS-2}	& 2.4	& 3.9	& 3.7	& 2.0	& 4.0	& 3.3	&	0.7	& 10.4 &	9.7	&  13.6	\\\hline
\textbf{TMS-1}	& 2.1	& 4.0	& 3.5	& 2.0	& 3.8	& 3.4	&	1.7	& 11.6 &	8.5	&  15.6	\\\hline
\textbf{TMS-2}	& 2.5	& 3.5	& 4.9	& 1.5	& 7.7	& 4.2	&	7.0	& 13.6 &	2.6	&  18.8	\\\hline
\textbf{TMS-4}	& 2.6	& 4.0	& 4.5	& 2.0	& 6.9	& 3.4	&	1.3	& 11.9 &	0.2	&  16.2	\\\hline
\textbf{TMS-3}	& 2.0	& 3.7	& 4.5	& 2.0	& 6.4	& 2.2	&	5.9	& 6.6  &	4.4	&  7.9	\\\hline
\end{tabular}
}
\caption{Relative difference of the transient test parameters measured during the test and those estimated from the corresponding model output.}
\label{tab:Model_Corr}
\end{table}

\FloatBarrier

\section{Equivalent Emissivity Model}\label{sec:EEM}

The accuracy of the TMS absolute measurements depends on the behaviour of the 4KCL and on the accuracy of the model with which we characterise it. In the Rayleigh Jeans regime, we can express the brightness temperature of a grey body as a function of its physical temperature $T$ and emissivity $\epsilon$ following eq.\,\ref{eq:Tb}. However, in practice this becomes more complicated, as we have designed the 4KCL based on directionally dependent absorbing geometries, and it is facing a feedhorn, which also collects energy from outside of the solid angle subtended by the target (spillover). The feedhorn itself also presents some ohmic loss and emission. Considering all of this, the feedhorn perceives a brightness temperature that follows eq.\,\ref{eq:pref}. In this section, we ignore the terms in eq.\,\ref{eq:pref} that are not directly related to the calibration target, and focus on obtaining an accurate model of the brightness temperature of the TMS 4KCL.

The frequency-dependent brightness temperature $T_B$ observed from a direction defined by $\vec{k}$, is obtained as 

\begin{equation}\label{eq:Tbcalc}
    T_B\left(\vec{k},\nu\right)=\left(1-r\left(\vec{k},\nu\right)\right)\int_{V}{P_{\rm abs}\left(u,\vec{k},\nu\right) T\left(u\right)\,du}
\end{equation}

where $r(\vec{k},\nu)$ denotes the reflection coefficient of the 4KCL ($\vec{k}$-dependent),   $P_{\rm abs}(u,\vec{k},\nu)$ is the normalised absorbed power density at a point $u$, and $T(u)$ is the physical temperature distribution in said point. The integral is calculated for the entire volume $V$ of the load. 

The evaluation of temperature and dissipated power distributions over the whole volume $V$ of the load can take considerable time.  If we  assume normal incidence and a temperature distribution predominant in the $z$-direction, we can approximate eq.\,\ref{eq:Tb} by
\begin{equation}\label{eq:Tbs}
    T_B\left(\vec{k},\nu\right) \approx \left(1-r\left(\vec{k},\nu\right)\right)\int_{l_z}{P'_{\rm abs}\left(z,\vec{k},\nu\right) T\left(z\right)\,dz},
\end{equation}
where $\mathrm{P'_{abs}(z,\vec{k})}$ is calculated as $\mathrm{P'_{abs}(z,\vec{k})}=\iint_{l_x,l_y}P_{abs}(u,\vec{k})dxdy$. 
The integral is then evaluated numerically, by dividing the volume of the load into $N$ subvolumes, according to height $z_i$:
\begin{equation}\label{eq:Tbsd}
    T_B\left(\vec{k},\nu\right) \approx \left(1-r\left(\vec{k},\nu\right)\right) \sum_{i=0}^{N-1} P'_{\rm abs}\left(z_i,\vec{k},\nu\right) T\left(z_i\right)\Delta z
\end{equation}

We have computed  $T_B$ at different frequencies from the results obtained with similar analysis of the dissipated power and thermal distributions as those described in sections\,\ref{sec:rfdesign} and \ref{sec:thsimulation}, respectively. We have divided the pyramids into 5 vertical partitions $z_i$, and then, we have computed the dissipated power $P_{\rm abs}(z_i,\nu)$ at 10, 12.5, 15, 17.5, and 20\,GHz in every subvolume, considering normal incidence. The simulation has included the TMS feedhorn in its nominal configuration, axis aligned and mouth placed at 1\,mm from the load.

\begin{figure}[t]
    \centering
    \subfloat{\includegraphics[width=0.6\textwidth]{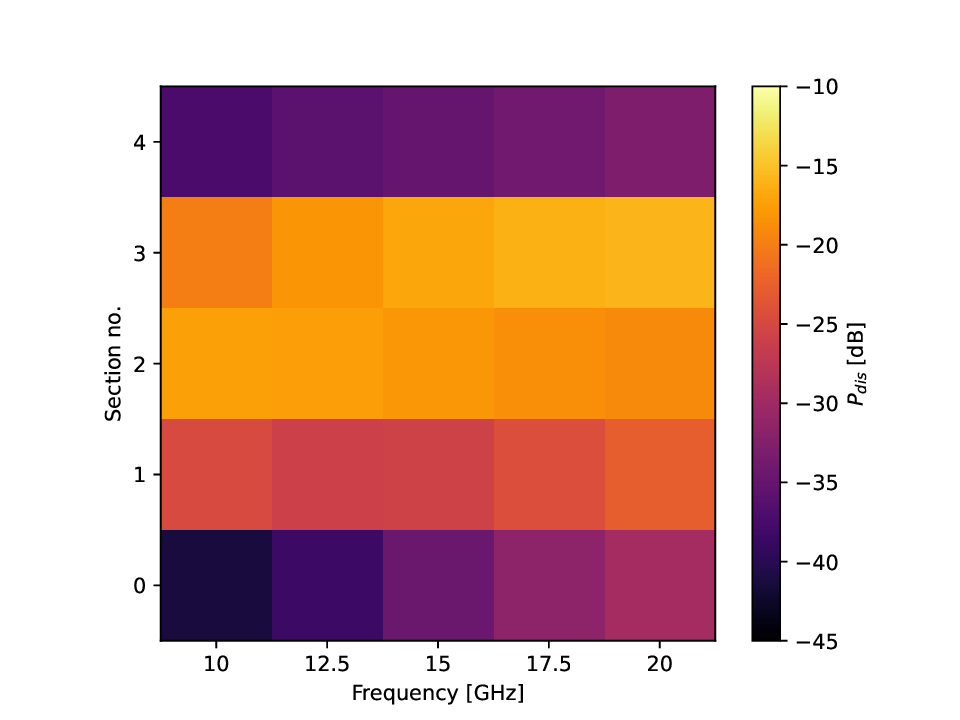}}
    \subfloat{\includegraphics[width=0.25\textwidth]{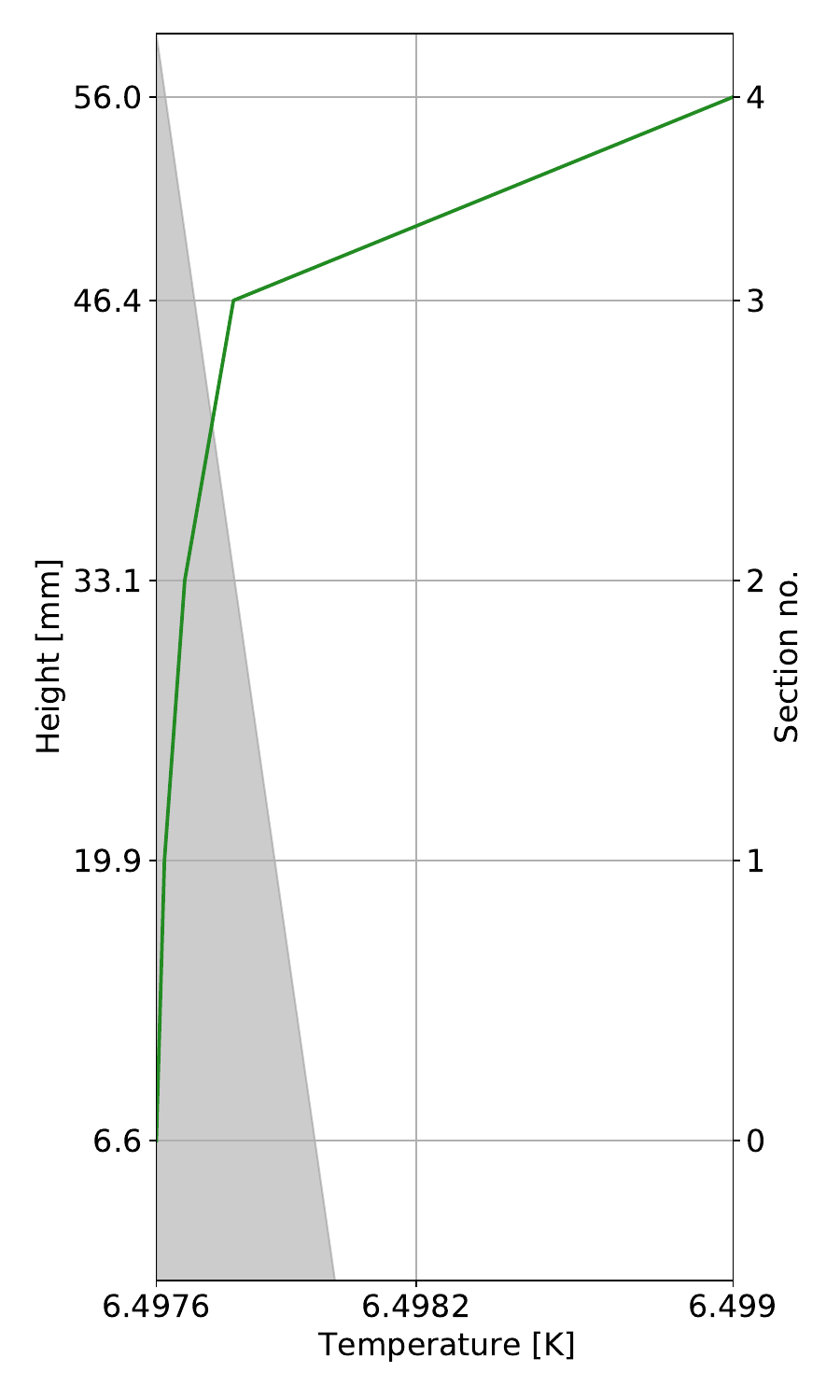}}
    \\
    \subfloat{\includegraphics[width=0.6\textwidth]{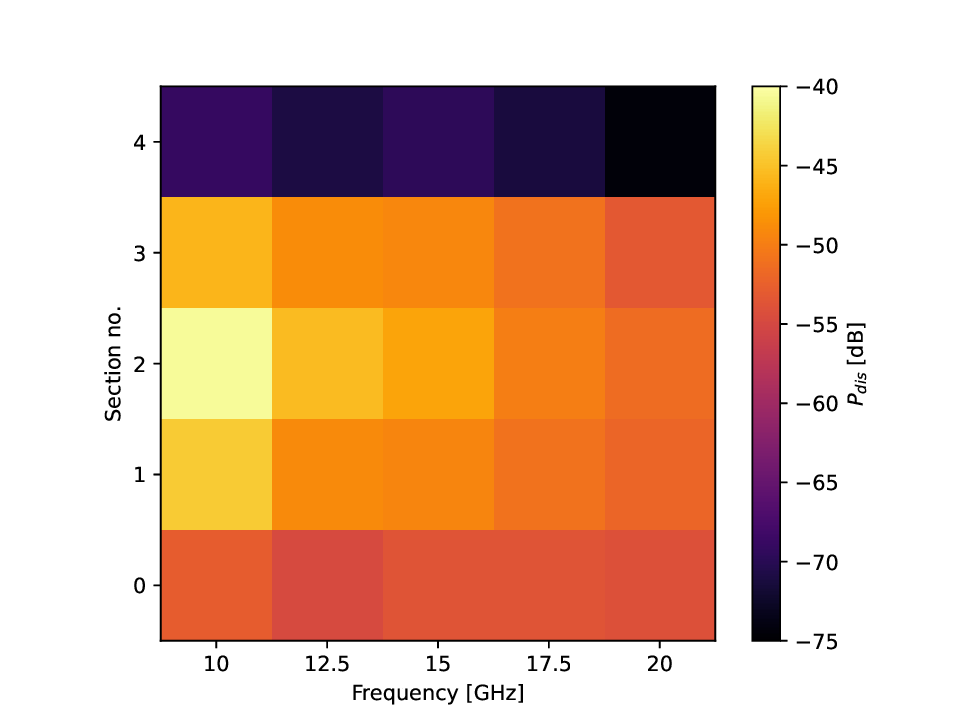}}
    \subfloat{\includegraphics[width=0.25\textwidth]{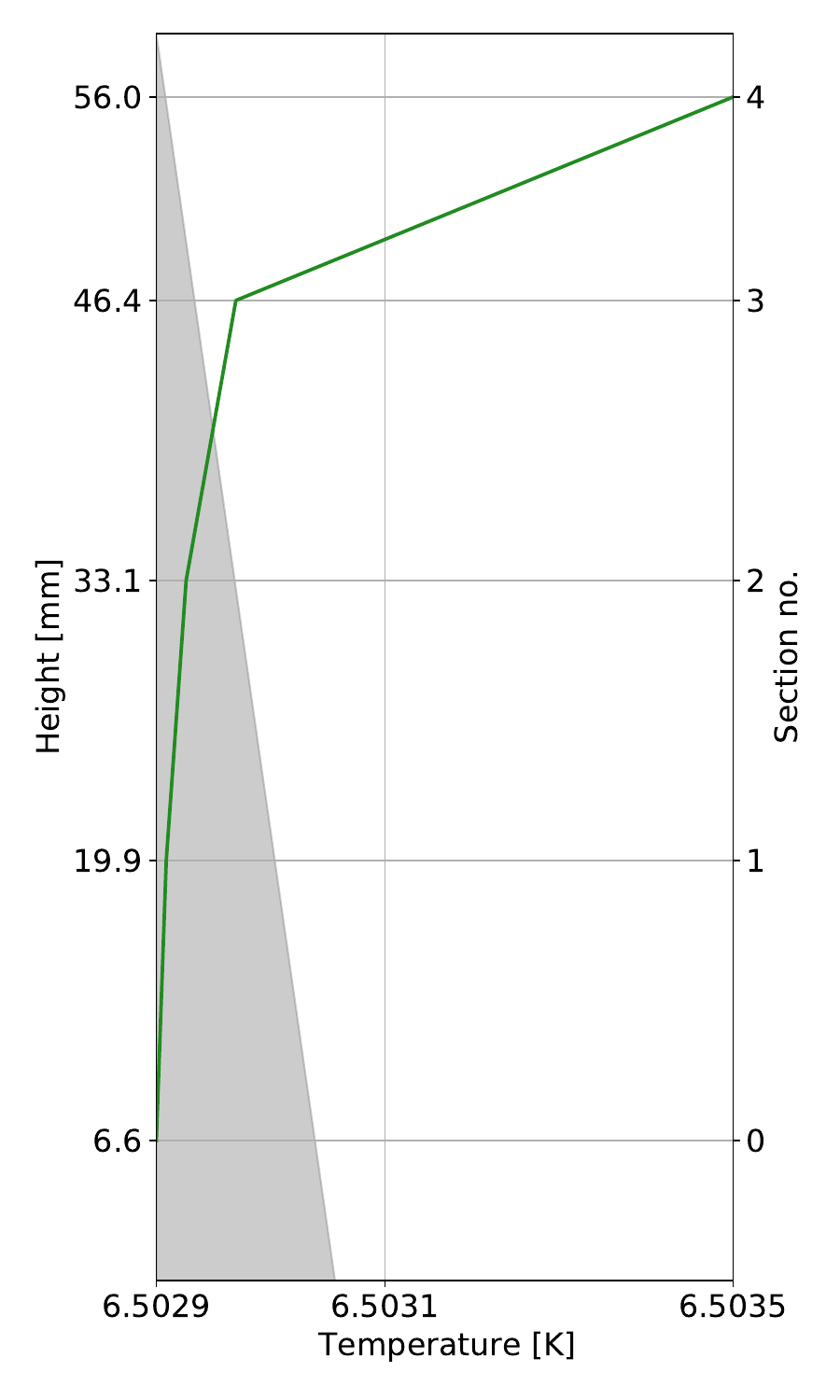}}
    \caption{Dissipated power  $P_{abs}(z_i,\nu)$ (\emph{left column}) and temperature (\emph{right column}) distributions along an inner (\emph{first row}) and outer (\emph{second row}) pyramidal element of the 4KCL. A grey shading representing the pyramid element has been included in the temperature plot.  $P_{abs}(z_i,\nu)$ is shown as a function of frequency $\nu$ (\emph{x-axis}) and height $z_i$ (\emph{y-axis}). The power distribution has been normalised to the total dissipated power (for each frequency).  The power dissipation process occurs mainly in the central part of the pyramid, while the contribution of the pyramid tip and base is quite reduced ($<-$20\,dB). There is a difference greater than 25\,dB between the maximum dissipated power in the inner and outer pyramidal elements. The temperature distribution has been calculated for the worst case of thermal imbalance between the load (6\,K) and the horn (20\,K). The tip represents the warmest point of the pyramidal element, which shows a thermal gradient below 2\,mK and 0.5\,mK for the central and the external element, respectively.}
    \label{fig:distpyr}
\end{figure}

From the thermal model of the 4KCL, we have computed the thermal distribution of the load for different cases, including the homogeneous case (with the load and the environment at the same temperature of 6\,K) and several cases of load/environment temperature imbalances. For the imbalance cases, we have considered horn temperatures of 10, 15, and 20\,K. We obtained significant thermal gradients ($\sim$mK) only for the 20\,K case, which represents the most extreme case of imbalance (worst expected case). Thus, this is the only scenario that we report below.

Figure\,\ref{fig:distpyr} shows a comparison of the dissipated power and the temperature distributions  along a single pyramid in the inner (\emph{top row}) and outer  (\emph{bottom row}) regions of the load. The results for the dissipated power are consistent with the previous results reported in section\,\ref{sec:rfdesign} (pyramidal elements sectioned into 7 parts). The power dissipation process occurs mainly in the central part of the pyramids for the whole frequency range, while the contribution of the pyramid tip and base is quite reduced ($<-$25\,dB). In addition, the power absorbed in the outer pyramids is much lower than in the inner pyramids, consistent with the illumination pattern of the TMS feedhorn (see figures\,\ref{fig:NF-freq10} and \ref{fig:NF-distance44}).

\begin{figure}[t]
    \centering
    \subfloat{\includegraphics[width=0.5\textwidth]{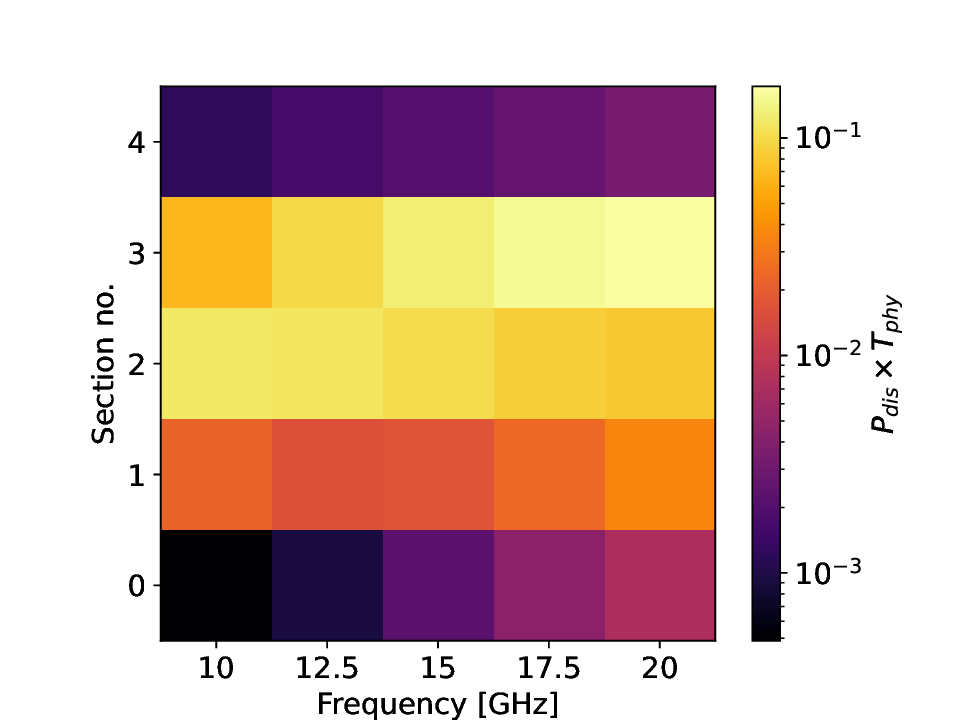}}
    \subfloat{\includegraphics[width=0.5\textwidth]{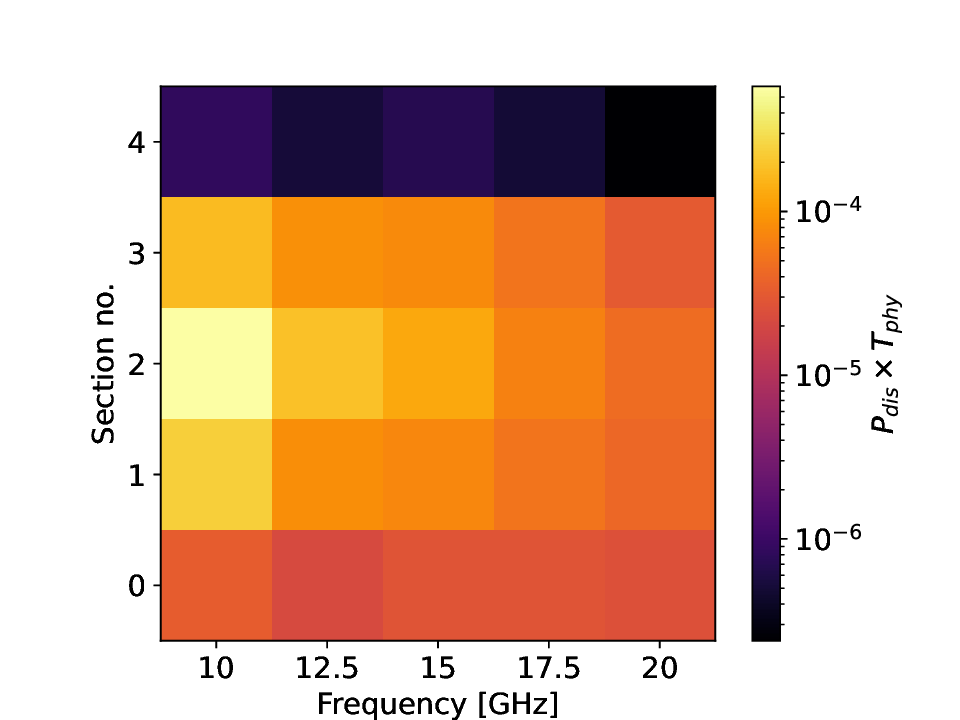}}
    \caption{Element-wise product of the dissipated power distribution $P_{\rm abs}(z_i,\nu)$ and the thermal distribution $T(z_i)$ along the pyramid for the worst case of load-environment imbalance, i.e., for a horn temperature of 20\,K. We have used a logarithmic colour scale to improve the visualisation of the contribution of each section of the pyramids to the total effective temperature $T_{\rm B}$ of the 4KCL. \emph{Left:} Element-wise product $P_{\rm abs}(z_i,\nu)\times T(z_i)$ distribution at the inner pyramid. \emph{Right:} Element-wise product $P_{\rm abs}(z_i,\nu)\times T(z_i)$ distribution at the outer pyramid.}
    \label{fig:Ptempdistpyr}
\end{figure}

Regarding the temperature distribution,  we found  thermal gradients that remain below 2\,mK in the inner pyramids, and below 0.5\,mK in the outer pyramids. However, observing the power distribution, we can assume that the warmest points of the pyramidal elements (the tips)  will have less weight than the intermediate sections in the effective temperature $T_{\rm B}$ of the 4KCL. Indeed, this effect can be seen in figure\,\ref{fig:Ptempdistpyr}, where we gather the element-wise product of the dissipated power and the thermal distributions shown in figure\,\ref{fig:distpyr} along the central (\emph{left}) and external (\emph{right}) pyramidal element.

\begin{figure}[t]
    \centering
\includegraphics[width=0.5\textwidth]{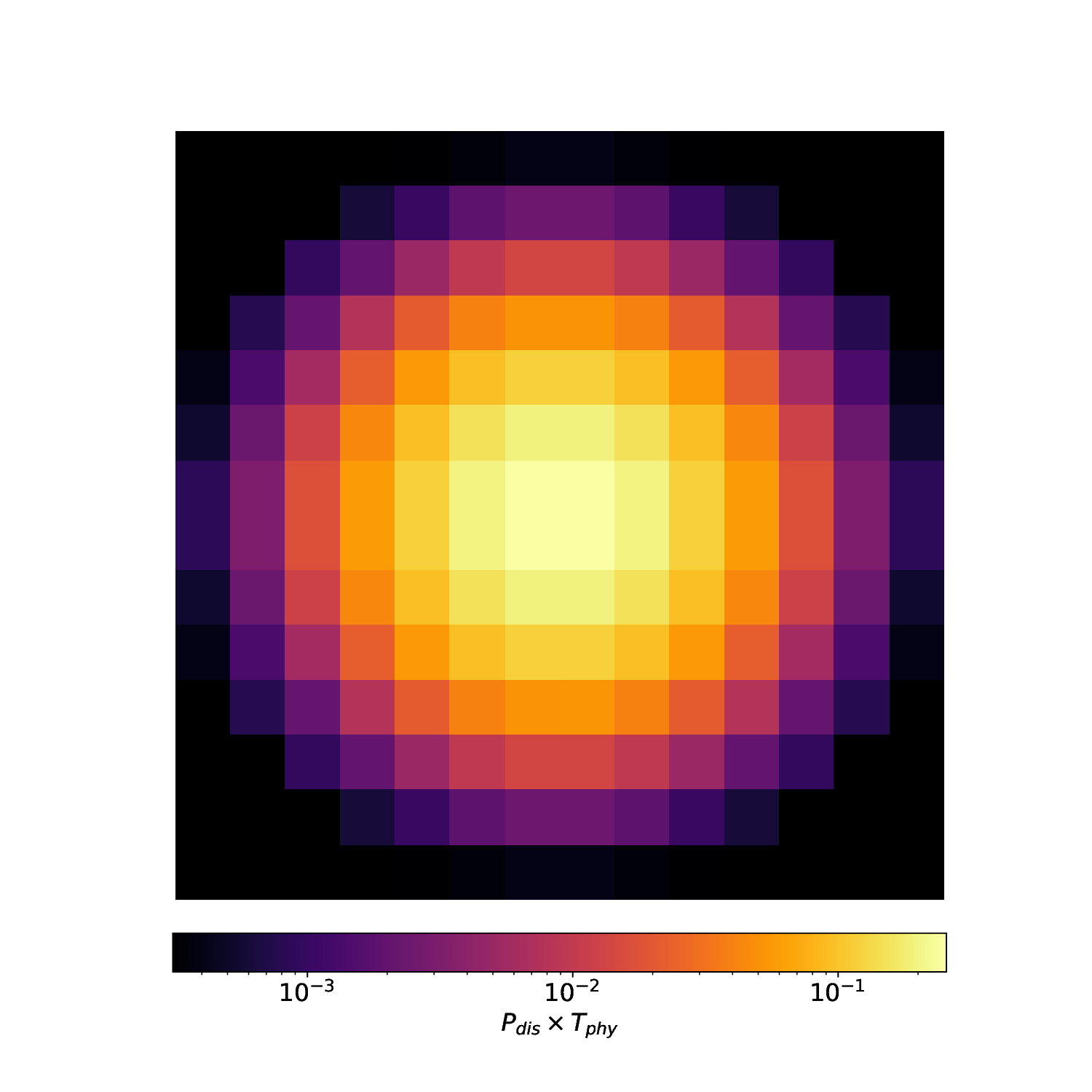}
    \caption{Global view of the element-wise product of the dissipated power distribution $P_{\rm abs}$ and the thermal distribution $T$ in the 4KCL at 15\,GHz for the worst case scenario (horn temperature of 20\,K).} 
    \label{fig:globalThPdis}
\end{figure}

The same effect is observed for the radial temperature distribution of the load. The slight differences in temperature between the central and external pyramids are counterbalanced by the radial power distribution (see figure\,\ref{fig:load-powerprofile-radial}). In figure\,\ref{fig:globalThPdis}, we gather the element-wise product of the dissipated power and the thermal distributions. Indeed, the global temperature of the external pyramids represent a very small contribution to the total effective temperature $T_{\rm B}$ of the load, as observed by the reference horn.

Considering the above, we can draw the following two conclusions: firstly, the system is very robust against systematic errors arising from the generation of thermal gradients when a thermal imbalance between the load and its environment occurs. Secondly, we can take the sensor at the base of the central pyramid as a reference, as in the radial temperature distribution the central pyramids have the most important contribution.

\begin{figure}[ht]
    \centering
    \includegraphics[width=0.8\textwidth]{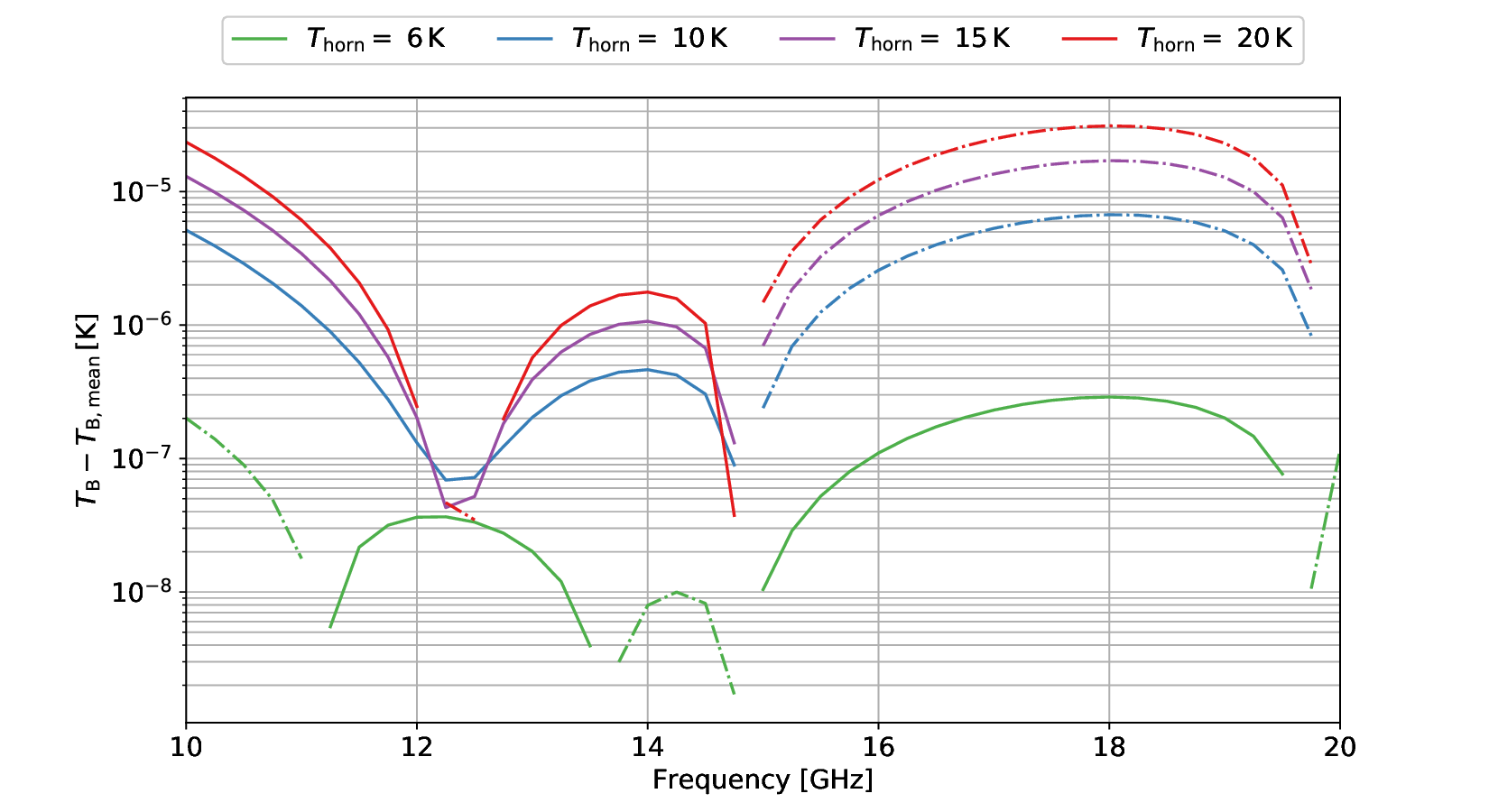}   
    \caption{Evaluation of the effective temperature $T_{\rm B}$ (eq.\,\ref{eq:Tb}) of the load as a function of horn temperature (K) within the TMS frequency range (GHz, \emph{x-axis}). For the data correlation, we have considered horn temperatures between 6 (\emph{case of load/ambient thermal balance}) and 20\,K. The results have been normalised to the mean value of $T_{\rm B}$ for each horn temperature considered, and a logarithmic colormap scale has been used in order to enhance small variations which are not otherwise observable. Positive and negative values are shown with \emph{solid} and \emph{dashed} lines, respectively. Mean values and peak-to-peak differences are gathered in Table\,\ref{tab:Teffpp}.}
    \label{fig:Teff}
\end{figure}

\begin{table}[]
\centering
\resizebox{0.6\textwidth}{!}{ 
\begin{tabular}{ccc}
\hline
\textbf{Horn temperature imbalance (K)} &
  \textbf{$\rm T_{ eff, mean}$  (K)} &
  \textbf{$\rm T_{eff, pp}$ ($\rm \mu$K)} \\\hline\hline
6                    & 6.196                & 0.474                \\
10                    & 6.281                & 11.434               \\
15                    & 6.390                & 29.022               \\
20                   & 6.502                & 52.769               \\
\hline
\end{tabular} 
}
\caption{Summary of the results obtained for the different scenarios of thermal imbalance (the first case, with the horn temperature at 6\,K, being the ideal case of thermal balance). For each case, we have included the mean value of effective temperature as seen by the feedhorn ($T_{\rm eff,mean}$) and the peak-to-peak difference within the 10--20\,GHz range ($T_{\rm eff,pp}$).}
\label{tab:Teffpp}
\end{table}

Finally, we present in figure\,\ref{fig:Teff} a visualisation that comprehensively summarises the frequency behaviour of the 4KCL, for the different thermal balance cases studied. We compare the $T_{\rm B}$ obtained from eq.\,\ref{eq:Tb} in the TMS band for the different expected scenarios of load/ambient thermal unbalance, starting from the ideal thermal balance situation (with the load and the horn at the same temperature of 6\,K). Mean values for each scenario have been removed. With this visualisation, we can see the small temperature variations within the bandwidth, which degrade the behaviour of the 4KCL with respect to the perfect blackbody. In nominal operation, the brightness temperature follows that of a blackbody with accuracy better than 0.5\,$\mathrm{\mu K}$. Thermal imbalance worsens the quality of the blackbody, increasing the degradation at high frequencies (up to $\mathrm{\sim 50\,\mu K}$ at horn temperatures of 20\,K). Mean values of the brightness temperature and peak-to-peak differences within the TMS bandwidth are gathered in Table\,\ref{tab:Teffpp}.

The EEM model presented in this section is of reasonable good quality and reliability, taking into account that it has been developed on the basis of laboratory measurements. Therefore, the uncertainty of the EEM model is mainly due to the uncertainty of the model fit that reproduce the thermal and RF experimental data, and to instrumental errors. Taking into account the close agreement between the load model and the experimental data, we can, thus, assume that in the worst case, instrumental accuracy (15--20\,mK) fully dominates the accuracy of the EEM model.

\section{Conclusions}\label{sec:conclusions}

We have presented the design, characterisation and test results of the 4\,K Cold Load subsystem for the Tenerife Microwave Spectrometer. The 4KCL consists of a bed of pyramids growing on an Aluminium baseplate of 164\,mm  diameter. The pyramids have height to width ratio of approximately 6: they consist of a thin layer of ECCOSORB CR\,117 absorber coated on pyramidal Aluminium cores. A coaxial radiation Aluminium shield, internally black-painted, surrounds the bed of pyramids to enhance emissivity and thermal conductivity.  The selected geometrical structure --- in combination with a careful selection of the materials --- ensure the absorptivity, and thus its emissivity, and the thermal homogeneity of the target.

The RF verification of the 4KCL entailed the measurement of the 4KCL specular and diffusive reflectivity (spillover).
We reported values of specular RL much better than the required $-$30\,dB over the frequency range between 8--24\,GHz, and also better than the $-$40\,dB design goal for most of the TMS band (10--20\,GHz). The reflectivity of the 4KCL was also characterised using the TMS nominal metamaterial feedhorn to simulate a similar experimental condition as that of the TMS operation, obtaining even better results for the RL. The tolerance of the design to non ideal alignment was tested measuring the Return Loss susceptibility to translational (up to $<$12\,mm ) and angular (tilts $>$5\textdegree) displacements.   
The effective spillover has been calculated from the diffusive reflectivity in the whole frequency range over the full sphere, and has been found to be always lower than -30\,dB, and lower than -40\,dB for frequencies above 13\,GHz. Spillover is hence even more negligible considering the effective operational conditions and the small mechanical gap between the shield and the feedhorn.

The thermal verification of the 4KCL entailed  20 thermal cycles on a prototype load of smaller size and 3 extra cycles on the nominal 4KCL, to assess tolerance against thermal stress; a thermal balance test was dedicated to validate the design and verify the predicted thermal model. After the 20 thermal cycles between 4\,K and room ambient temperature, the visual inspection of the dummy did not reveal any anomalies. The same is true after the nominal 4KCL underwent three complete thermal cycles. In addition, steady state tests allowed to characterise the temperature distribution of the load. We reported overall gradient over the load of about 25\,mK in operational conditions, in compliance with the technical requirements.  Thermal imbalance tests and transient tests demonstrated that the 4KCL thermal homogeneity is preserved  even in temperature imbalance conditions, between the load and the feedhorn --- up to 20\,K --- or in the presence of temperature fluctuations.

The 4KCL is one of the most critical components of an experiment like TMS, whose goal is absolute measurements of the sky's brightness temperature. 
In this work, we have therefore presented an equivalent emissivity model for the 4KCL subsystem, developed combining laboratory measurements and simulations, for different operational scenarios of thermal imbalance between antenna, environment, and the 4KCL. This allowed us to demonstrate the excellent behaviour of the 4KCL, even in the most extreme cases of thermal imbalance (up to 14\,K).

Several activities related to the 4KCL subsystem remain to be completed before the commissioning of the TMS experiment, including its integration and functional verification. We intend to reproduce the INAF-OAS  thermo-mechanical verification tests results in the TMS facilities at the IAC. In addition, we will setup a fine-tuning of the radiometric model of the TMS integrated instrument in its nominal operating conditions.

\acknowledgments

 The TMS experiment is being developed by the Instituto de Astrofisica de Canarias (IAC), with an instrumental participation from the INAF group (Bologna, Italy), the University of Milano (Italy), and the Universidad Politecnica de Cartagena (UPCT). Partial financial support is provided by the Spanish Ministry of Science and Innovation (MICINN), under the projects AYA2017-84185-P, IACA15-BE-3707, EQC2018-004918-P, PID2020-120514GB-I00, ICT2022-007828, and the FEDER Agreement INSIDE-OOCC (ICTS-2019-03-IAC-12). We also acknowledge financial support of  the Severo Ochoa Programs SEV-2015-0548 and CEX2019-000920-S.  We acknowledge the support of all the technicians, engineers, scientists and administrative staff of the IAC and QUIJOTE experiment.

\bibliographystyle{unsrt}
\bibliography{biblio}

\end{document}